\documentclass[11pt]{article}
\pdfoutput=1
\usepackage{epsfig}
\usepackage[usenames,dvipsnames,svgnames,table]{xcolor}
\usepackage[normalem]{ulem}
\usepackage{amsmath,amssymb}
\usepackage{arydshln}
\newcommand{\qcd}{QCD}
\newcommand{\sm}{SM}
\newcommand{\mssm}{MSSM}
\newcommand{\nnlo}{NNLO}

\newcommand{\lhc}{LHC}
\newcommand{\api}{\frac{\alpha_s}{\pi}}

\newcommand{\mgluino}{m_{\tilde g}}
\newcommand{\mtop}{m_t}
\newcommand{\msquark}{m_{\tilde q}}
\newcounter{notecount}
\def\wt{\widetilde}

\def\be{\begin{equation}}
\def\ee{\end{equation}}
\newcommand{\bea}{\begin{eqnarray}}
\newcommand{\eea}{\end{eqnarray}}
\newcommand{\bdm}{\begin{displaymath}}
\newcommand{\edm}{\end{displaymath}}
\long\def\symbolfootnote[#1]#2{\begingroup%
\def\thefootnote{\fnsymbol{footnote}}\footnote[#1]{#2}\endgroup}

\def\WG{LHC-HXSWG}

\def\sq2{\sqrt{2}}
\def\drbar{\overline{\rm DR}}
\def\msbar{\overline{\rm MS}}

\def\smallsm{\scriptscriptstyle{\rm SM}}
\def\tb{\tan\beta}

\def\gl{\tilde{g}}
\def\mg{m_{\gl}}

\def\x1g{x_{1}}

\def\FH{{\tt FeynHiggs}}
\def\sushi{{\tt SusHi}}

\def\PB{{\tt POWHEG BOX}}

\def\hmass{125.5}
\def\lss{{\em light-stop} scenario}
\providecommand{\lsim}
{\;\raisebox{-.3em}{$\stackrel{\displaystyle <}{\sim}$}\;}
\providecommand{\gsim}
{\;\raisebox{-.3em}{$\stackrel{\displaystyle >}{\sim}$}\;}

\def\matbp{$\ma$--$\,\tb$ plane}

\newcommand{\susy}{SUSY}

\def\muzeroF{\bar{\mu}_{\scriptscriptstyle F}}
\def\muzeroR{\bar{\mu}_{\scriptscriptstyle R}}
\def\Nrep{N_{\!\scriptscriptstyle R}}
\def\pdfas{PDF+$\alpha_s$}

\newcommand{\as}{\alpha_s}
\newcommand{\oas}{{\cal O}(\as)}
\newcommand{\smallz}{{\scriptscriptstyle Z}} 
\newcommand{\smallw}{{\scriptscriptstyle W}} %
\newcommand{\smallH}{{\scriptscriptstyle H}} %
\newcommand{\smallr}{{\scriptscriptstyle R}} %
\newcommand{\smalla}{{\scriptscriptstyle A}} %
\newcommand{\mz}{m_\smallz}
\newcommand{\mw}{m_\smallw}
\newcommand{\mh}{m_h}
\newcommand{\mH}{m_\smallH}
\newcommand{\ma}{m_\smalla}
\newcommand{\muF}{\mu_{\scriptscriptstyle F}}

\newcommand{\muR}{\mu_\smallr}

\newcommand{\msusy}{M_S}

\def\mt{m_t}
\def\stu{\tilde{t}_1}
\def\std{\tilde{t}_2}

\def\mb{m_b}
\def\mbp{M_b}
\def\Db{\Delta_b}
\def\yb{Y_b}
\def\ybp{\yb^\phi}
\def\ybeff{\wt \yb}
\def\ybeffp{\ybeff^{\,\phi}}
\def\sbu{\tilde{b}_1}

\def\bu{m_{\tilde{b}_1}^2}
\def\bd{m_{\tilde{b}_2}^2}

\begin{document}

\begin{titlepage}


\mbox{}
\vspace*{-2cm}

{\flushright{
        \begin{minipage}{5.1cm}
          CERN-PH-TH/2014-048,\\ 
          DESY 14-040, IFUM 1024-FT,\\
          LPN14-061, MCnet-14-07,\\
          WUB/14-01	
        \end{minipage}        }

}
\renewcommand{\thefootnote}{\fnsymbol{footnote}}
\vskip 1.5cm
\begin{center}
{\LARGE\bf Towards precise predictions \\ [7pt]
for Higgs-boson production in the MSSM}
\vskip 1.0cm
{\Large  E.~Bagnaschi$^{\,a,b}$, R.V.~Harlander$^{\,c}$, S.~Liebler$^{\,d}$, 
H.~Mantler$^{\,e}$, P.~Slavich$^{\,a,b}$ \\ [7pt]
and A.~Vicini$^{\,f}$}
\vspace*{8mm} \\ {\sl ${}^a$ LPTHE, UPMC Univ.~Paris 06, 
  Sorbonne Universit\'es, 4 Place Jussieu, F-75252 Paris, France}
\vspace*{2mm}\\
{\sl ${}^b$ LPTHE, CNRS, 4 Place Jussieu, F-75252 Paris, France }
\vspace*{2mm}\\
{\sl ${}^c$ Fachbereich C, Bergische Universit\"at Wuppertal,
  Gau\ss stra\ss e 20, D-42119 Wuppertal, Germany}
\vspace*{2mm}\\ {\sl ${}^d$ Universit\"at Hamburg, Luruper Chaussee
  149, D-22761 Hamburg, Germany}
\vspace*{2mm}\\{\sl ${}^e$ CERN, Theory Division, CH-1211 Geneva 23,
  Switzerland}
\vspace*{2mm}\\
{\sl ${}^f$
    Dipartimento di Fisica, Universit\`a di Milano and
    INFN, Sezione di Milano,\\
    Via Celoria 16, I--20133 Milano, Italy}
\end{center}
\symbolfootnote[0]{{\tt e-mail addresses:}} 
\symbolfootnote[0]{{\tt bagnaschi@lpthe.jussieu.fr,
    robert.harlander@uni-wuppertal.de, stefan.liebler@desy.de,}}
\symbolfootnote[0]{{\tt hendrik.mantler@cern.ch,
    slavich@lpthe.jussieu.fr, alessandro.vicini@mi.infn.it}}

\vskip 1cm

\begin{abstract}
We study the production of scalar and pseudoscalar Higgs bosons via
gluon fusion and bottom-quark annihilation in the MSSM. Relying on the
NNLO-QCD calculation implemented in the public code \sushi, we provide
precise predictions for the Higgs-production cross section in six
benchmark scenarios compatible with the LHC searches. We also provide
a detailed discussion of the sources of theoretical uncertainty in our
calculation. We examine the dependence of the cross section on the
renormalization and factorization scales, on the precise definition of
the Higgs-bottom coupling and on the choice of PDFs, as well as the
uncertainties associated to our incomplete knowledge of the SUSY
contributions through NNLO. In particular, a potentially large
uncertainty originates from uncomputed higher-order QCD corrections to
the bottom-quark contributions to gluon fusion.
\end{abstract}
\vfill
\end{titlepage}    
\setcounter{footnote}{0}


\section{Introduction}

The recent discovery of a Higgs boson with mass around \hmass\ GeV by
the ATLAS and CMS experiments at the Large Hadron Collider
(LHC)~\cite{ATLASHiggs,CMSHiggs} puts new emphasis on the need for
precise theoretical predictions for Higgs production and decay rates,
both in the Standard Model (SM) and in plausible extensions of the
latter such as the Minimal Supersymmetric Standard Model (MSSM). The
current status of these calculations is summarized in the reports of
the LHC Higgs Cross Section Working Group
(\WG)~\cite{Dittmaier:2011ti,Dittmaier:2012vm,Heinemeyer:2013tqa}.

In the SM, the main mechanism for Higgs production at hadron colliders
is gluon fusion \cite{H2gQCD0}, where the coupling of the gluons to
the Higgs is mediated by loops of heavy quarks, primarily top and
bottom.  The knowledge of this process includes: the next-to-leading
order (NLO) QCD contributions~\cite{H2gQCD1} computed for arbitrary
values of the Higgs and quark masses~\cite{SDGZ,HK,babis1,ABDV}; the
next-to-next-to-leading order (NNLO) QCD contributions due to
top-quark loops, in the heavy-top
limit~\cite{H2gQCD2,Anastasiou:2002yz} and including finite top-mass
effects~\cite{H2gQCD3}; soft-gluon resummation effects \cite{softglu}
and estimates of the next-to-next-to-next-to-leading order (NNNLO) QCD
contributions~\cite{nnnlo}; the first-order electroweak (EW)
contributions
\cite{DjG,ABDV0,Degrassi:2004mx,APSU,oai:arXiv.org:1007.1891} and
estimates of the mixed QCD-EW contributions~\cite{qcdew}.

The Higgs sector of the MSSM consists of two $SU(2)$ doublets, $H_1$
and $H_2$, whose relative contribution to electroweak symmetry
breaking is determined by the ratio of vacuum expectation values of
their neutral components, $\tb\equiv v_2/v_1$. The spectrum of
physical Higgs bosons is richer than in the SM, consisting of two
neutral scalars, $h$ and $H$, one neutral pseudoscalar, $A$, and two
charged scalars, $H^\pm$. The couplings of the MSSM Higgs bosons to
matter fermions differ from those of the SM Higgs, and they can be
considerably enhanced or suppressed depending on $\tb$. As in the
SM, one of the most important production mechanisms for the neutral
Higgs bosons is gluon fusion, mediated by loops involving the top and
bottom quarks and their superpartners, the stop and sbottom
squarks. However, for intermediate to large values of $\tb$
bottom-quark annihilation can become the dominant production mechanism
for the neutral Higgs bosons that have enhanced couplings to down-type
fermions.

If the third-generation squarks have masses around one TeV or even
larger, their contributions to the gluon-fusion process are
suppressed, and a sufficiently accurate determination of the cross
section can be achieved by rescaling the SM results for the top- and
bottom-quark contributions by appropriate Higgs-quark effective
couplings. If, on the other hand, some of the squarks have masses of
the order of a few hundred GeV -- a scenario not yet excluded by the
direct searches at the LHC -- a precise calculation of the
contributions to the gluon-fusion cross section from diagrams
involving squarks becomes mandatory.
The NLO-QCD contributions to scalar production arising from diagrams
with colored scalars and gluons were first computed in the
vanishing-Higgs-mass limit (VHML) in ref.~\cite{Dawson:1996xz}, and
the full Higgs-mass dependence was included in later calculations
\cite{babis1,ABDV,MS}. For what concerns pseudoscalar production, the
NLO-QCD contributions arising from diagrams with quarks and gluons are
known~\cite{SDGZ,HK,babis1,ABDV} while diagrams involving only squarks
and gluons do not contribute to the gluon-fusion process due to the
structure of the pseudoscalar couplings to squarks.
In contrast, a full calculation of the contributions to either scalar
or pseudoscalar production arising from two-loop diagrams with quarks,
squarks and gluinos -- which can involve up to five different particle
masses -- is still missing. Calculations based on a combination of
analytic and numerical methods were presented in
refs.~\cite{babis2,spiraDb}, but neither explicit analytic formulae
nor public computer codes implementing the results of those
calculations have been made available so far.

Approximate results for the quark-squark-gluino contributions can
however be obtained assuming the presence of some hierarchy between
the Higgs mass and the masses of the particles running in the
loops. If the Higgs boson is lighter than all the particles in the
loops, it is possible to expand the result in powers of the Higgs
mass, with the first term in the expansion corresponding to the
VHML. This limit was adopted in
refs.~\cite{Harlander:2003bb,Harlander:2004tp,Degrassi:2008zj} for the
calculation of the top-stop-gluino contributions to scalar production
and in refs.~\cite{Harlander:2005if,Degrassi:2011vq} for the analogous
calculation of pseudoscalar
production. Refs.~\cite{Degrassi:2008zj,Degrassi:2011vq} also discussed
the reliability of the VHML by considering the next term in the
expansion in the Higgs mass.

While an expansion in the Higgs mass is a viable approximation in the
computation of the top-stop-gluino contributions to the production of
the lightest scalar $h$, it might not be applicable to the production
of the heaviest scalar $H$ and of the pseudoscalar $A$, if their mass
is comparable to the mass of the top quark. Moreover, an expansion in
the Higgs mass is certainly useless in the calculation of the
bottom-sbottom-gluino contributions, due to the presence of a light
bottom quark. All of these limitations can, however, be overcome with
an expansion in inverse powers of the superparticle masses. Since it
does not assume any hierarchy between the Higgs mass and the mass of
the quark in the loop, such an expansion is applicable to both
top-stop-gluino and bottom-sbottom-gluino contributions, as long as
the squarks and the gluino are heavier than the considered Higgs boson
and the top quark. Results for scalar production based on an expansion
in the superparticle masses were presented in
refs.~\cite{Degrassi:2010eu,Harlander:2010wr,Degrassi:2012vt}, and
analogous results for pseudoscalar production were presented in
ref.~\cite{Degrassi:2011vq}.

In order to improve the accuracy of the MSSM prediction for the
gluon-fusion cross section, and to allow for a meaningful comparison
with the SM prediction, several contributions beyond the NLO in QCD
should be included. The NNLO-QCD contributions to scalar production
arising from diagrams with top quarks and the subset of EW
contributions arising from diagrams with light quarks can be obtained
from the corresponding SM results with an appropriate rescaling of the
Higgs couplings to quarks and to gauge bosons. The NNLO-QCD top-quark
contributions to pseudoscalar production have also been
computed~\cite{pseudonnlo}.
Approximate results beyond the NLO in QCD also exist for the
contributions of diagrams involving superparticles. A first estimate
of the NNLO-QCD contributions of diagrams involving stop squarks was
presented in ref.~\cite{Harlander:2003kf}, and an approximate
calculation of those contributions, assuming the VHML and specific
hierarchies among the superparticle masses, was recently presented in
refs.~\cite{Pak:2010cu,Pak:2012xr}. Furthermore, a subset of
potentially large $\tan\beta$-enhanced contributions from diagrams
involving sbottom-gluino or stop-chargino loops can be resummed in the
LO cross section by means of an effective Higgs-bottom
coupling~\cite{effL,GHS,deltab2l}.

In a significant part of the MSSM parameter space, the couplings of
the heavier neutral Higgs bosons $H$ and $A$ to bottom quarks are
enhanced by $\tan\beta$ with respect to the corresponding coupling of
the SM Higgs, while their couplings to top quarks are suppressed by
$\tan\beta$. When that is the case, the bottom-quark contributions to
the gluon-fusion process -- which for a SM-like Higgs with mass around
$125.5$~GeV amount to roughly $7\%$ of the cross section -- can dominate
over the top-quark contributions.  The bottom-quark contributions are
subject to large QCD corrections enhanced by powers of
$\ln(m_\phi^2/m_b^2)$, where $\phi$ denotes a generic Higgs boson, and
so far they have been computed only at the
NLO~\cite{SDGZ,HK,babis1,ABDV}. As a result, the uncomputed
higher-order QCD corrections to the bottom-quark contributions can
become the dominant source of uncertainty in the cross section for the
production of heavy MSSM Higgs bosons in gluon fusion.

On the other hand, as mentioned earlier, when the couplings to bottom
quarks are sufficiently enhanced the production of MSSM Higgs bosons
through bottom-quark annihilation dominates over gluon fusion.  In the
four-flavor scheme (4FS), where one does not consider the bottom
quarks as partons in the proton, the process is initiated by two
gluons or by a light quark-antiquark pair, and the cross section is
known at the NLO in QCD~\cite{4FS}. In the five-flavor scheme (5FS),
where the bottom quarks are in the initial partonic state, the cross
section is known up to the NNLO in
QCD~\cite{5FSNLO,Harlander:2003ai}. The use of bottom-quark parton
density functions (PDFs) in the 5FS allows to resum terms enhanced by
$\ln(m_\phi^2/m_b^2)$ that would arise in the 4FS when one or both
bottom quarks are collinear to the incoming partons. As in the case of
gluon fusion, the $\tan\beta$-enhanced contributions from diagrams
involving superpartners can be resummed in the LO result by means of
an effective Higgs-bottom coupling. The remaining one-loop
contributions from superpartners have been found to be
small~\cite{bbH1loop}.

A considerable effort has been devoted over the years to making the
existing calculations of Higgs production available to the physics
community in the form of public computer codes. In the case of the SM,
NNLO-QCD predictions of the total cross section for gluon fusion,
including various refinements such as EW corrections and finite
top-mass effects, are provided, e.g., by {\tt HIGLU}~\cite{higlu},
{\tt ggh@nnlo}~\cite{ggh@nnlo}, {\tt HNNLO}~\cite{HNNLO} and {\tt
  iHixs}~\cite{iHixs}. The code {\tt bbh@nnlo}~\cite{bbh@nnlo}
provides instead a NNLO-QCD prediction of the total cross section for
Higgs production in bottom-quark annihilation in the 5FS. For what
concerns the production of MSSM Higgs bosons via gluon fusion, {\tt
  HIGLU} implements the results of ref.~\cite{MS} for the NLO-QCD
contributions arising from diagrams with squarks and gluons, as well
as the results of refs.~\cite{GHS,deltab2l} for the resummation of the
$\tan\beta$-enhanced squark contributions in an effective Higgs-bottom
coupling.

More recently, two codes that compute the cross section for Higgs
production including approximate results for the contributions of
diagrams with quarks, squarks and gluinos have become available. As
described in ref.~\cite{Bagnaschi:2011tu}, the
NLO-QCD~\cite{ABDV,Degrassi:2008zj, Degrassi:2011vq, Degrassi:2010eu,
  Degrassi:2012vt} and EW~\cite{ABDV0,oai:arXiv.org:1007.1891}
contributions to Higgs-boson production via gluon fusion in the SM and
in the MSSM have been implemented in a module for the so-called
\PB~\cite{POWHEG}, a framework for consistently matching NLO-QCD
computations of matrix elements with parton-shower Monte Carlo
generators, avoiding double counting and preserving the NLO accuracy
of the calculation.
The code \sushi~\cite{Harlander:2012pb}, on the other hand, computes
the cross section for Higgs-boson production in both gluon fusion and
bottom-quark annihilation, in the SM and in the MSSM. In the case of
gluon fusion, \sushi\ includes the exact results of ref.~\cite{HK} for
the NLO-QCD contributions of two-loop diagrams with top and bottom
quarks, and the approximate results of refs.~\cite{Harlander:2004tp,
  Degrassi:2011vq,Degrassi:2012vt} and
refs.~\cite{Degrassi:2011vq,Degrassi:2010eu} for the NLO-QCD
contributions of two-loop diagrams with stop and sbottom squarks,
respectively. The NLO-QCD contributions of one-loop diagrams with
emission of an additional parton are taken from
ref.~\cite{Harlander:2010wr}. The NNLO-QCD contributions from diagrams
with top quarks are included via a call to {\tt ggh@nnlo}, and the
corresponding contributions from diagrams with stop squarks are
estimated following ref.~\cite{Harlander:2003kf}. Finally, the known
SM results for the EW
contributions~\cite{ABDV0,APSU,oai:arXiv.org:1007.1891} are adapted to
the MSSM by rescaling the Higgs couplings to top quarks and to gauge
bosons. In the case of bottom-quark annihilation, \sushi\ obtains from
{\tt bbh@nnlo} the NNLO-QCD result valid in the SM, then rescales it
by an effective Higgs-bottom coupling that accounts for the
$\tan\beta$-enhanced squark contributions~\cite{effL,GHS}.

In this paper we use \sushi\ for a precise study of scalar and
pseudoscalar Higgs production in the MSSM. In section
\ref{sec:xsection} we present predictions for the total inclusive
cross section for Higgs production in six benchmark scenarios
compatible with the LHC results, focusing in particular on a scenario
with relatively light stops where the effect of the SUSY contributions
can be significant. In section \ref{sec:uncertainty} we provide a
detailed discussion of the sources of theoretical uncertainty in the
calculation of the total cross section for Higgs-boson production in
the MSSM. We examine the dependence of the cross sections for gluon
fusion and bottom-quark annihilation on the renormalization and
factorization scales, on the precise definition of the Higgs-bottom
coupling and on the choice of PDFs, as well as the uncertainty
associated to our incomplete knowledge of the SUSY contributions
through NNLO. In particular, we point out a potentially large
uncertainty arising from uncomputed higher-order QCD corrections to
the bottom-quark contributions to gluon fusion, which can affect the
interpretation of the searches for the MSSM Higgs bosons in scenarios
where their couplings to bottom quarks are enhanced with respect to
the SM. In section \ref{sec:conclusions} we present our
conclusions. Finally, in the appendix we list the cross sections and
uncertainties for the production of the three neutral Higgs bosons in
selected points of the parameter space for the six benchmark
scenarios.


\section{Higgs-boson production in viable MSSM scenarios}
\label{sec:xsection}

The discovery of a neutral scalar with mass around \hmass\ GeV puts
the studies of the Higgs sector of the MSSM in an entirely new
perspective. In order to remain viable, a point in the MSSM parameter
space must now not only pass all the experimental bounds on
superparticle masses, but also lead to the prediction of a scalar with
mass, production cross section and decay rates compatible with those
measured at the LHC. In particular, the relatively large mass of the
SM-like scalar discovered at the LHC implies either stop masses of the
order of 3 TeV -- which would result in a negligible stop contribution
to the production cross section -- or a large value of the left-right
mixing term in the stop mass matrix (see, e.g.,
refs.~\cite{Heinemeyer:2011aa,Arbey:2011ab}). In the latter case, at
least one of the stops could have a mass as low as a few hundred GeV,
and induce a significant contribution to the gluon-fusion cross
section.

In view of these considerations, we will focus on the set of MSSM
scenarios compatible with the LHC findings that has recently been
proposed in ref.~\cite{Carena:2013qia}. We will study the effect of
the different contributions to the total cross section for the
production of the MSSM Higgs bosons, relying on the approximate
NNLO-QCD calculations implemented in \sushi.


\subsection{The benchmark scenarios}
\label{sec:benchmarks}

The SM parameters entering our calculations include the $Z$-boson mass
$\mz = 91.1876$~GeV, the $W$-boson mass $\mw = 80.398$~GeV, the Fermi
constant $G_F = 1.16637\!\times\!10^{-5}$ and the strong coupling
constant $\alpha_s(\mz) = 0.119$.\footnote{The SM inputs agreed upon by the 
\WG\ are listed on the group's website~\cite{SMinputs}.} For the masses of
the top and bottom quarks we take the pole mass $\mt = 173.2$~GeV
\cite{CDF:2013jga} and the SM running mass (in the $\msbar$ scheme)
$\mb(\mb) = 4.16$~GeV~\cite{botmass}.

At the tree level, the MSSM neutral scalar masses $\mh$ and $\mH$ and
the scalar mixing angle $\alpha$ can be computed in terms of $\mz$,
$\tb$ and the pseudoscalar mass $\ma$ only. However, the radiative
corrections to the tree-level predictions can be substantial, and they
bring along a dependence on all of the other MSSM parameters. To
compute the masses and the couplings of Higgs bosons and
superparticles in a given point of the MSSM parameter space we use the
public code \FH~\cite{Heinemeyer:1998yj}, which includes the full
one-loop~\cite{Frank:2006yh} and dominant
two-loop~\cite{Heinemeyer:1998np,noistop,noisbot,Heinemeyer:2004xw,
  Heinemeyer:2007aq} corrections to the neutral Higgs masses. Since
the theoretical uncertainty of the Higgs-mass calculation in \FH\ has
been estimated to be of the order of 3
GeV~\cite{th-error},\footnote{\,To reduce this uncertainty, it would
  be necessary to include in the mass calculation the remaining
  two-loop effects~\cite{martin-2l} and at least the dominant
  three-loop effects~\cite{Martin:2007pg,HS-3l}. Note also that there
  is an additional uncertainty of approximately $1$~GeV stemming from
  the uncertainty of the SM input parameters, especially $\mt$.} we
consider as phenomenologically acceptable the points in the MSSM
parameter space where \FH\ predicts the existence of a scalar with
mass between $122.5$~GeV and $128.5$~GeV and with approximately
SM-like couplings to gauge bosons.

In addition to $\tb$ and $\ma$, the MSSM parameters most relevant to
the prediction of the masses and production cross sections of the
Higgs bosons are: the soft SUSY-breaking masses for the stop and
sbottom squarks, which for simplicity we set all equal to a common
mass parameter $\msusy$; the soft SUSY-breaking gluino mass $\mg$; the
soft SUSY-breaking Higgs-squark-squark couplings $A_t$ and $A_b$; the
superpotential Higgs-mass parameter $\mu$. In our convention for the
sign of the latter, the left-right mixing terms in the stop and
sbottom mass matrices are $X_t \equiv A_t - \mu\cot\beta$ and $X_b
\equiv A_b - \mu\tan\beta$, respectively.
It should be noted that in our analysis the soft SUSY-breaking squark
masses and trilinear couplings are expressed in an ``on-shell'' (OS)
renormalization scheme, as described in
refs.~\cite{Heinemeyer:1998np,noistop} for the stop sector and in
refs.~\cite{noisbot,Heinemeyer:2004xw,Degrassi:2010eu} for the sbottom
sector.  Since the two-loop calculation of the Higgs masses
implemented in \FH\ and the NLO-QCD calculation of the production
cross section implemented in \sushi\ employ the same OS scheme, the
input values of the soft SUSY-breaking parameters can be passed
seamlessly from the Higgs-mass calculation to the cross-section
calculation. Concerning the parameters $\tb$, $\mu$ and $\ma$, their
definition is relevant to the Higgs-mass calculation only. In
particular, $\tb$ and $\mu$ are expressed in the $\drbar$ scheme, at a
renormalization scale that \FH\ takes by default equal to $\mt$, while
$\ma$ is identified with the pole mass of the pseudoscalar. Finally,
the choice of renormalization scheme for $\mg$ amounts to a
higher-order effect, because the gluino mass enters only the two-loop
part of the corrections.

\begin{table}[t]
\begin{center}
\begin{tabular}{|c|c|c|c|c|}
\hline
Scenario & $\msusy$~[GeV] & $X_t$~[GeV] & $\mu$~[GeV] & $M_2$~[GeV]\\
\hline
\hline
$\mh^{\rm max}$ & $1000$ & $2000$ & $200$ & $200$ \\
$\mh^{\rm mod+}$ & $1000$ & $1500$ & $200$ & $200$ \\
$\mh^{\rm mod-}$ & $1000$ & $-1900~~\,$ & $200$ & $200$ \\
{\em light stop} & $500$ & $1000$ & $400$ & $400$ \\
{\em light stau} & $1000$ & $1600$ & $500$ & $200$ \\
{\em tau-phobic} & $1500$ & $3675$ & $2000$ & $200$ \\
\hline
\end{tabular}
\end{center}
\vspace*{-1mm}
\caption{Choices of MSSM parameters for the benchmark scenarios
  proposed in ref.~\cite{Carena:2013qia}.}
\label{tab:scenarios}
\end{table}

A detailed description of the six benchmark scenarios adopted in our
analysis can be found in the paper where they were originally
proposed, ref.~\cite{Carena:2013qia}. All of the scenarios are
characterized by relatively large values of the ratio $X_t/\msusy$,
ensuring that the mass of the SM-like Higgs falls within the required
range without the need for extremely heavy stops. In addition, the
masses of the gluino and of the first-two-generation squarks are set
to 1.5~TeV, large enough to evade the current
ATLAS~\cite{ATLAS:sbotpm,ginoATLAS} and
CMS~\cite{CMS:sbot0,CMS:sbotpm,ginoCMS} bounds.  The prescriptions of
ref.~\cite{Carena:2013qia} for the parameters $\msusy$, $X_t$, $\mu$
and for the soft SUSY-breaking wino mass $M_2$ are listed in table
\ref{tab:scenarios}. We vary the parameters $\tb$ and $\ma$ within the
ranges
\be
\label{tbma}
2 \,\leq\, \tb \,\leq\, 50~,~~~~~~~~~~~~~
90~{\rm GeV} \,\leq\, \ma \,\leq\, 1~{\rm TeV}~.
\ee
In all scenarios the Higgs-sbottom-sbottom coupling $A_b$ is set equal
to $A_t$, the left-right mixing of the first-two-generation squarks is
neglected and the bino mass $M_1$ is obtained from the GUT relation
$M_1/M_2 = (5/3)(\mz^2/\mw^2-1)$, with the exception of the fourth
scenario where we set $M_1 = 340$~GeV.\footnote{\,The choice $M_1 =
  350$ originally proposed in ref.~\cite{Carena:2013qia} would result
  in a stop LSP for $\tb \gsim 20$.} Finally, the choices of
ref.~\cite{Carena:2013qia} for the soft SUSY-breaking parameters in
the slepton sector have a very small impact on the predictions for the
Higgs masses and production cross sections, therefore we do not report
them here.

The fourth scenario in table \ref{tab:scenarios}, denoted as {\em
  light stop}, deserves a special discussion.  In this scenario the
two stop masses are $324$~GeV and $672$~GeV; the sbottom masses depend
on $\tb$, but the lightest sbottom is always heavier than $450$~GeV,
while the heaviest one is always lighter than $550$~GeV. With such
relatively low masses, loops involving squarks can give a sizable
contribution to the cross section for Higgs production, but we have to
worry about the exclusion bounds from the LHC.  Indeed, the ATLAS and
CMS collaborations have presented preliminary results for the searches
of direct stop- and sbottom-pair production, based on the full 8-TeV
data sample, considering the decay chains

\begin{center}
  $\stu ~\rightarrow~ t \,\chi^0_1 ~\rightarrow~
  b\,W\,\chi^0_1$~~~~\cite{ATLAS:stop,CMS:stop}~,
~~~~~~~
  $\stu ~\rightarrow~ b \,\chi^\pm_1 ~\rightarrow~
  b\,W\,\chi^0_1$~~~~\cite{ATLAS:stop,CMS:stop}~,
~~~~~~~
  $\stu ~\rightarrow~ c \,\chi^0_1$~~~~\cite{ATLAS:stopc,CMS:stopc}~,\\~\\
~~~~~~~  $\sbu ~\rightarrow~ b \,\chi^0_1$~~~~\cite{CMS:sbot0,ATLAS:stopsbot0}~,
~~~~~~~~~~~~~~~~~~~~~~
  $\sbu ~\rightarrow~ t \,\chi^\pm_1 ~\rightarrow~
  t\,W\,\chi^0_1$~~~~\cite{ATLAS:sbotpm,CMS:sbotpm}~.
\end{center}

The allowed values of the stop and sbottom masses depend on the
chargino and neutralino masses, as well as on the branching ratios for
the different squark decays. With the choice of parameters in table
\ref{tab:scenarios}, $M_2 = \mu = 400$ GeV, together with $M_1 = 340$
GeV, the masses for the lightest chargino and neutralino have a mild
dependence on $\tb$, but they stay within the ranges $m_{\chi^\pm_1}
\approx 341$ -- $346$ GeV and $m_{\chi^0_1} \approx 316$ -- $320$ GeV
for $\tb>10$.  In this case the lightest stop decays almost entirely
through the loop-induced, flavor-violating channel $\stu \rightarrow c
\,\chi^0_1$. This channel has been investigated by
ATLAS~\cite{ATLAS:stopc} and CMS~\cite{CMS:stopc}, but the resulting
bounds only reach to values of $m_{\stu}$ around $250$~GeV. For the
lightest sbottom, the two-body decays $\sbu \rightarrow \stu\,W$ and
$\sbu \rightarrow b\,\chi^0_j$ (with $j$ up to 3 or 4) are
kinematically open. The direct decay of $\sbu$ to the lightest
neutralino would be constrained by the searches in
refs.~\cite{CMS:sbot0,ATLAS:stopsbot0}, but $i)$ that channel is never
dominant in the considered range of parameters and $ii)$ the
experimental bounds only reach to values of $m_{\chi^0_1}$ below $280$
GeV. Finally, the heaviest stop and sbottom can decay through a
multitude of channels, and their direct decays to $\chi^0_1$ or
$\chi^\pm_1$ are significantly suppressed.


\subsection{Cross section for Higgs production}
\label{sec:xsec-xsec}

We are now ready to present our precise predictions for the production
of MSSM Higgs bosons at the LHC. As mentioned earlier, we rely on the
code \sushi,\footnote{For a detailed description of the cross-section
  calculation implemented in \sushi\ we refer to the code's
  manual~\cite{Harlander:2012pb}.}  which includes all of the
available NLO-QCD contributions to the gluon-fusion process,
supplemented with the known SM results for the NNLO-QCD contributions
in the heavy-top limit and for the EW contributions (both adapted to
the MSSM by appropriately rescaling the Higgs couplings).
While the results implemented in \sushi\ for the NNLO-QCD top
contributions are strictly valid only for a Higgs mass below the top
threshold, $m_\phi < 2\, \mt$, a comparison with the NLO results
suggests that they provide a decent approximation also for larger
values of the Higgs mass~\cite{Spira:1997dg,Harlander:2003xy}.
The NNLO-QCD contributions from stop loops are
estimated following ref.~\cite{Harlander:2003kf}, i.e., neglecting the
contributions of three-loop diagrams but retaining the NNLO
contributions that arise from the product of lower-order terms. We
have also checked that, when all of the NNLO-QCD contributions are
omitted, the results of \sushi\ for the gluon-fusion cross section
agree with those of the calculation implemented in the
\PB~\cite{Bagnaschi:2011tu}, which includes the same NLO-QCD and EW
contributions. For what concerns the bottom-quark annihilation
process, \sushi\ includes the NNLO-QCD results valid in the SM within
the 5FS, also rescaled by the effective Higgs-bottom couplings of the
MSSM.

In our study, we fix the center-of-mass energy of the proton-proton
collisions to $8$~TeV. While the numerical value of the total cross
section for Higgs production does obviously depend on the collision
energy, we have checked that the relative importance of the various
contributions to the production processes and their qualitative
behavior over the MSSM parameter space do not change substantially if
we set the energy to $13$~TeV. By default, we use the MSTW2008 set of
PDFs~\cite{Martin:2009iq}, and we fix the renormalization and
factorization scales entering the gluon-fusion cross section to
$\muR=\muF=m_\phi/2$~~\cite{Anastasiou:2002yz,Anastasiou:2005qj},
where $\phi=\{h,H,A\}$ denotes the considered Higgs boson. For
bottom-quark annihilation, the central values of the scales are chosen
as $\muR=m_\phi$ and
$\muF=m_\phi/4$~~\cite{5FSNLO,Harlander:2003ai,Boos:2003yi}.
In the calculation of the gluon-fusion cross section we relate the
bottom Yukawa coupling to the pole mass $\mbp$, computed at the
three-loop level~\cite{mb3loop} from the input value for the running
mass, $\mb(\mb)$. In the case of bottom-quark annihilation, on the
other hand, we relate the bottom Yukawa coupling to $\mb(m_\phi)$, in
turn obtained from $\mb(\mb)$ via four-loop renormalization-group
evolution~\cite{runmb}. In both cases, the $\tb$-enhanced SUSY
corrections to the relation between mass and Yukawa coupling of the
bottom quark are included following refs.~\cite{effL,GHS}.
The theoretical uncertainties associated to the choice of PDFs, to the
variation of the renormalization and factorization scales and to the
definition of the bottom Yukawa coupling will be discussed in detail in
section~\ref{sec:uncertainty}.

\begin{figure}[p]
\begin{center}
\includegraphics[angle=270,width=0.49\textwidth]{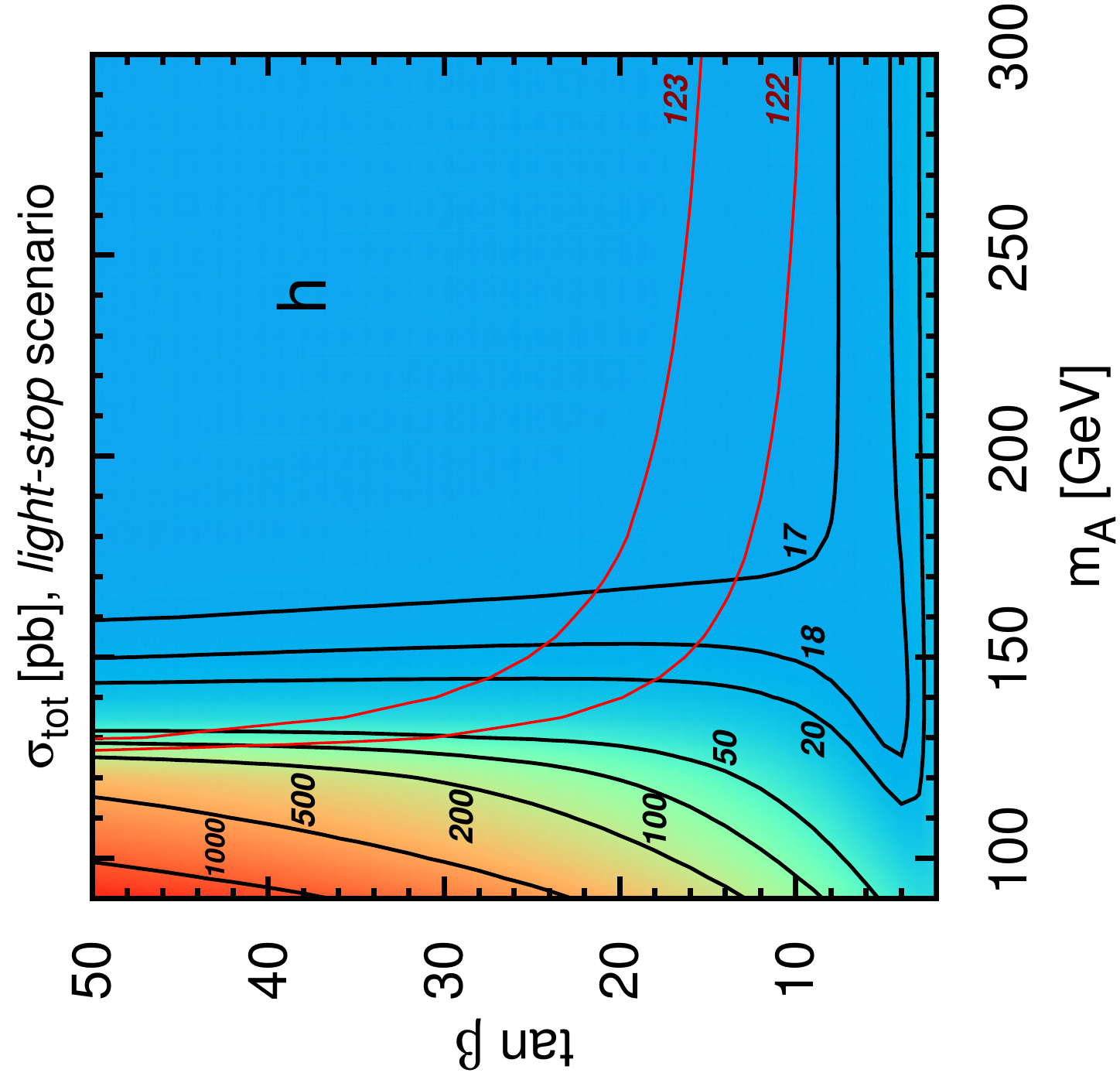}~~~
\includegraphics[angle=270,width=0.49\textwidth]{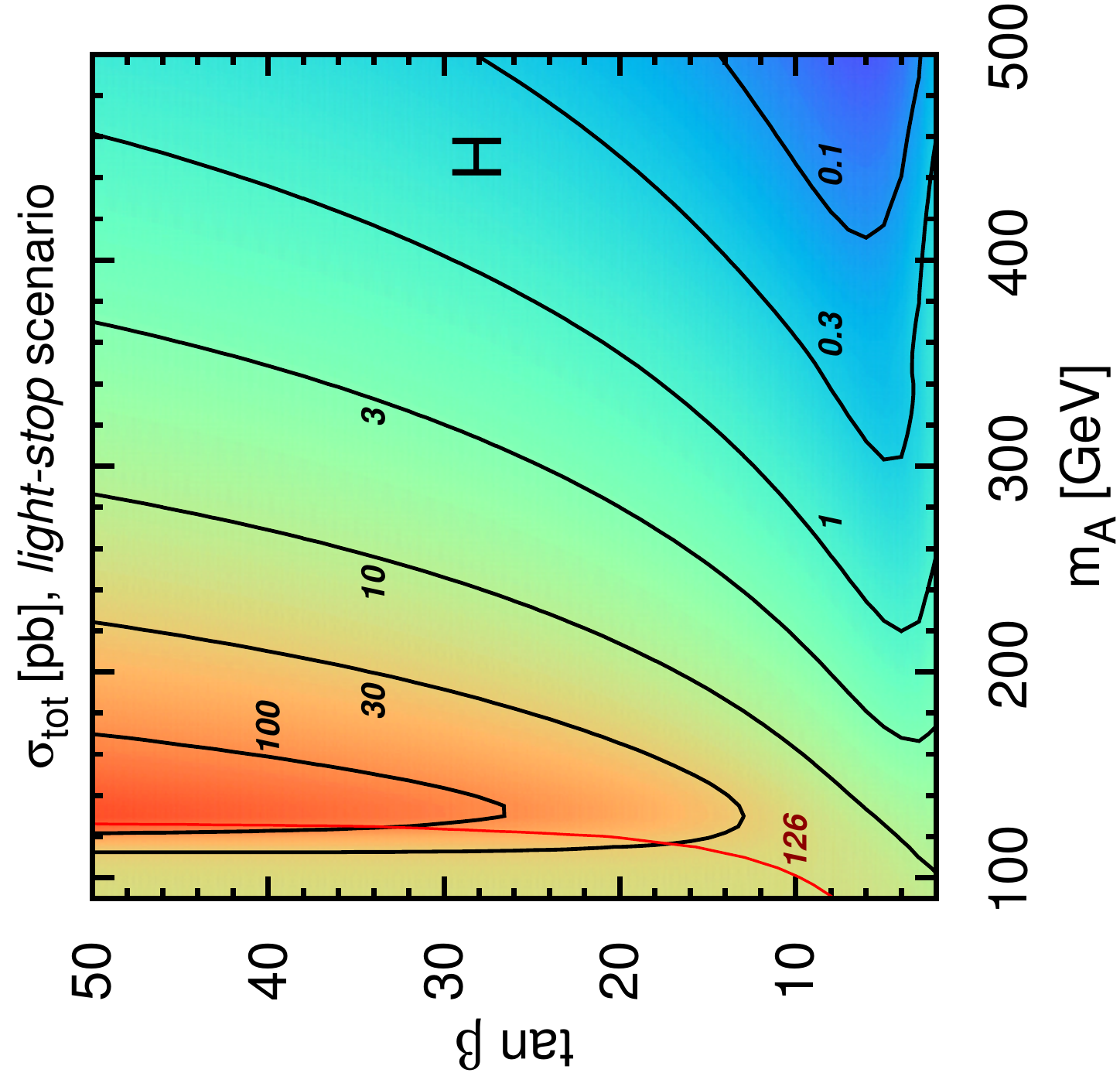}
\end{center}
\vspace*{-0.3cm}
\caption{Total cross section in picobarn (pb) for the production of $h$ (left)
  and $H$ (right), as a function of $\ma$ and $\tb$ in the \lss. The
  solid red lines are contours of equal mass for each scalar.}
\label{fig:totxs-scalar}
\end{figure}
\begin{figure}[p]
\begin{center}
\includegraphics[angle=270,width=0.49\textwidth]{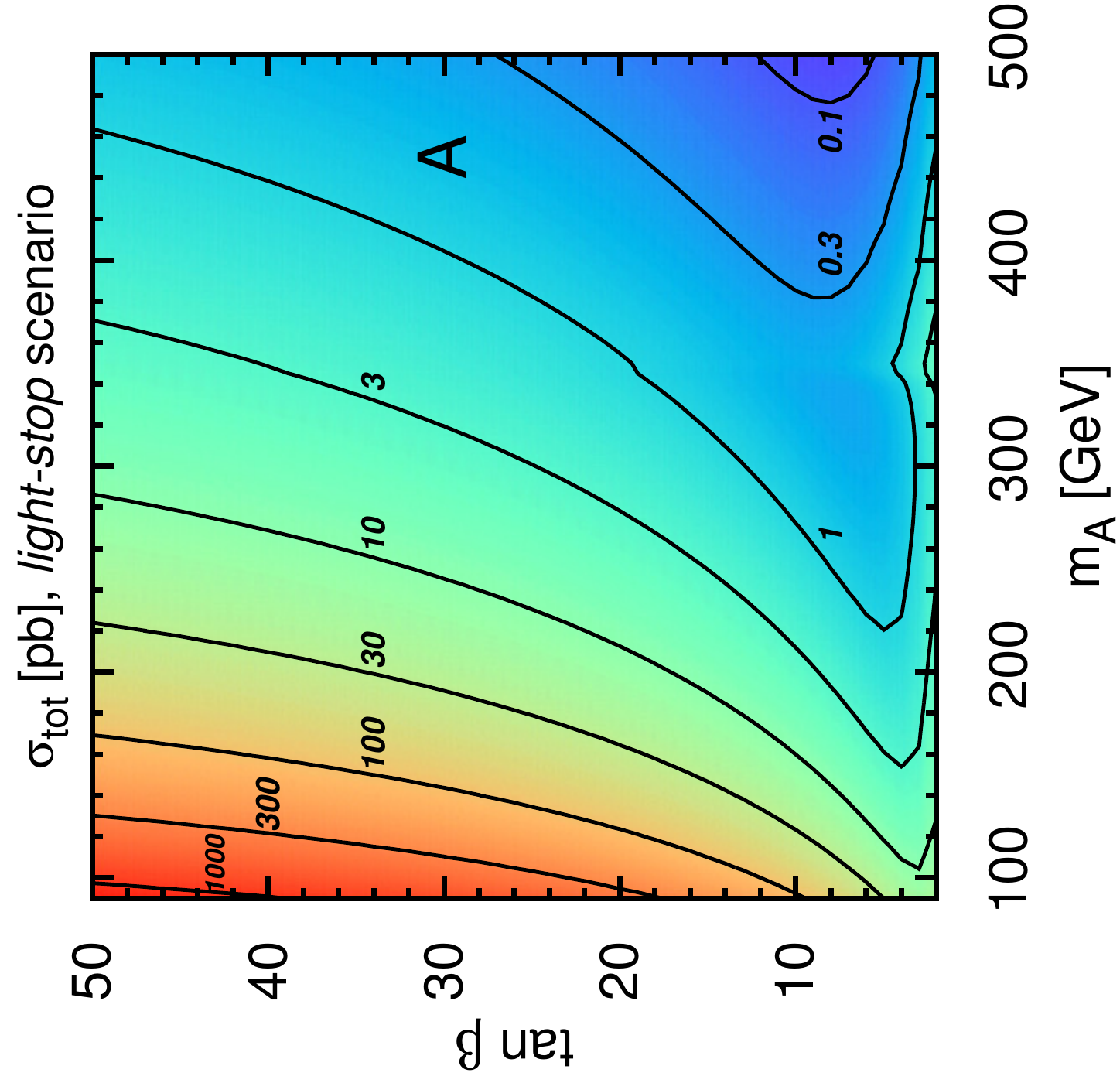}
\end{center}
\vspace*{-0.3cm}
\caption{Same as figure \ref{fig:totxs-scalar} for the production of the
  pseudoscalar $A$.}
\label{fig:totxs-pseudo}
\end{figure}

In figures~\ref{fig:totxs-scalar} and~\ref{fig:totxs-pseudo} we show
the total cross section -- i.e., the sum of gluon fusion and
bottom-quark annihilation -- for the production of the scalars ($h,
H$) and of the pseudoscalar ($A$), respectively, as contour plots in
the \matbp. For the other MSSM parameters, we adopt the
\lss\ described in section~\ref{sec:benchmarks}.  Tables for the
numerical values of the cross section (and the corresponding
uncertainties) in all of the six benchmark scenarios are given in the
appendix.  In the two plots of figure~\ref{fig:totxs-scalar},
referring to $h$ (left) and $H$ (right) production, the red lines are
contours of equal mass for the corresponding scalar. In this scenario,
the prediction for the mass of the lightest scalar reaches a maximum
of $123.8$~GeV at large $\tb$. The heaviest-scalar mass grows with
$\ma$, and we show only the contour corresponding to $126$~GeV to
avoid clutter (for large $\ma$, the contours are roughly at
$\mH\approx\ma$ and independent of $\tb$). The $x$-axis of the plot
for $h$ production ends at $\ma=300$~GeV because, for larger values,
the cross section becomes essentially independent of $\ma$. The
$x$-axis of the plots for $H$ and $A$ ends at $\ma=500$~GeV because
the expansion in the SUSY masses used to approximate the two-loop
squark contributions in \sushi\ becomes unreliable when the Higgs mass
approaches the lowest squark-mass threshold, which in the
\lss\ corresponds to $2\,m_{\stu}\approx\, 650$ GeV. The theoretical
uncertainty associated with this approximation will be discussed in
section~\ref{sec:susyerr}.

The qualitative behavior of the cross sections in
figures~\ref{fig:totxs-scalar} and~\ref{fig:totxs-pseudo} can be
easily interpreted considering the relations between the scalar and
pseudoscalar masses in the MSSM Higgs sector, and how each of the
Higgs bosons couples to the top and bottom quarks (the squark
contributions are generally sub-dominant, as will be discussed
below). In the so-called decoupling limit, $\ma \gg \mz$, the lightest
scalar $h$ has SM-like couplings to quarks, while its mass is
essentially independent of $\ma$ and, for $\tb\gsim 10$, depends only
weakly on $\tb$. The cross section for $h$ production (left plot in
figure~\ref{fig:totxs-scalar}) varies very little in this region, and
differs from the SM result for a Higgs boson of equal mass only
because of the squark contributions to the gluon-fusion process.  For
$\ma \lsim 130$~GeV, on the other hand, the couplings of $h$ to top
(bottom) quarks are non-standard, being suppressed (enhanced) by
$\tb$. In this narrow region the total cross section for $h$
production is dominated by the contributions of the diagrams that
involve the Higgs-bottom coupling, and it grows significantly with
$\tb$.

The behavior of the cross section for $H$ production in the
\matbp\ (right plot in figure~\ref{fig:totxs-scalar}) is different
from -- and somewhat complementary to -- the one for $h$
production. In the strip where $\ma \lsim 130$~GeV, the heaviest
scalar has a mass around $125$ GeV and significant couplings to both
top and bottom quarks, and the cross section for its production grows
with $\tb$. For larger $\ma$, on the other hand, $\mH$ grows together
with $\ma$, and the couplings of $H$ to top (bottom) quarks are
suppressed (enhanced) by $\tb$. The total cross section for $H$
production is therefore dominated, already for moderate $\tb$, by the
contributions of the diagrams that involve the Higgs-bottom
coupling. The latter grow significantly with $\tb$, but decrease with
$\ma$, being suppressed by powers of the ratio $\mb^2/\mH^2$. Finally,
the pseudoscalar couplings to top (bottom) quarks are suppressed
(enhanced) by $\tb$ for all values of $\ma$. Therefore, the behavior
of the cross section for $A$ production in the \matbp, see
figure~\ref{fig:totxs-pseudo}, resembles the behavior of $h$
production when $\ma \lsim 130$~GeV, and the one of $H$ production
for larger $\ma$: in both cases, the cross section grows with $\tb$,
but decreases with $\ma$.

\begin{figure}[p]
\begin{center}
\includegraphics[angle=270,width=0.49\textwidth]{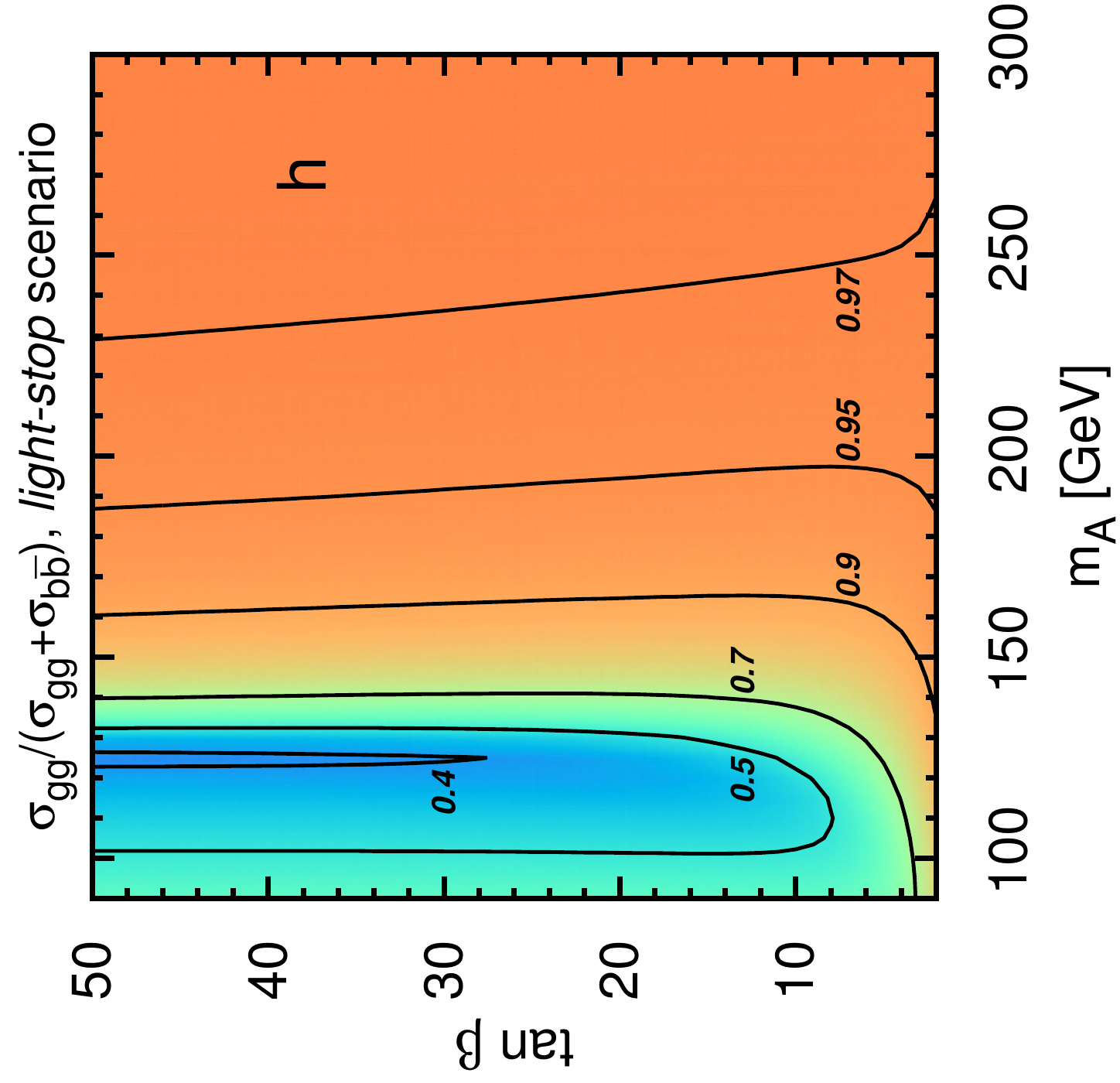}~~~
\includegraphics[angle=270,width=0.49\textwidth]{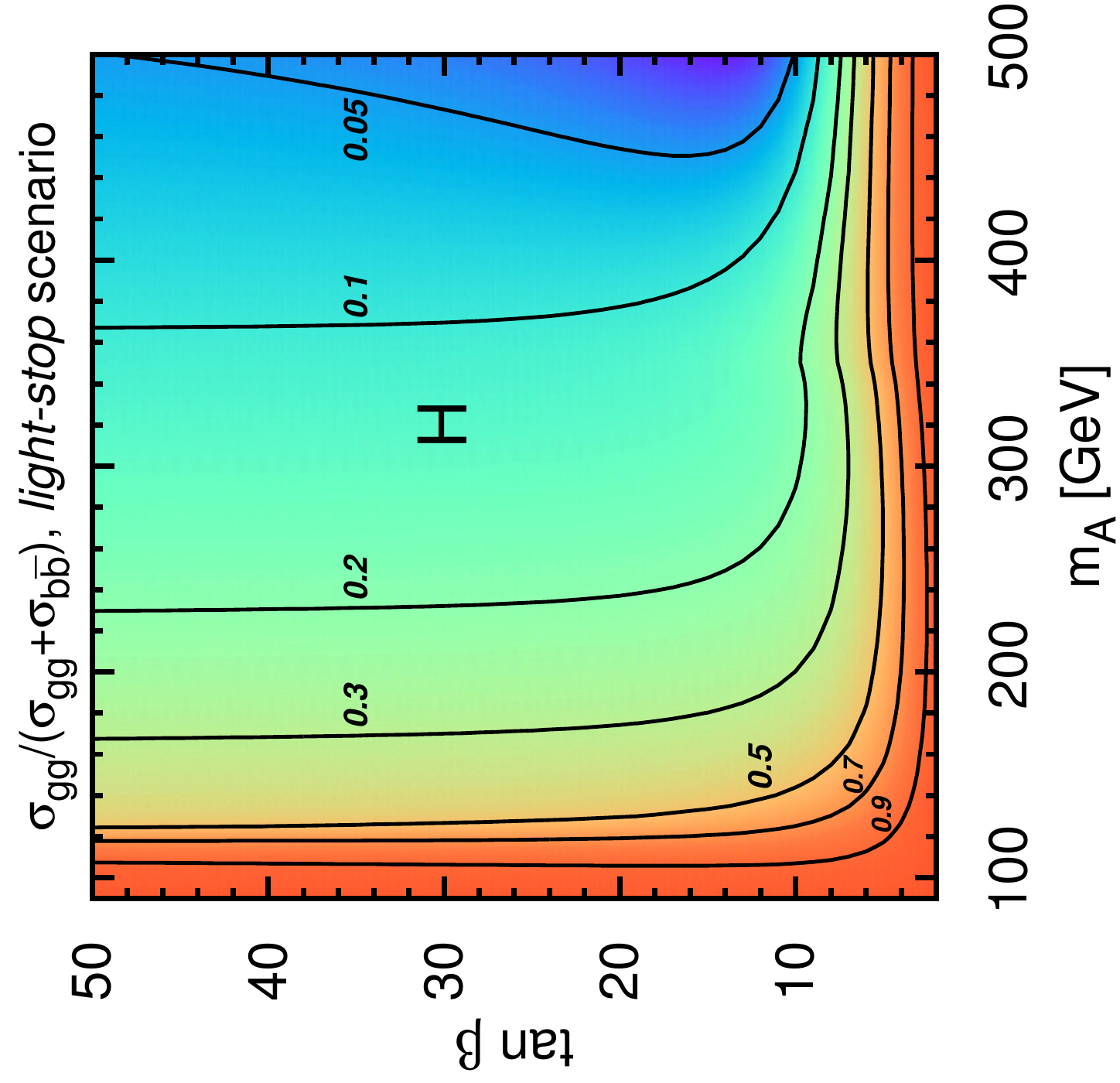}
\end{center}
\vspace*{-0.3cm}
\caption{Ratio of gluon-fusion cross section over total cross section
  for the production of $h$ (left) and $H$ (right), as a function of
  $\ma$ and $\tb$ in the \lss.}
\label{fig:ratio-scalar}
\end{figure}
\begin{figure}[p]
\begin{center}
\includegraphics[angle=270,width=0.49\textwidth]{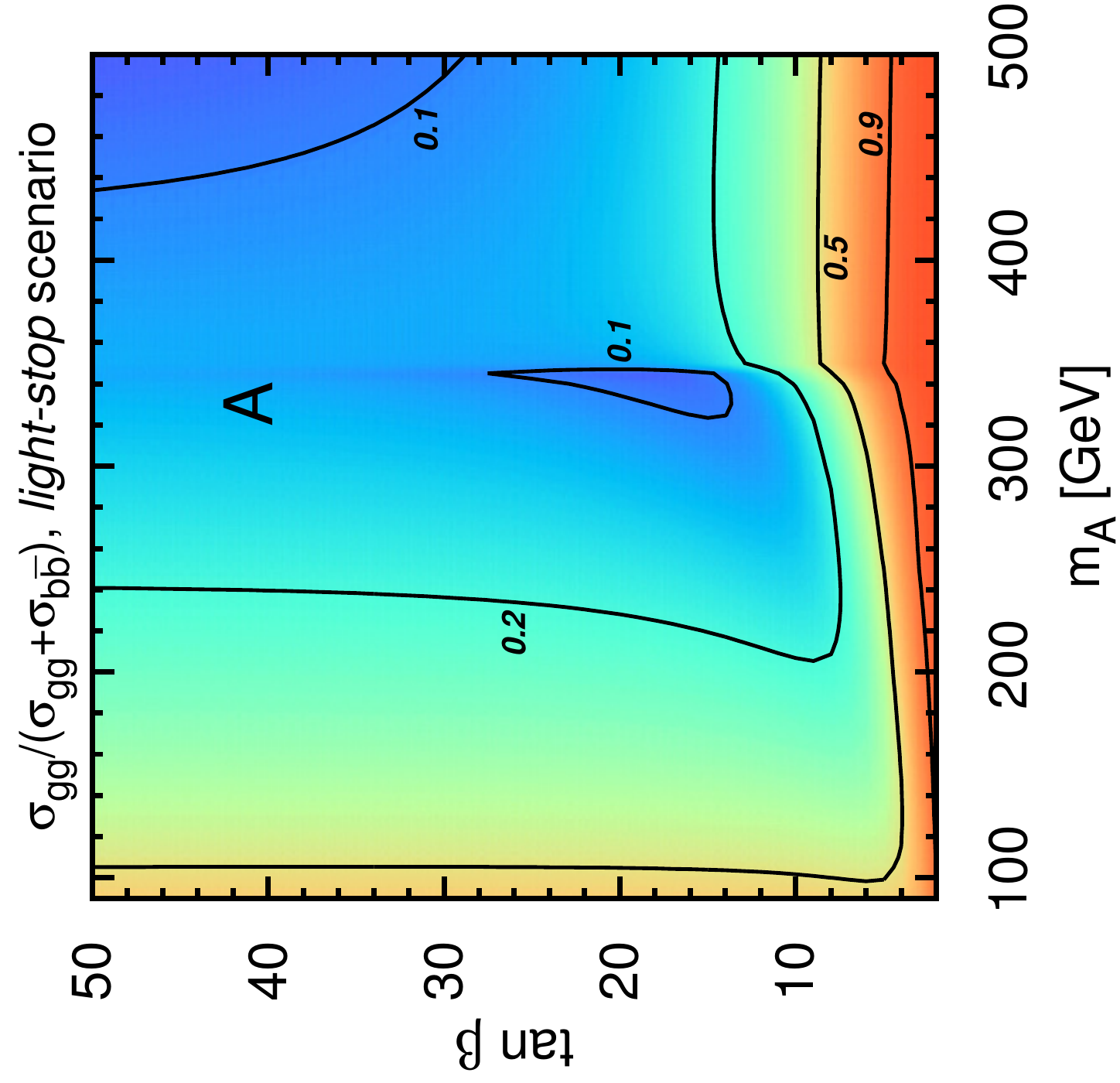}
\end{center}
\vspace*{-0.3cm}
\caption{Same as figure \ref{fig:ratio-scalar} for the production of
  the pseudoscalar $A$.}
\label{fig:ratio-pseudo}
\end{figure}

To disentangle the effects of the two main production channels for the
MSSM Higgs bosons, we show in figures~\ref{fig:ratio-scalar}
and~\ref{fig:ratio-pseudo} the ratio between the gluon-fusion cross
section and the sum of gluon-fusion and bottom-quark-annihilation
cross sections in the \lss, again as contour plots in the
\matbp. Predictably, the plots reflect the behavior of the coupling
of the considered Higgs boson to bottom quarks.  The left plot in
figure~\ref{fig:ratio-scalar} shows that, when $\ma$ is large enough
that the couplings of the lightest scalar are SM-like, gluon fusion is
by far the dominant process for $h$ production, and the contribution
of bottom-quark annihilation amounts only to a few percent. Only in
the strip with $\ma \lsim 130$~GeV and $\tb \gsim 8$, where the
coupling of $h$ to bottom quarks is sufficiently enhanced by $\tb$,
does bottom-quark annihilation become the dominant
process. Conversely, bottom-quark annihilation gives the largest
contribution to the cross section for $H$ production (right plot in
figure~\ref{fig:ratio-scalar}) when $\ma \gsim 130$~GeV and $\tb \gsim
6$, while in the case of $A$ production
(figure~\ref{fig:ratio-pseudo}) the cross section is dominated by
bottom-quark annihilation already for $\ma\gsim 100$~GeV, as long as
$\tb \gsim 5\,$--$\,8$.

\begin{figure}[p]
\begin{center}
\includegraphics[angle=270,width=0.49\textwidth]{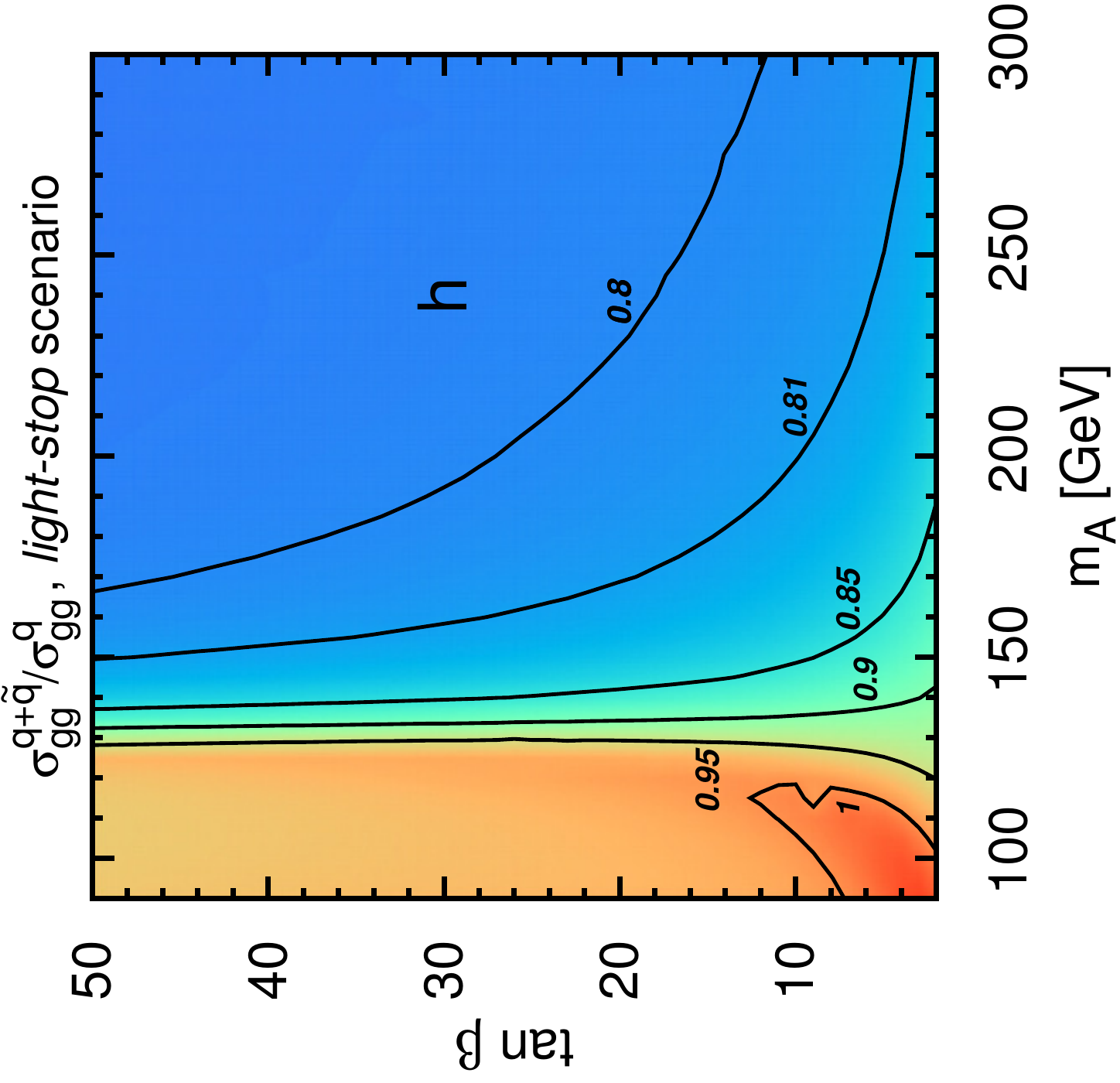}~~~
\includegraphics[angle=270,width=0.49\textwidth]{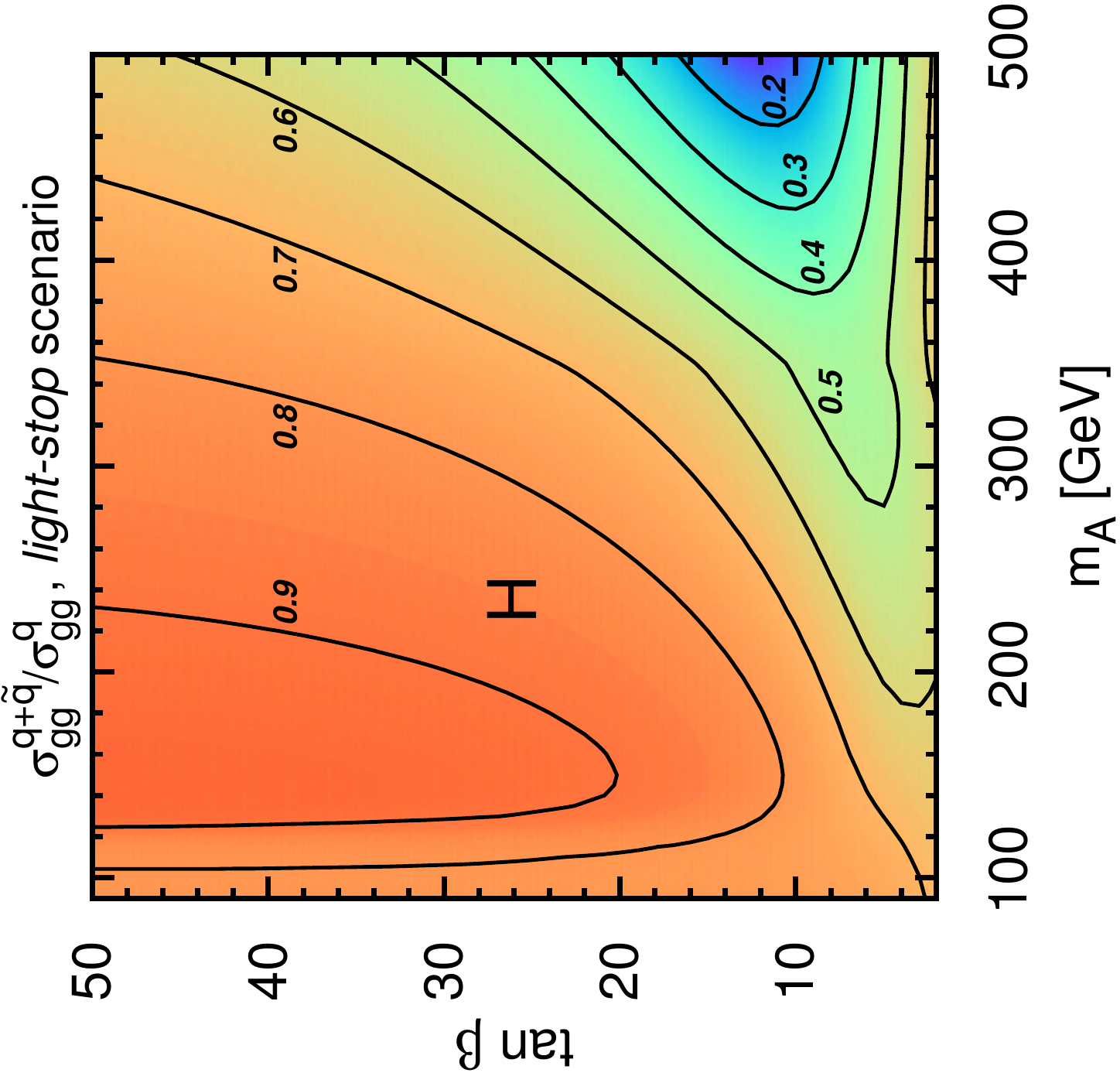}
\end{center}
\vspace*{-0.3cm}
\caption{Ratio of gluon-fusion cross section for the production of 
$h$ (left) and $H$ (right) over the corresponding cross section
neglecting squark contributions, as a function of $\ma$ and $\tb$ in
the \lss.}
\label{fig:squark-scalar}
\end{figure}
\begin{figure}[p]
\begin{center}
\includegraphics[angle=270,width=0.49\textwidth]{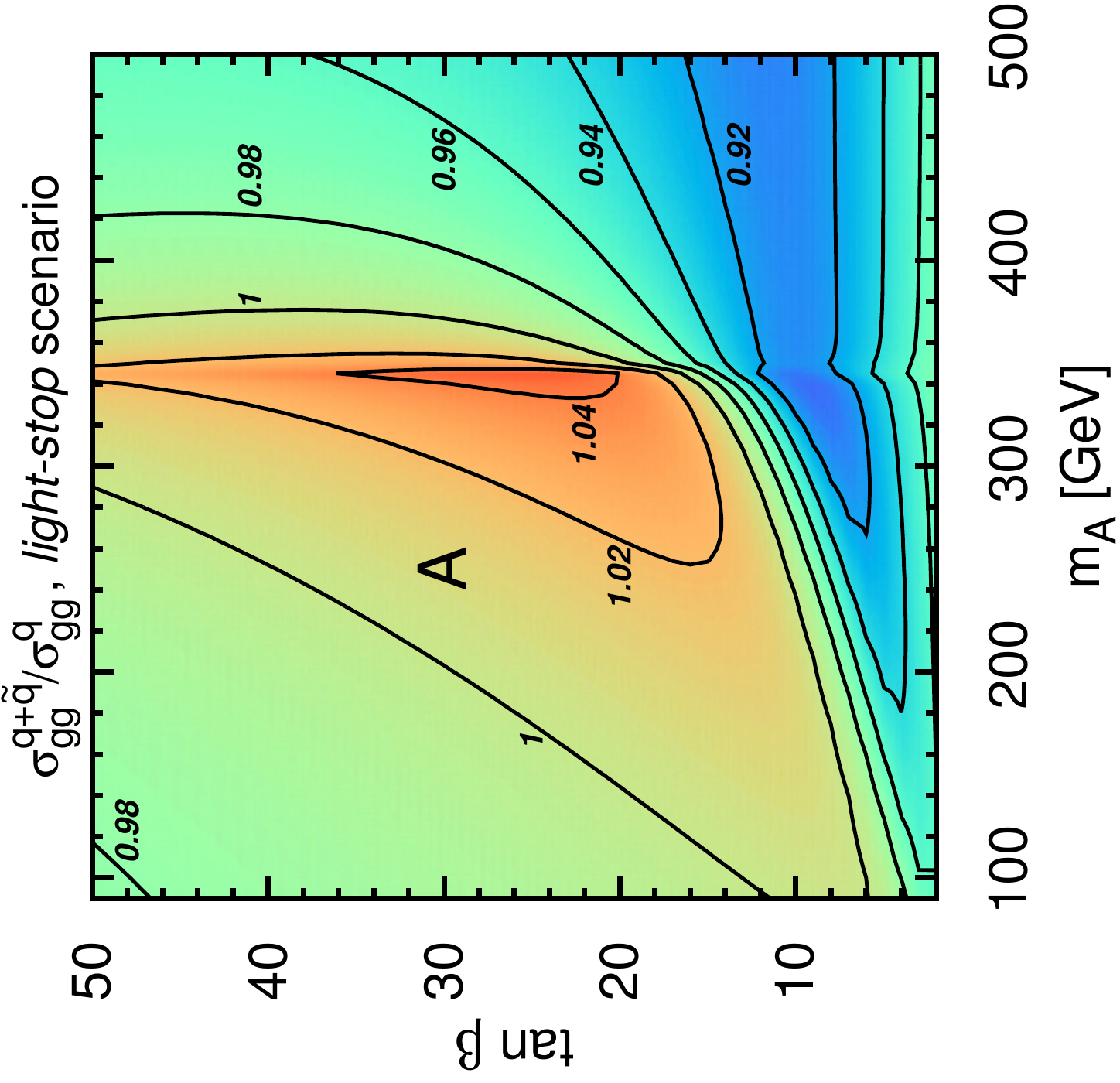}
\end{center}
\vspace*{-0.3cm}
\caption{Same as figure \ref{fig:squark-scalar} for the production of
  the pseudoscalar $A$.}
\label{fig:squark-pseudo}
\end{figure}

To assess the relevance of the squark contributions to the
gluon-fusion cross section in the \lss, we show in
figures~\ref{fig:squark-scalar} and~\ref{fig:squark-pseudo} the ratio
of the total gluon-fusion cross section over the cross section
computed including only the contributions of quarks (with appropriate
rescaling of the Higgs-quark couplings). The left plot of
figure~\ref{fig:squark-scalar} shows that -- in this scenario
characterized by relatively light squarks -- the interference between
the top and stop contributions can reduce the cross section for $h$
production by as much as $20\%$ in the decoupling region with large
$\ma$ and $\tb$. Remarkably, in this region the partial NNLO-QCD
contributions from stop loops that we include following
ref.~\cite{Harlander:2003kf} account by themselves for a $6\%$
suppression of the cross section.  The theoretical uncertainty
associated to these contributions will be discussed in
section~\ref{sec:susyerr}. For what concerns $H$ production (right
plot of figure~\ref{fig:squark-scalar}), the squark contributions
reduce the cross section by up to $30\%$ for low values of $\ma$, and
the suppression becomes even stronger with increasing pseudoscalar
mass. In particular, near the lower-right corner of the plot, where
$\ma\gsim 420$~GeV and $\tb$ ranges between $6$ and $20$, the
interference between the quark and squark contributions induce a
suppression of the cross section by $70\,$--$\,80\%$. In this region
the top contribution is suppressed by $\tb$, while the bottom
contribution is suppressed by $\mb^2/\mH^2$ and only moderately
enhanced by $\tb$, so they both become comparable in size with the
stop contribution. The resulting gluon-fusion cross section is rather
small, of the order of a few femtobarns. Finally,
figure~\ref{fig:squark-pseudo} shows that, in the case of $A$
production, the effect of the squark contributions on the cross
section for gluon fusion in the \lss\ is always less than $10\%$. This
is due to the fact that the pseudoscalar couples only to two different
squark-mass eigenstates, while gluons couple only to pairs of the same
squarks. Therefore, there is no squark contribution to the
gluon-fusion process at the LO, and the whole effect in
figure~\ref{fig:squark-pseudo} arises from two-loop diagrams.

\begin{figure}[t]
\begin{center}
\includegraphics[angle=270,width=0.49\textwidth]{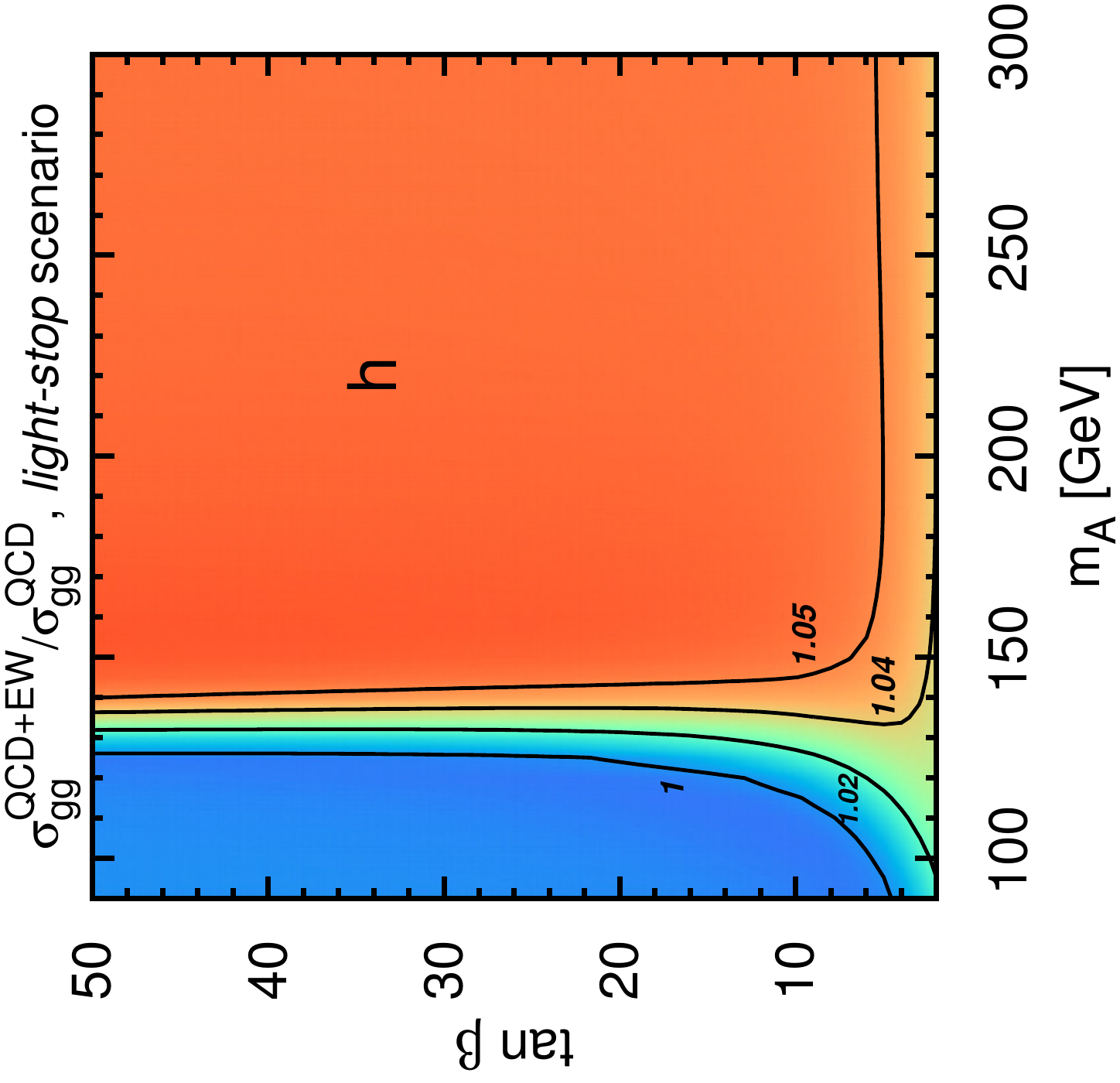}~~~
\includegraphics[angle=270,width=0.49\textwidth]{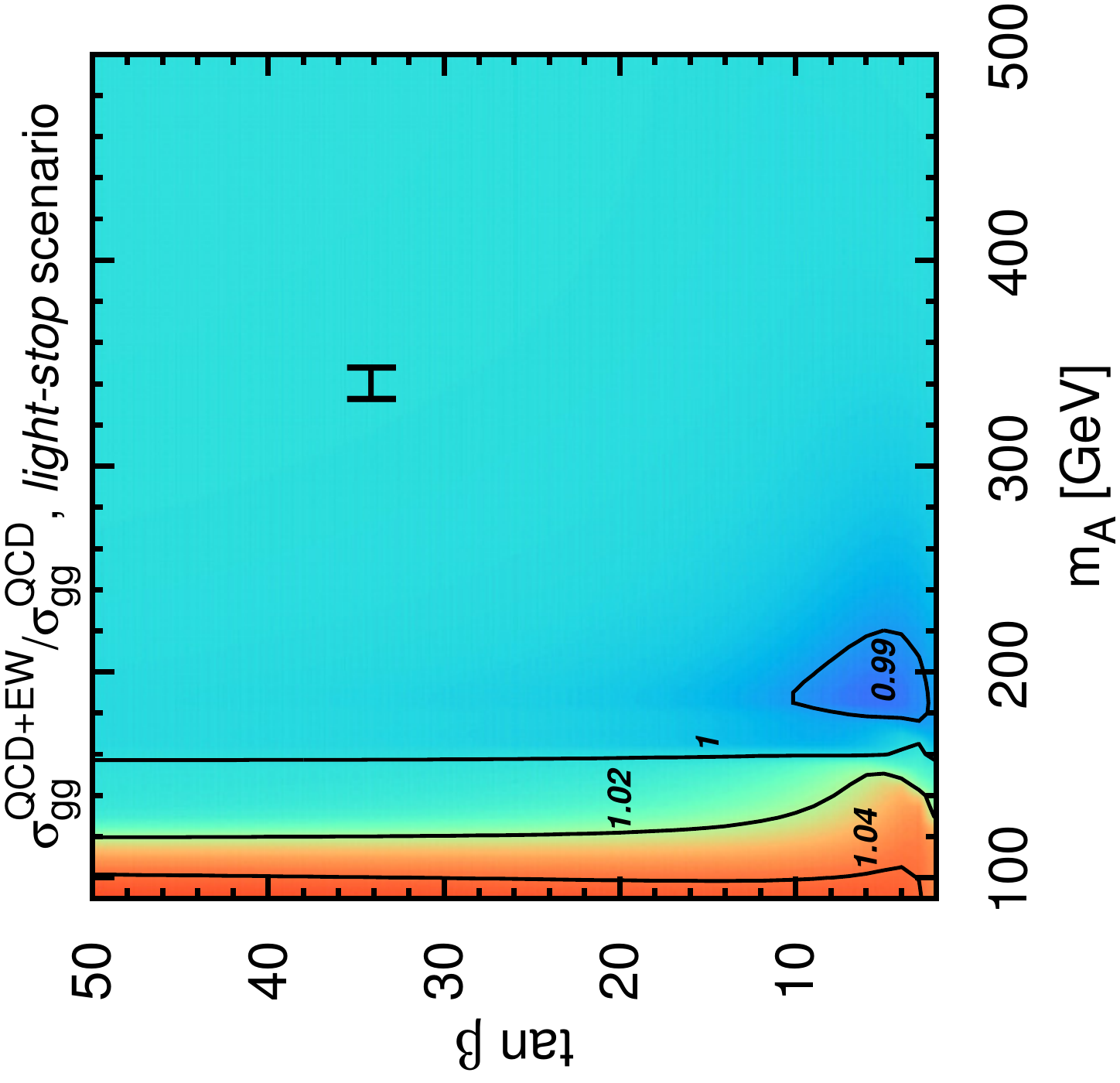}
\end{center}
\vspace*{-0.3cm}
\caption{Ratio of gluon-fusion cross section for the production of 
$h$ (left) and $H$ (right) over the corresponding cross section
neglecting EW contributions, as a function of $\ma$ and $\tb$ in
the \lss.}
\label{fig:ew-scalar}
\end{figure}

For a SM Higgs boson sufficiently lighter than the top threshold, the
EW corrections to gluon fusion are well
approximated~\cite{APSU,oai:arXiv.org:1007.1891} by the contributions
of two-loop diagrams in which the Higgs couples to EW gauge bosons,
which in turn couple to the gluons via a loop of light quarks
(including the bottom). In \sushi, these contributions are
incorporated in the MSSM calculation of the gluon-fusion cross section
by rescaling the two-loop EW amplitude given in
ref.~\cite{oai:arXiv.org:1007.1891} with the appropriate Higgs-gauge
boson couplings.\footnote{\,In fact, \sushi\ implements two
alternative procedures for including the EW contributions in the total
cross section for gluon fusion. We follow the one described in
eq.~(37) of the code's manual~\cite{Harlander:2012pb}.} In
figure~\ref{fig:ew-scalar} we investigate the impact of the
light-quark EW contributions on the production of the scalars $h$ and
$H$, plotting the ratio of the gluon-fusion cross sections computed
with and without those contributions, in the \matbp\ for the \lss.
The figure shows that the EW corrections tend to increase the cross
section, and their impact depends mainly on the strength of the
coupling of the considered scalar to gauge bosons. In the case of $h$
production (left plot) the EW corrections become fairly constant,
around $6\%$, in the region of sufficiently large $\ma$ where the
lightest scalar has SM-like couplings. Conversely, in the case of $H$
production (right plot) the EW corrections reach a comparable value
only in the strip of very low $\ma$, and they quickly drop below $1\%$
as soon as $\ma\gsim150$~GeV. On the other hand, since the
pseudoscalar does not couple to two gauge bosons at tree level, there
are no EW contributions from light-quark loops to its production.

For what concerns the remaining sources of EW corrections to gluon
fusion, those arising from two-loop diagrams involving top quarks are
known to be small for a SM-like Higgs with mass around $125$
GeV~\cite{APSU}, while in the case of $H$ and $A$ they are suppressed
in most of the parameter space by the small (or vanishing) Higgs
couplings to top quarks and to gauge bosons. On the other hand, the EW
corrections involving the bottom Yukawa coupling, which have not yet
been computed because they are negligible for the SM Higgs, could
become relevant for the production of $H$ and $A$. In addition, a full
computation of the EW corrections should include the contributions of
diagrams involving superparticles. The non-decoupling SUSY effects
that dominate at large $\tb$ are indeed included in an effective
Higgs-bottom coupling, as discussed in section \ref{sec:deltab}, but
the remaining contributions, so far uncomputed, could become relevant
if some of the superparticles are relatively light.

Results for the Higgs-production cross section in the other benchmark
scenarios listed in table~\ref{tab:scenarios} can be found in the
appendix. In the four scenarios denoted as $\mh^{\rm max}$, $\mh^{\rm
mod+}$, $\mh^{\rm mod-}$ and {\em light stau}, the couplings of the
Higgs bosons to top and bottom quarks and to gauge bosons are rather
similar to the ones in the \lss. Thus, the discussion given above for
the qualitative behavior in the \matbp\ of the total cross section, of
the EW corrections and of the relative importance of gluon fusion and
bottom-quark annihilation applies to those four scenarios as
well. However, all of the third-generation squarks have masses around
$1$~TeV, therefore the impact of the SUSY contributions on the
gluon-fusion cross section is considerably smaller than in the case of
the \lss. The suppression of the cross section for $h$ production in
the decoupling limit never goes beyond $6\%$. For what concerns $H$
production, the effect of the interference between quark and squark
contributions becomes significant only for very large $\ma$ and
moderate $\tb$, where the gluon-fusion cross section is tiny
anyway. The largest effect, a suppression by $30\,$--$\,40\%$, is
found in the {\em light-stau} scenario for $\ma \gsim 850$~GeV and
$10\lsim\tb\lsim20$, where the cross section is of the order of a
tenth of a femtobarn. The SUSY contributions to $A$ production,
already small in the \lss\ because they only arise at two loops, are
further suppressed in the $\mh^{\rm max}$, $\mh^{\rm mod+}$, $\mh^{\rm
mod-}$ and {\em light-stau} scenarios.

In the last scenario in table~\ref{tab:scenarios}, denoted as {\em
tau-phobic}, the MSSM parameters are arranged in such a way that, for
certain values of $\ma$ and $\tb$, the radiative corrections to the
$(1,2)$ element of the CP-even Higgs mass matrix suppress
significantly the mixing angle $\alpha$, so that the coupling of $h$
to taus -- which is proportional to $\sin\alpha$ -- is in turn
suppressed with respect to its SM value. However, the couplings of the
scalars to top and bottom quarks are modified as well, in particular
the coupling of $h$ to bottom quarks is suppressed.  As a result, in
the {\em tau-phobic} scenario the behavior in the \matbp\ of the
various contributions to the Higgs-production cross section differs
from the one found in the other scenarios. The total cross section for
$h$ production shows some enhancement with $\tb$ even for large values
of $\ma$, while for small $\ma$ the total cross section for $H$
production has a milder dependence on $\tb$ than in the other
scenarios. Also, the suppression of the $h$ coupling to bottom quarks
makes the contribution of bottom-quark annihilation to $h$ production
smaller than in the other scenarios. Finally, the {\em tau-phobic}
scenario is characterized by third-generation squark masses around
$1.5$~TeV, and by a value of the superpotential Higgs-mass parameter,
$\mu=2$~TeV, much larger than in the other scenarios. Since $\mu$
enters the couplings of the Higgs bosons to squarks, the impact of the
SUSY contributions on the cross section for scalar production is --
despite the heavier squarks -- somewhat larger than in the $\mh^{\rm
max}$, $\mh^{\rm mod+}$, $\mh^{\rm mod-}$ and {\em light-stau}
scenarios, and in the case of pseudoscalar production it is even
larger than in the \lss. 


\section{Sources of theoretical uncertainty}
\label{sec:uncertainty}

Like any other quantity evaluated perturbatively, the cross sections
for Higgs production in gluon fusion and bottom-quark annihilation
suffer from an intrinsic theoretical uncertainty due to the
truncation at finite order in the coupling constants. Typically, the
residual dependence on the renormalization and factorization scales is
used as an estimate of this uncertainty. In section~\ref{sec:scale} we
discuss our study of the scale dependence of the cross sections.

In addition, there are sources of uncertainty that are more specific
to the Higgs-production processes considered in this paper. As we
discuss in section~\ref{sec:bottomyuk}, one of the most important
sources of uncertainty in the production of Higgs bosons with
non-standard couplings to quarks is the dependence of the cross
section on the precise definition of the bottom-quark mass and Yukawa
coupling. The numerical difference between the pole bottom mass and
the running mass computed at a scale of the order of the Higgs mass is
more than $40\%$, and -- in a fixed-order calculation of the cross
sections -- the effect of such a large variation cannot be compensated
by the large logarithms that are induced at NLO by counterterm
contributions.  Furthermore, it is well known that the relation
between the bottom mass and the corresponding Yukawa coupling is
affected by potentially large, $\tb$-enhanced SUSY corrections that
must be properly resummed. The dependence of the cross sections on the
details of the resummation procedure constitutes a further source of
uncertainty.

In section~\ref{sec:pdfuncertainty} we discuss the uncertainties associated
to the choice of PDF sets.  We also investigate the issue of
consistency between the pre-defined value of the bottom mass in the
PDFs and the value of the mass used to extract the bottom Yukawa
coupling.

Finally, in section~\ref{sec:susyerr} we discuss two sources of
uncertainty arising from our incomplete knowledge of the SUSY
contributions to gluon fusion. In particular, we assess the validity
of the expansion in inverse powers of the SUSY masses used to
approximate the contributions of two-loop diagrams involving
superparticles. We also estimate the uncertainty associated to the
fact that \sushi\ does not include the contributions of three-loop
diagrams involving superparticles.


\subsection{Scale dependence of the cross section}
\label{sec:scale}

In this section we study the dependence of the cross section for Higgs
production on the renormalization scale $\muR$ at which the relevant
couplings in the partonic cross section are expressed, and on the
factorization scale $\muF$ entering both the PDFs and the partonic
cross section.  We recall that, although the complete result for the
hadronic cross section does not depend on $\muR$ and $\muF$, its
approximation at a given perturbative order retains a dependence on
those scales, which is formally one order higher than the accuracy of
the calculation. In a given calculation at fixed order, the two scales
are arbitrary, and they are typically fixed at some central values
$\muzeroR$ and $\muzeroF$ characteristic of the hard scattering
process. The variation of the scales around their central values
provides an estimate of the size of the uncomputed higher-order
contributions.

We discuss separately the cases of gluon fusion (section
\ref{sec:uncert-gluon-fusion}) and of bottom-quark annihilation
(section \ref{sec:uncert-bottom-fusion}). In the former, $\muR$
denotes the scale at which we express the strong gauge coupling
entering the partonic cross section already at the LO, while in the
latter it denotes the scale at which we express both the bottom Yukawa
coupling entering at the LO and the strong gauge coupling entering at
the NLO. We postpone to section \ref{sec:bottomyuk} a discussion of
the dependence of the gluon-fusion cross section on the scale at which
we express the bottom Yukawa coupling.

\subsubsection{Gluon fusion}
\label{sec:uncert-gluon-fusion}

The natural hard scale in the production of a Higgs boson $\phi$ is
obviously of the order of $m_{\phi}$.  In our study of gluon fusion we
take $\muzeroR=\muzeroF=m_\phi/2$ as central values for the
renormalization and factorization scales, because, with this choice,
the cross section shows a reduced sensitivity to scale variations and
an improved convergence of the perturbative
expansion~\cite{Anastasiou:2002yz}.  Moreover, it has been observed
that this choice allows to mimic the effects of soft-gluon resummation
in the total cross section~\cite{Anastasiou:2005qj}.

We study the impact of the scale variation around the central choice
$(\muzeroR,\muzeroF)$ following the \WG\
prescription~\cite{Dittmaier:2011ti}: we consider seven combinations
of renormalization and factorization scales, defined as the set
$C_\mu$ of the pairs $(\muR,\muF)$ obtainable from the two sets
$\muR=\{m_\phi/4,\,m_\phi/2,\,m_\phi\}$ and
$\muF=\{m_\phi/4,\,m_\phi/2,\,m_\phi\}$, with the additional
constraint that $1/2 \leq \muR/\muF \leq 2$ (i.e., we treat the
variations of the ratio $\muR/\muF$ on the same footing as the
variations of the individual scales, discarding the two pairs where
the ratio varies by a factor of four around its central value).  We
then determine the maximal and minimal values of the cross section on
the set $C_\mu$,
\begin{equation}
\label{eq:sigmapm}
\sigma^-~\equiv~ \min_{(\muR,\,\muF)\,\in \,C_\mu} \{\sigma(\muR,\muF)
\}~,~~~~~~~
\sigma^+~\equiv~ \max_{(\muR,\,\muF)\,\in\, C_\mu} \{\sigma(\muR,\muF) \}~,
\end{equation}
and define the relative scale uncertainty of the cross section as
$\Delta_\mu \equiv \Delta^+_\mu - \Delta^-_\mu$, where
\begin{equation}
\label{eq:deltamu}
\Delta^+_\mu ~\equiv~
\frac{\sigma^+-\sigma(\muzeroR,\muzeroF)}{\sigma(\muzeroR,\muzeroF)}~~,
~~~~~~~
\Delta^-_\mu ~\equiv~
\frac{\sigma^--\sigma(\muzeroR,\muzeroF)}{\sigma(\muzeroR,\muzeroF)}~~.
\end{equation}

\begin{figure}[p]
\begin{center}
\includegraphics[width=0.49\textwidth]{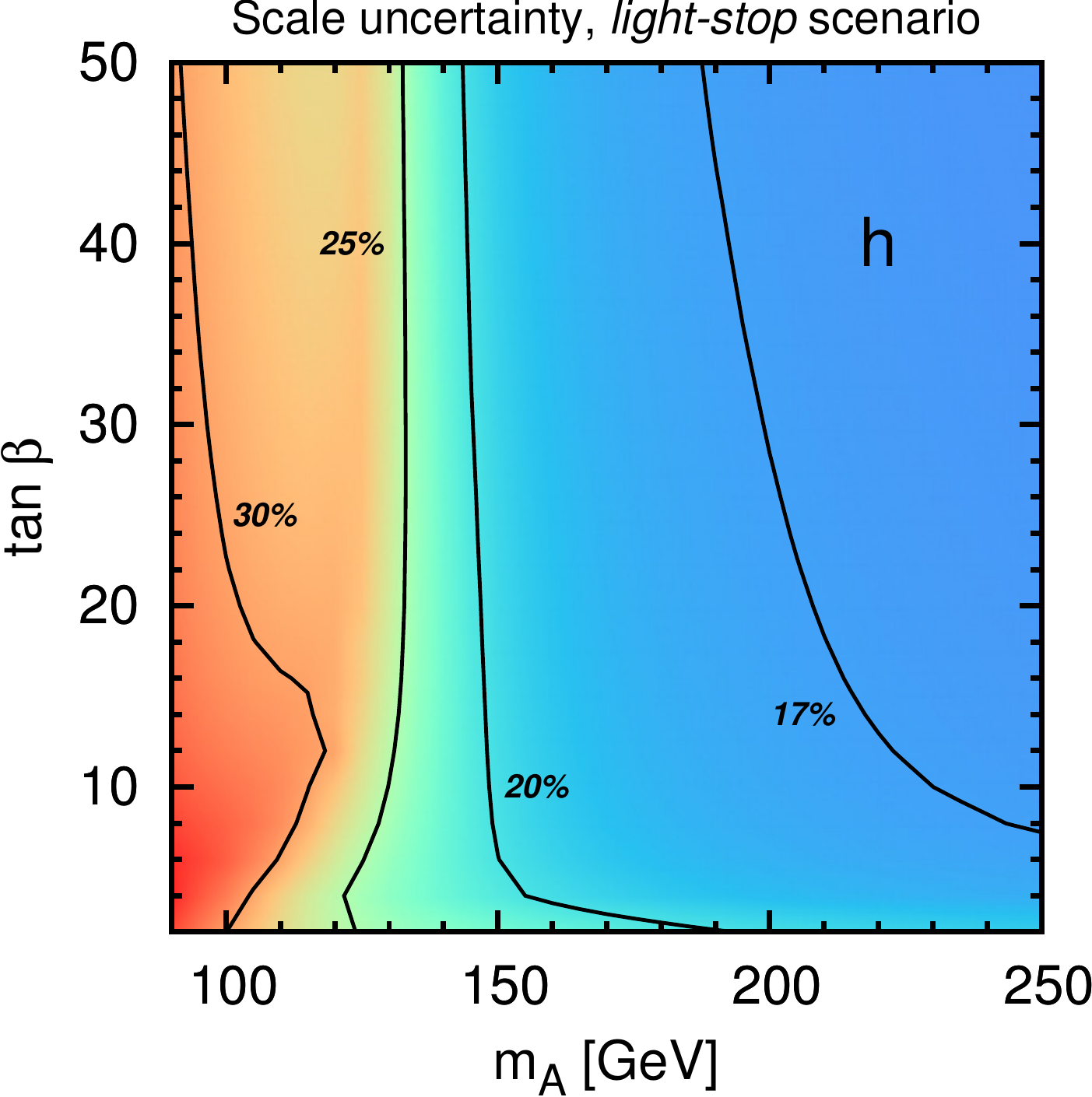}~~~
\includegraphics[width=0.49\textwidth]{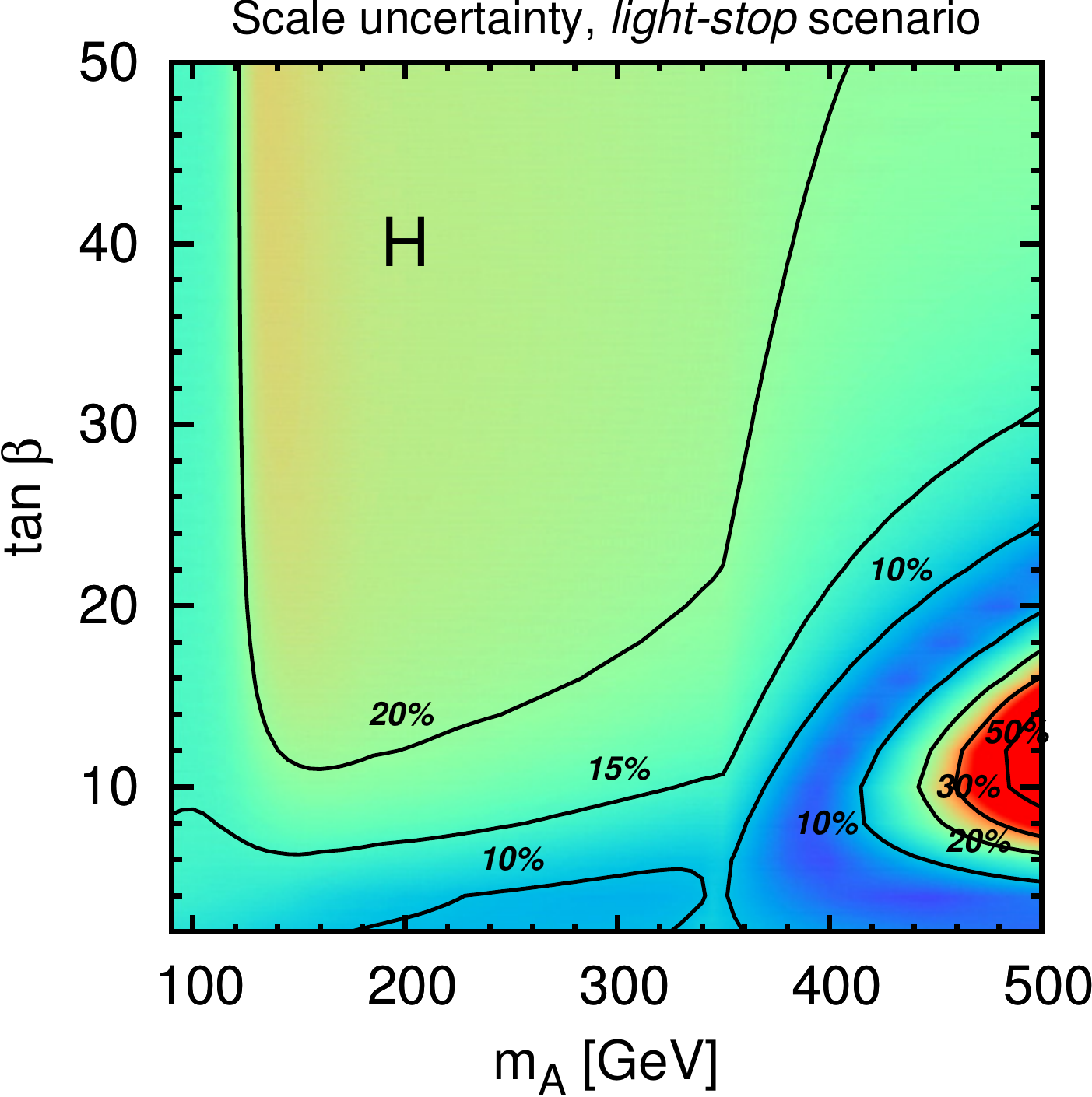}
\end{center}
\vspace*{-0.3cm}
\caption{Relative scale uncertainty $\Delta_\mu$ (in percent) for $h$
  production (left) and $H$ production (right) in gluon fusion in the \lss.}
\label{fig:ggh-light-heavy-murmuf}
\end{figure}
\begin{figure}[p]
\begin{center}
\includegraphics[width=0.49\textwidth]{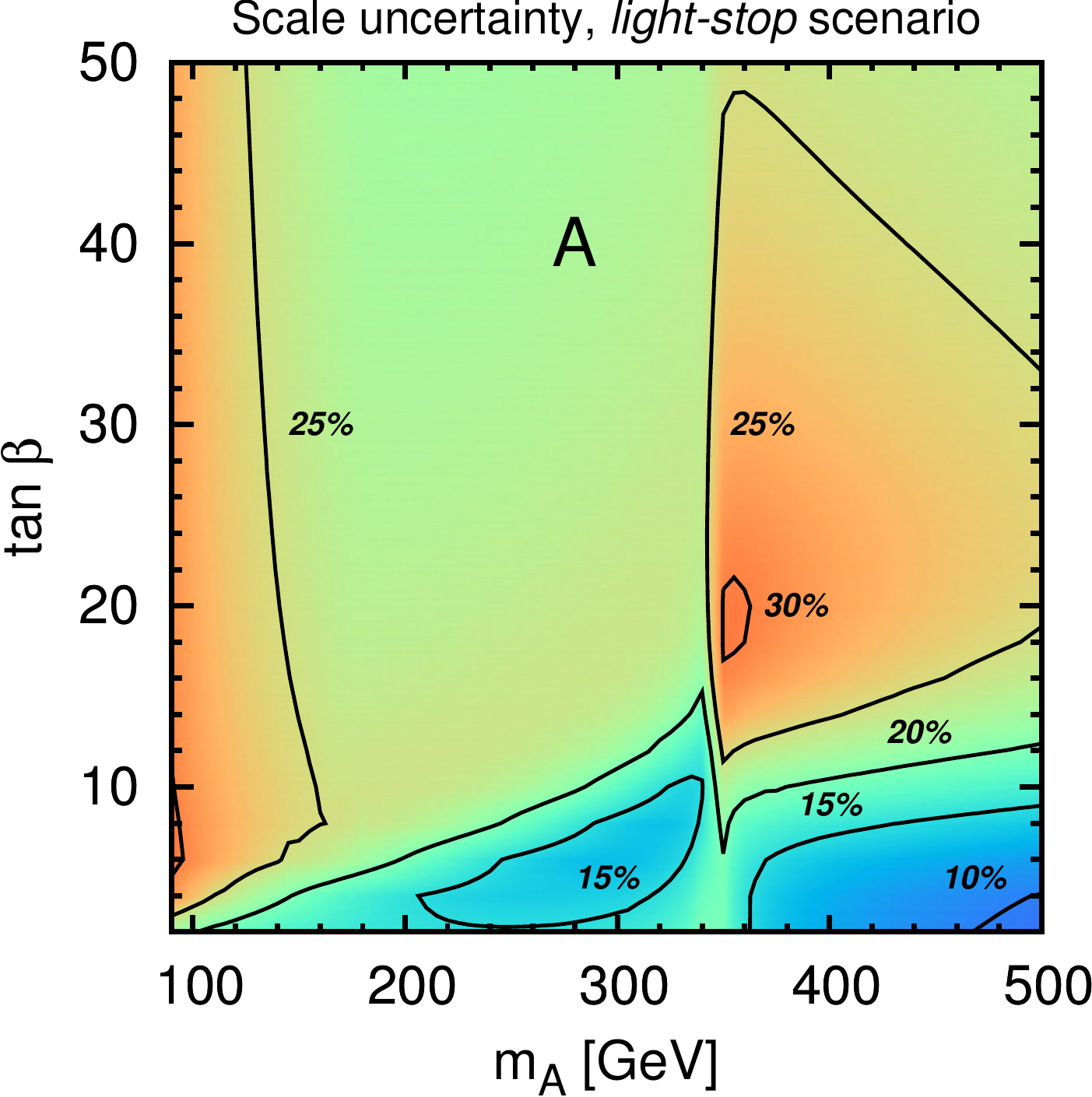}
\end{center}
\vspace*{-0.3cm}
\caption{Same as figure \ref{fig:ggh-light-heavy-murmuf} for the production of
  the pseudoscalar $A$.}
\label{fig:ggh-odd-murmuf}
\end{figure}

In figures~\ref{fig:ggh-light-heavy-murmuf}
and~\ref{fig:ggh-odd-murmuf} we show the contours of equal
$\Delta_\mu$ for scalar and pseudoscalar production in the \matbp,
fixing the MSSM parameters as in the \lss.  
The qualitative features of the plots can be understood by considering
that the top, bottom, SUSY and EW contributions to the gluon-fusion
cross section are known at different orders in the perturbative
expansion. In particular, the top contribution is included in \sushi\
with full mass dependence through ${\cal O}(\alpha_s^3)$ (i.e., NLO)
and in the VHML at ${\cal O}(\alpha_s^4)$ (i.e, NNLO).  Its residual
scale dependence amounts to an ${\cal O}(\alpha_s^5)$ effect, with the
exception of some mass-dependent effects at ${\cal O}(\alpha_s^4)$,
which are known to be numerically small~\cite{H2gQCD3}.  The bottom
and sbottom contributions are included at the NLO and they account for
an ${\cal O}(\alpha_s^4)$ effect. The stop contributions are included
through the NNLO, see section~\ref{sec:susyerr}, but their effect on
scale dependence is also of ${\cal O}(\alpha_s^4)$ because we neglect
the genuine three-loop terms. Finally, while the EW corrections are
computed at ${\cal O}( \alpha \alpha_s^2)$, their inclusion as a fully
factorized term at the NLO causes their effect on scale variation to
be of ${\cal O}( \alpha \alpha_s^4)$, numerically very small.
As a consequence of the varying accuracy of the different
contributions, the scale uncertainty for the production of a given
Higgs boson depends on which contribution plays the dominant role in
the considered region of the \matbp. The uncertainty is lowest, around
$10\,$--$\,20\%$, where the top contribution dominates: this is the
case for $h$ production (left plot in
figure~\ref{fig:ggh-light-heavy-murmuf}) in the decoupling region,
where the uncertainty stabilizes to roughly $16\%$ at large $\ma$
(i.e., slightly smaller than the $18\%$ we obtain for the same Higgs
mass in the \sm); for $H$ production (right plot in
figure~\ref{fig:ggh-light-heavy-murmuf}) in the strip with $\ma \lsim
120$~GeV, as well as when $\tb\lsim 10$ and $\ma \lsim 400$~GeV; for
$A$ production (figure~\ref{fig:ggh-odd-murmuf}) in the strip with
$\tb\lsim 10$. In contrast, the scale uncertainty exceeds $20\%$ in
the regions where the bottom contribution is enhanced or downright
dominant: at large $\tb$ for $H$ and $A$ production, and at small
$\ma$ for $h$ production.

The plots for $H$ and $A$ production in
figures~\ref{fig:ggh-light-heavy-murmuf} and~\ref{fig:ggh-odd-murmuf}
show additional structures. In the case of $H$ production, the scale
uncertainty becomes very large for $8 \lsim \tb \lsim 16$ and $\ma
\gsim 460$~GeV. As appears from figure~\ref{fig:squark-scalar}, this
region is characterized by a significant cancellation between the top,
bottom and stop contributions to the gluon-fusion amplitude, resulting
in a very small NLO cross section and an enhanced sensitivity to
higher-order effects.  In the case of $A$ production, the structure
visible for $\ma \approx 350$~GeV is associated to the cusp-like
behavior of the top contribution to the gluon-fusion amplitude around
the threshold $\ma = 2\,\mt$. Another feature of $H$ and $A$
production, partially overshadowed by the structures described above,
is a tendency towards smaller scale uncertainties for larger
pseudoscalar (and hence scalar) masses. This is due to the fact that
the strong gauge coupling -- which controls the size of the
higher-order effects that we are estimating -- is evaluated at a scale
proportional to the mass of the considered Higgs boson, and gets
smaller when the scale increases.

The other scenarios were studied following the same procedure, and the
results are qualitatively similar.  For $h$ production, the
scale dependence in the decoupling region is similar to, or even
bigger than, the one in the \sm. For $H$ production, due to the
different interplay of quark and squark contributions, the
cancellations that in the \lss\ cause the region of very large
uncertainty for $8 \lsim \tb \lsim 16$ and $\ma \gsim 460$~GeV occur
at higher values of $\ma$.

Finally, a study of independent variations of the renormalization and
factorization scales shows that, in a large fraction of the parameter
space, the former yield a much larger uncertainty than the latter.
The factorization-scale uncertainty is smaller in size than the
renormalization-scale uncertainty already at the LO, and it is further
reduced by the inclusion of higher-order terms.

\subsubsection{Bottom-quark annihilation}
\label{sec:uncert-bottom-fusion}

\begin{figure}[p]
\begin{center}
\includegraphics[width=0.49\textwidth]{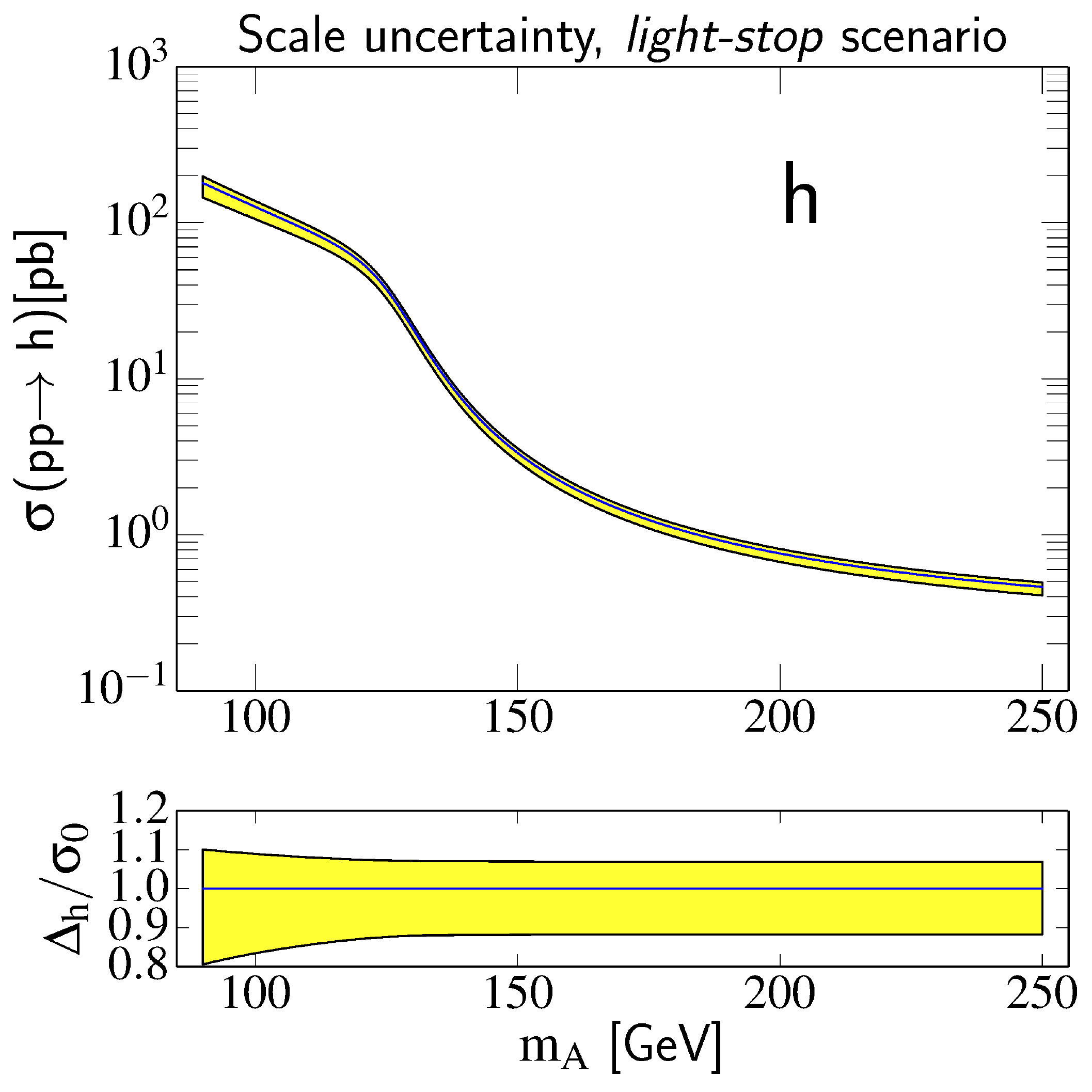}~~~~~
\includegraphics[width=0.49\textwidth]{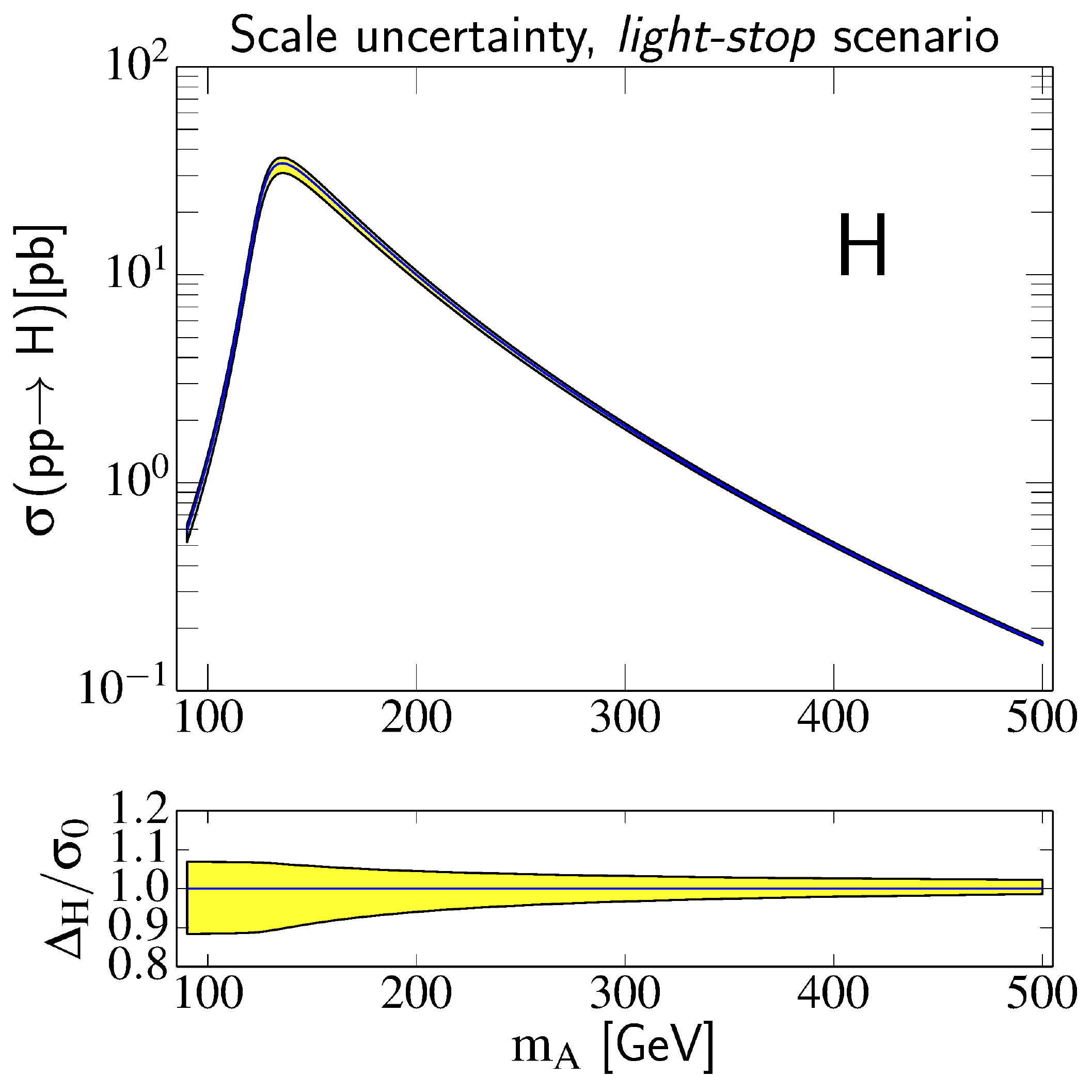}
\end{center}
\vspace*{-0.3cm}
\caption{Scale uncertainty of the cross section for $h$ production
  (left) and $H$ production (right) in bottom-quark annihilation, in
  the \lss\ with $\tb = 20$.}
\label{fig:bbh-light-heavy-murmuf}
\end{figure}
\begin{figure}[p]
\begin{center}
\includegraphics[width=0.49\textwidth]{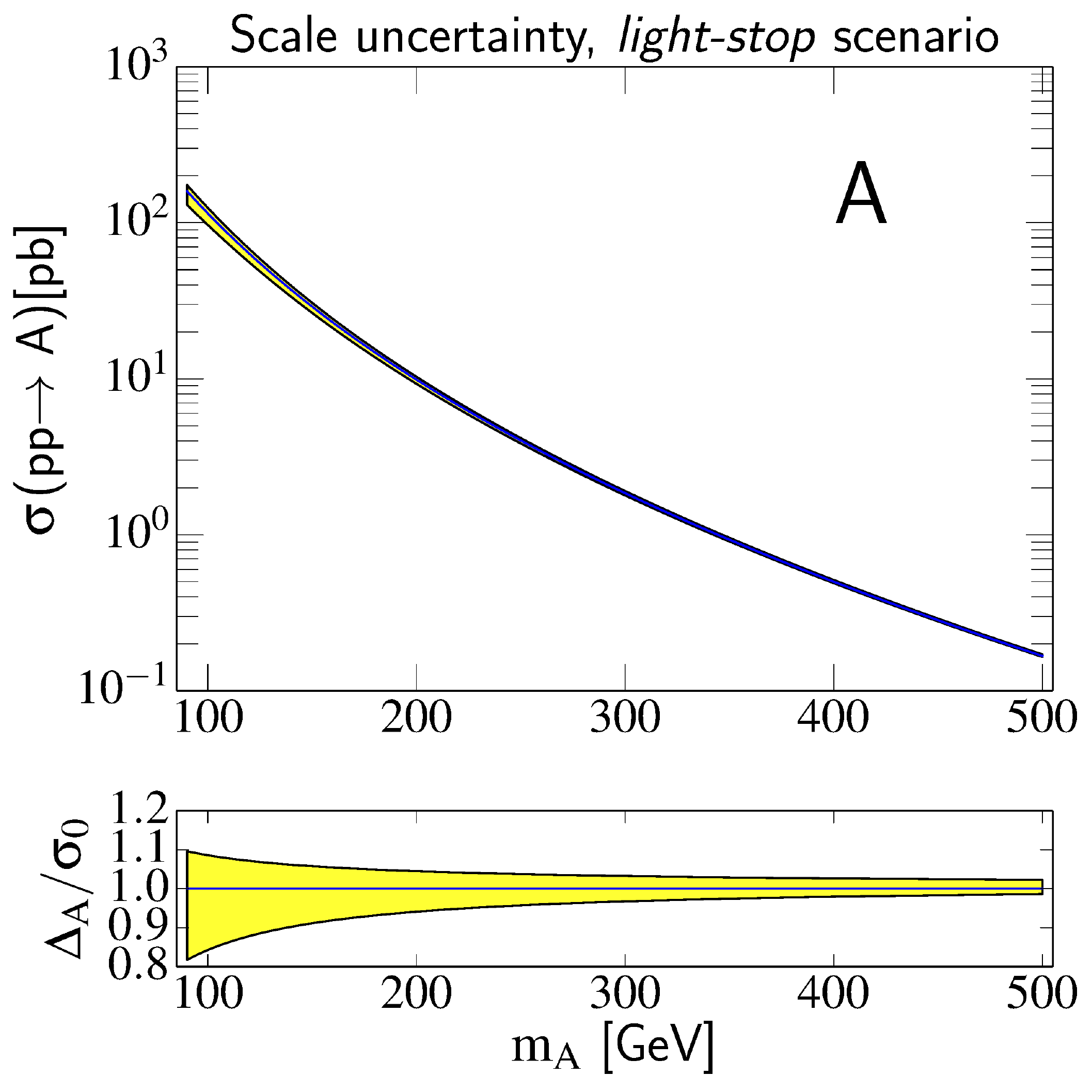}
\end{center}
\vspace*{-0.3cm}
\caption{Same as figure \ref{fig:bbh-light-heavy-murmuf} for the
  production of the pseudoscalar $A$.}
\label{fig:bbh-odd-murmuf}
\end{figure}

In \sushi, the cross section for Higgs production in bottom-quark
annihilation is implemented at NNLO-QCD in the 5FS. Our default choice
for the central scales is $\muzeroR = m_{\phi}$ and $\muzeroF =
m_{\phi}/4$, following the observation that radiative corrections are
particularly small for this value of the factorization
scale~\cite{5FSNLO,Harlander:2003ai,Boos:2003yi}. To study the
uncertainty associated to the variation of the scales, we consider
seven combinations corresponding to all possible pairings of
$\muR=\{m_\phi/2,m_\phi,2\,m_\phi\}$ and
$\muF=\{m_\phi/8,m_\phi/4,m_\phi/2\}$, with the additional constraint
that $2 \leq \muR/\muF \leq 8$ (again, we discard the two pairs with
the largest variation of $\muR/\muF$ around its central value, which
in this case is $4\,$).  We then determine the scale uncertainty
$\Delta_\mu$ in analogy to eqs.~(\ref{eq:sigmapm}) and
(\ref{eq:deltamu}).

Differently from the case of gluon fusion, the scale uncertainty of
bottom-quark annihilation depends very weakly on $\tb$. This is due to
the fact that, in eq.~(\ref{eq:deltamu}), the $\tb$-dependence of the
cross section via the effective Higgs-bottom coupling cancels out in
the ratio, leaving only a mild, indirect dependence -- only for scalar
production -- via the value of the Higgs mass that determines $\muR$ and
$\muF$. 

In figures~\ref{fig:bbh-light-heavy-murmuf}
and~\ref{fig:bbh-odd-murmuf} we show the scale dependence of the cross
section for scalar and pseudoscalar production, respectively, as a
function of $\ma$ in the \lss\ with $\tb=20$. In the upper part of
each plot, the solid line denotes the cross section for bottom-quark
annihilation computed with the central scale choice
$(\muzeroR,\muzeroF)$, while the yellow band around the solid line is
delimited by the maximal and minimal cross sections $\sigma^+$ and
$\sigma^-$, defined in analogy to eq.~(\ref{eq:sigmapm}). The lower
part of each plot shows the relative variation of the cross section
with respect to the central value (i.e., the total width of the yellow
band corresponds to $\Delta_\mu$). While the values of the total cross
section do of course depend on the chosen benchmark scenario, the
relative scale variation is essentially the same in all scenarios, due
to the above-mentioned cancellation of the dependence on the effective
Higgs-bottom coupling.

The left plot in figure~\ref{fig:bbh-light-heavy-murmuf} shows that
the relative scale uncertainty of the cross section for $h$ production
can be as large as $30\%$ for low values of $\ma$, then it stabilizes
to roughly $18\%$ in the decoupling region where $\mh$ becomes
independent of $\ma$. In contrast, the relative scale uncertainty of
the cross section for the production of $H$ (right plot in
figure~\ref{fig:bbh-light-heavy-murmuf}) and $A$
(figure~\ref{fig:bbh-odd-murmuf}) decreases as $\ma$ (and hence $\mH$)
increases. As already mentioned for the case of gluon fusion, this
behavior is due to the fact that the higher-order effects that we are
estimating are controlled by the strong gauge coupling, and the latter
decreases when the scale at which it is computed, which is
proportional to the Higgs mass, increases.

Finally, an independent variation of the renormalization and
factorization scales shows that, in this case, the dominant
uncertainty is given by the dependence on the factorization scale.


\subsection{Definition of the Higgs-bottom coupling}
\label{sec:bottomyuk}

In the production of a SM-like Higgs boson, the contribution of
bottom-quark annihilation and the effect of the bottom-quark loops in
gluon fusion amount to a few percent of the total cross
section. Therefore, in that case the theoretical uncertainty
associated to the definition of the Higgs coupling to bottom quarks is
negligible compared to other sources of uncertainty. On the other
hand, this uncertainty becomes significant in scenarios where the
Higgs-bottom coupling is enhanced with respect to its SM counterpart,
$\yb^{\smallsm} = \sqrt{2}\,\mb/v$ (here $v \approx 246$ GeV). In the
MSSM the tree-level couplings of the neutral Higgs bosons to bottom
quarks are modified as follows:
\be
\label{couplings}
\yb^h ~=~ -\frac{\sin\alpha}{\cos\beta}~\yb^{\smallsm}~,~~~~~~~
\yb^\smallH ~=~ \frac{\cos\alpha}{\cos\beta}~\yb^{\smallsm}~,~~~~~~~
\yb^\smalla ~=~ \tb~\yb^{\smallsm}~,
\ee
where $\alpha$ is the mixing angle in the CP-even Higgs sector.  In
the decoupling limit, $\ma \gg \mz$, the mixing angle simplifies to
$\alpha \approx \beta -\pi/2$, so that the coupling of $h$ to bottom
quarks is SM-like, while the couplings of $H$ and $A$ are both
enhanced by $\tb$.

In this section we discuss two issues that affect the precise
definition of the Higgs-bottom couplings: the first concerns the
choice of renormalization scheme -- and scale -- for the bottom mass
from which the couplings are extracted; the second concerns
higher-order effects in the procedure through which the $\tb$-enhanced
SUSY contributions are resummed in effective Higgs-bottom couplings.

\subsubsection{Scheme and scale dependence of the bottom mass}
\label{sec:mub}

The parameter $\mb$ enters the expression for the gluon-fusion
amplitude with two distinct roles: as the actual mass of the bottom
quarks running in the loops, and as a proxy for the Higgs-bottom
coupling $\ybp$, where $\phi = \{h,H,A\}$. The numerical value of
$\mb$ depends strongly on the renormalization scheme and scale: an
$\msbar$ mass $\mb(\mb) = 4.16$~GeV corresponds to a pole mass
$\mbp=4.92$~GeV at three-loop level, whereas evolving $\mb(\mb)$ up to
a scale of the order of the typical energy of the gluon-fusion process
decreases significantly its value. For example, if we evolve at
four-loop level the bottom mass up to the scale at which we express
the strong gauge coupling, $\muR=m_\phi/2$, we obtain $\mb(m_\phi/2) =
2.93$~GeV for $m_\phi = 125$~GeV. While any change in the definition
of the bottom mass and Yukawa coupling entering the one-loop part of
the amplitude is formally compensated for, up to higher orders, by
counterterm contributions in the two-loop part, the numerical impact
of such strong variations on the prediction for the gluon-fusion cross
section can be significant.

To illustrate this point, we identify the mass of the bottom quarks in
the loops with the pole mass $\mbp$, and consider the dependence of
the gluon-fusion cross section on the prescription for the
Higgs-bottom coupling $\ybp$, focusing on $\phi = \{h,H\}$. In the
\lss\ with $\ma=130$~GeV and $\tb=40$, where both Higgs scalars are
relatively light and have enhanced couplings to the bottom quark, the
effect of extracting $\ybp$ from the $\msbar$ mass $\mb(\mb)$ instead
of the pole mass $\mbp$ leads to a $17\%$ decrease in the cross
section for $h$ production, and a $24\%$ decrease in the cross section
for $H$ production. The use of $\mb(m_\phi/2)$ would instead decrease
the cross section for $h$ production by $34\%$, and the one for $H$
production by $51\%$, with respect to the values obtained with $\mbp$.
As a second example, we take the \lss\ with $\ma=300$~GeV and
$\tb=10$, where the lightest scalar $h$ has SM-like couplings to
quarks. In this case the cross section for $h$ production varies by
less than $2\%$ when choosing among the three options discussed above
for the definition of $\yb^h$. For the heaviest scalar $H$, on the
other hand, the changes in the cross section relative to the value
derived with $\mbp$ amount to $-22\%$ and $-50\%$ when $\yb^\smallH$
is extracted from $\mb(\mb)$ and $\mb(\mH/2)$, respectively.

The strong sensitivity of the production of non-standard Higgs bosons
on the choice of renormalization scheme (and scale) for the bottom
mass and Yukawa coupling has been discussed in the past, see
e.g.~refs.~\cite{SDGZ,iHixs,Baglio:2010ae}.  However, unlike many
other processes for which there are theoretical arguments in favor of
one or the other choice, for Higgs production in gluon fusion we are
not aware of any such arguments that go beyond heuristic. As was
already noted in ref.~\cite{SDGZ}, the options of relating $\ybp$ to
$\mbp$ or to $\mb(\mb)$ might seem preferable to the one of using
$\mb(m_\phi/2)$, in that they lead to smaller two-loop contributions.
If in the one-loop part of the amplitude for scalar production we
identify the mass of the bottom quark with $\mbp$ and the bottom
Yukawa coupling with $\mb(\mu_b)$, where $\mu_b$ is a generic
renormalization scale, the contribution of diagrams with bottom quarks
and gluons to the two-loop part of the amplitude reads
\be
\label{eq:Ab2loop}
{\cal A}_b^{2\ell}(\tau) ~~\propto~~ C_F\,\left[{\cal F}_{C_F} (\tau)
+ {\cal F}_{1/2}^{1\ell}(\tau)\,\left(1-\frac34\,\ln\frac{\mb^2}{\mu_b^2}
\right)\right] \,+~ C_A\,{\cal F}_{C_A} (\tau)~,
\ee
where $C_F= 4/3$ and $C_A=3$ are color factors, $\tau =
4\,\mb^2/m_\phi^2$, and we omit an overall multiplicative
factor. Truncating the functions at the first order in an expansion in
powers of $\tau$, one finds~\cite{ABDV}
\bea
\label{eq:Ab1loop}
{\cal F}_{1/2}^{1\ell}(\tau) & = & 
-2\,\tau\,\left(1-\frac14\,L_{ b\phi}^2\right)\,+~{\cal O}(\tau^2)~,\\
\label{eq:FCF}
{\cal F}_{C_F} (\tau) & = & -\tau\,\left[
5 + \frac95\, \zeta_2^2-\zeta_3-(3+\zeta_2+4\,\zeta_3)\,L_{b\phi}
+\zeta_2\,L_{ b\phi}^2 + \frac14\,L_{ b\phi}^3+\frac1{48}\,L_{ b\phi}^4\right]
\,+~{\cal O}(\tau^2)~,\\
\label{eq:FCA}
{\cal F}_{C_A} (\tau) & = & -\tau\,\left[
3 - \frac85\, \zeta_2^2-3\,\zeta_3+3\,\zeta_3\,L_{b\phi}
-\frac14(1+2\,\zeta_2)\,L_{ b\phi}^2 -\frac1{48}\,L_{ b\phi}^4\right]
\,+~{\cal O}(\tau^2)~,
\eea
with 
\be
L_{ b\phi} ~\equiv~ \ln(-4/\tau) ~=~ \ln(m_\phi^2/\mb^2) -i\,\pi~. 
\ee

The equations above show that the two-loop bottom contribution to the
gluon-fusion amplitude contains powers of $\ln(m_\phi^2/\mb^2)$, and
that the choice $\mu_b = \mb$ does eliminate some of the
logarithmically enhanced terms. Similarly, relating the coupling
entering the one-loop part of the amplitude to the pole mass $\mbp$
eliminates the whole piece proportional to ${\cal
  F}_{1/2}^{1\ell}(\tau)$ in eq.~(\ref{eq:Ab2loop}). Each of the two
remaining terms, $C_F\,{\cal F}_{C_F}(\tau)$ and $C_A\,{\cal
  F}_{C_A}(\tau)$, also contains powers of $\ln(m_\phi^2/\mb^2)$, but
for realistic values of $m_\phi$ the two terms largely cancel out
against each other, resulting in a small two-loop contribution from
bottom quarks.
However, such cancellation should be considered accidental: there is
no argument suggesting that it persists at higher orders in QCD, or
that it is motivated by some physical property of the bottom
contribution to gluon fusion. To illustrate this point, we can
consider the case of Higgs decay to two photons: the one-loop bottom
contribution to the amplitude has the same structure as the
corresponding contribution to gluon fusion, but the two-loop
bottom-gluon contribution is obtained from eq.~(\ref{eq:Ab2loop}) by
dropping the term proportional to $C_A$, which originates from
diagrams with three- and four-gluon interactions. In that case no
significant cancellation occurs, and the amplitude is not minimized
when $\ybp$ is extracted from $\mb(\mb)$ or $\mbp$. In fact, it was
also noted in ref.~\cite{SDGZ} that the two-loop bottom-gluon
contribution to the amplitude for Higgs decay to photons is minimized
when the one-loop contribution is fully expressed in terms of
$\mb(m_\phi/2)$.

\begin{figure}[p]
\begin{center}
\includegraphics[angle=270,width=0.49\textwidth]{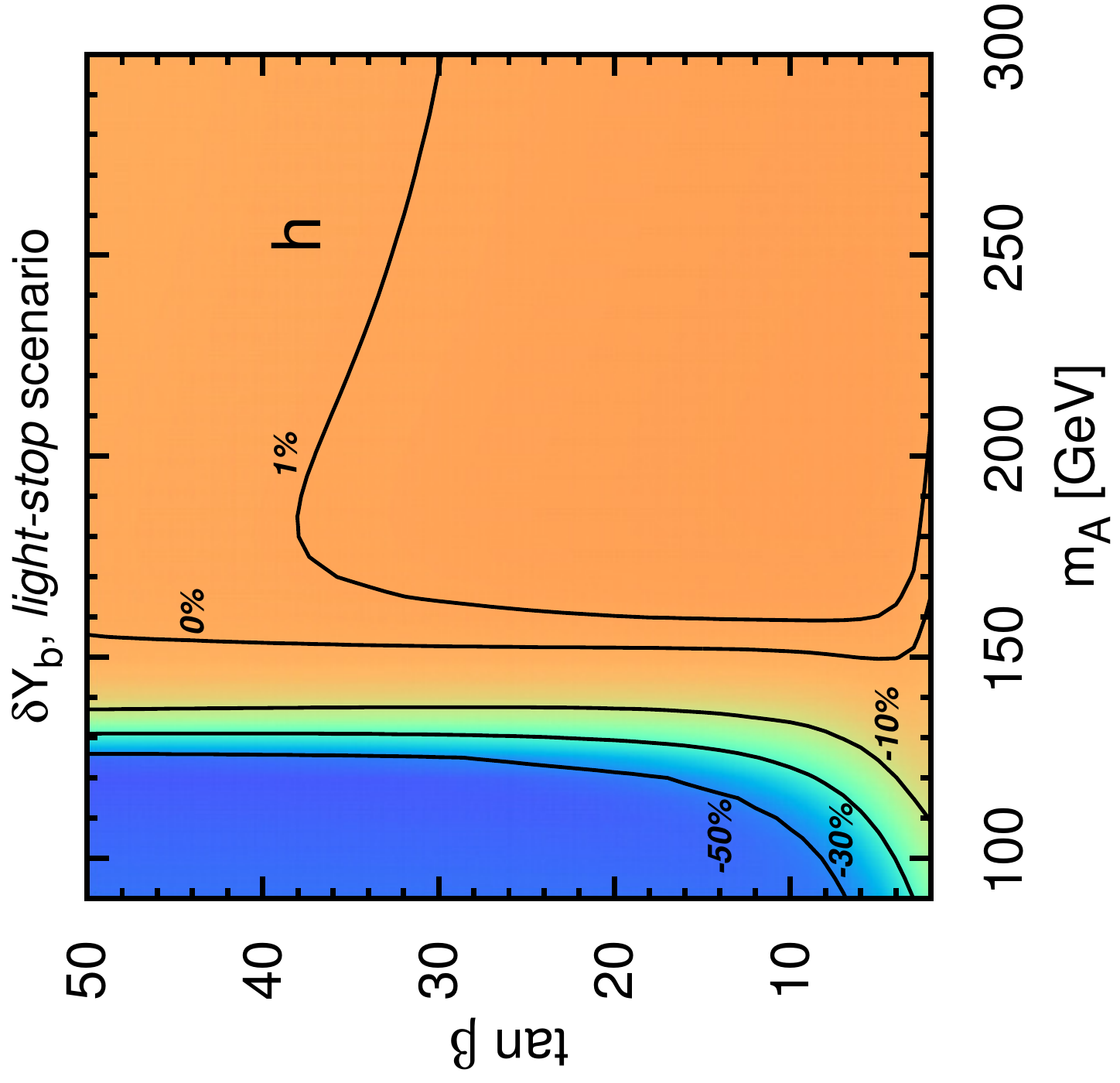}~~~
\includegraphics[angle=270,width=0.49\textwidth]{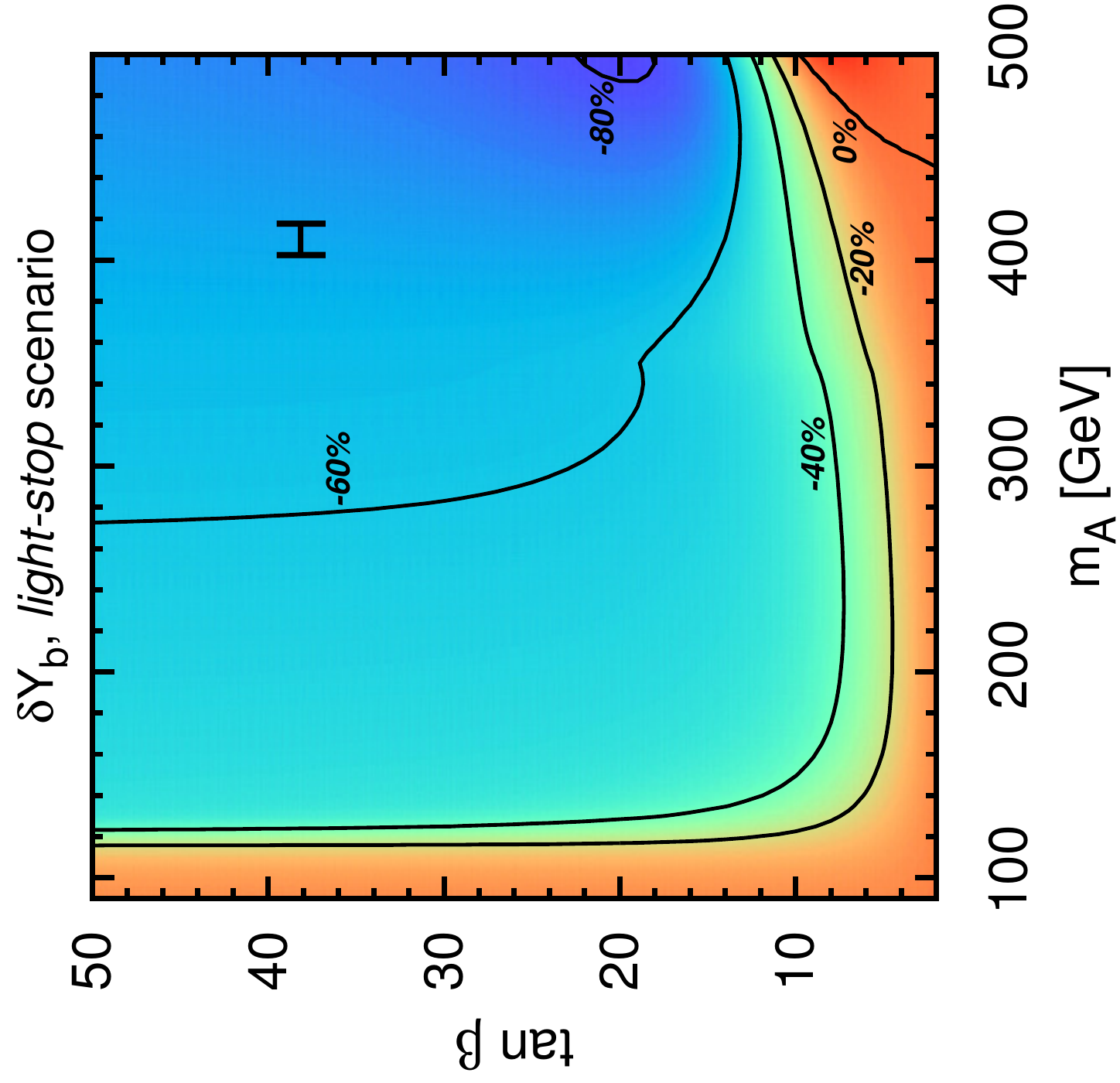}
\end{center}
\vspace*{-0.3cm}
\caption{Variation (in percent) of the gluon-fusion cross section for
  the production of $h$ (left) and $H$ (right) when the Higgs-bottom
  coupling $\ybp$ is extracted from $m_b(m_\phi/2)$ instead of $\mbp$,
  as a function of $\ma$ and $\tb$ in the \lss.}
\label{fig:bottom-scalar}
\end{figure}
\begin{figure}[p]
\begin{center}
\includegraphics[angle=270,width=0.49\textwidth]{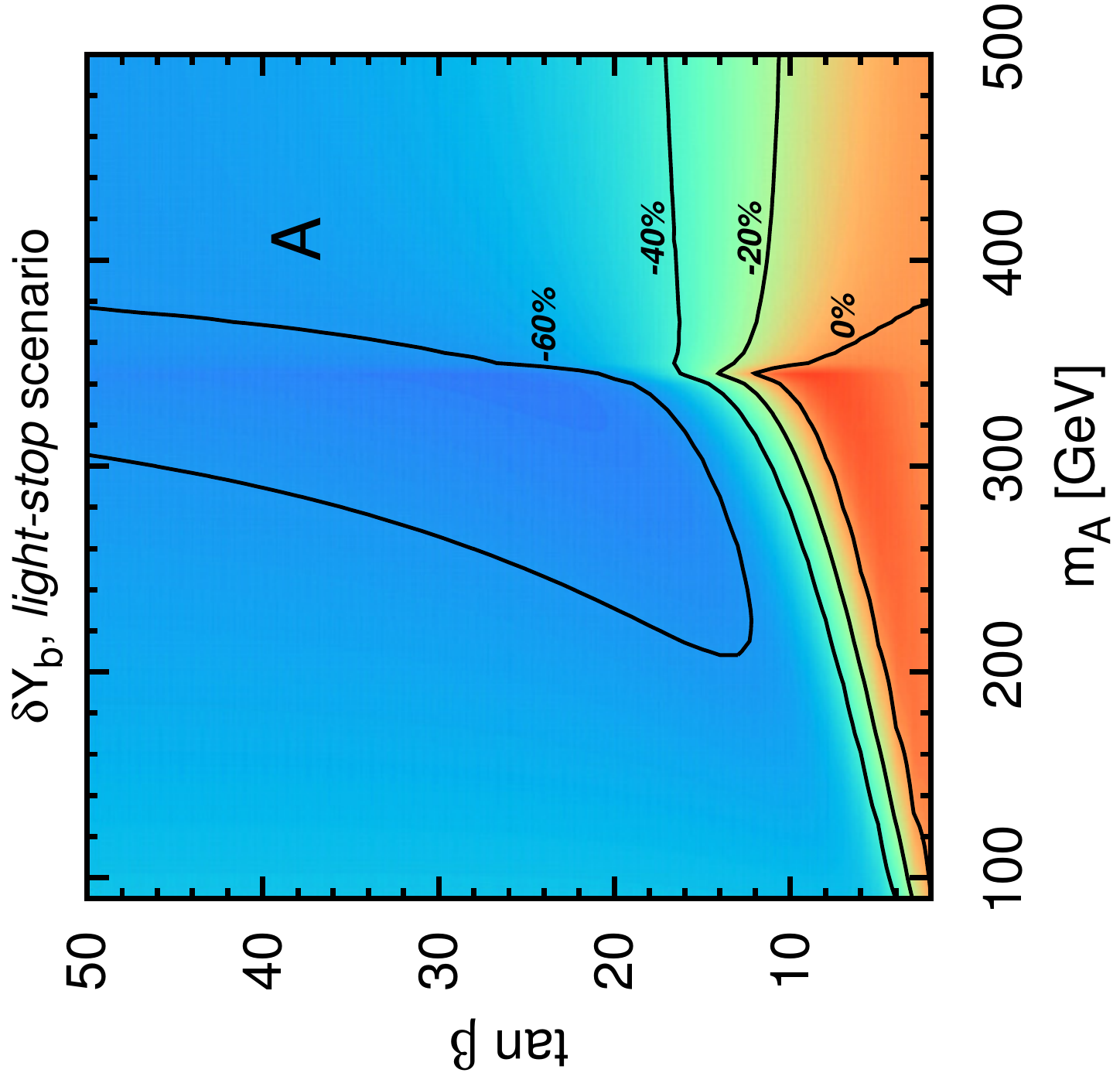}
\end{center}
\vspace*{-0.3cm}
\caption{Same as figure \ref{fig:bottom-scalar} for the production of
  the pseudoscalar $A$.}
\label{fig:bottom-pseudo}
\end{figure}

In the case of the Higgs coupling to photons, the problems related to
the ambiguity in the definition of $\ybp$ have been solved with a
resummation of the leading and next-to-leading logarithms of the ratio
$m_\phi^2/\mb^2$~\cite{resum}. Until a similar calculation is
performed for the Higgs coupling to gluons, there is no obvious reason
to favor one choice of renormalization scheme (and scale) for the
bottom Yukawa coupling over the others. In our study we choose to
relate the coupling to the pole mass $\mbp$, and we consider the
difference between the results obtained using $\mbp$ and those
obtained using $\mb(m_\phi/2)$ as a measure of the uncertainty
associated with the uncomputed higher-order QCD corrections. For the
production of a SM-like Higgs with mass around $125.5$~GeV, this
procedure -- also advocated by the \WG\ in
ref.~\cite{Dittmaier:2011ti} -- results in an uncertainty of
$1\,$--$\,2\%$ in the gluon-fusion cross section. On the other hand,
as we show in figures \ref{fig:bottom-scalar} and
\ref{fig:bottom-pseudo} for scalar and pseudoscalar production in the
\lss, the cross section could be reduced by more than $60\%$ in the
regions of the \matbp\ where the gluon-fusion process is dominated by
the bottom-quark contribution. It is however worth recalling that, as
shown in figures \ref{fig:ratio-scalar} and \ref{fig:ratio-pseudo}, in
such regions the total cross section for Higgs production is dominated
by bottom-quark annihilation. In the 5FS, the cross section for the
latter process is known at the NNLO in
QCD~\cite{5FSNLO,Harlander:2003ai}, and it is free of large logarithms
of the ratio $m_\phi^2/\mb^2$ when $\ybp$ is related to
$\mb(m_\phi)$. The theoretical uncertainty of the cross section for
bottom-quark annihilation associated to reasonable variations around
this scale choice is already included in the uncertainty bands shown
in figures~\ref{fig:bbh-light-heavy-murmuf}
and~\ref{fig:bbh-odd-murmuf} in the previous section.

\subsubsection{Resummation of {\boldmath $\tan\beta$}-enhanced
  corrections}
\label{sec:deltab}

It is well known that, in the MSSM, loop diagrams involving
superparticles induce $\tan\beta$-enhanced corrections to the
couplings of the Higgs bosons to bottom quarks~\cite{hrs}.  If all
superparticles are considerably heavier than the Higgs bosons they can
be integrated out of the MSSM Lagrangian, leaving behind a
two-Higgs-doublet model with effective Higgs-bottom couplings
\be
\label{effcoup}
\ybeff^{h} ~=~ \frac{\yb^{h}}{1+\Db}
\left(1-\Db\,\frac{\cot\alpha}{\tan\beta}\right),~~~\,
\ybeff^{\smallH} ~=~ \frac{\yb^{\smallH}}{1+\Db}
\left(1+\Db\,\frac{\tan\alpha}{\tan\beta}\right),~~~\,
\ybeff^{\smalla} ~=~ \frac{\yb^{\smalla}}{1+\Db}
\left(1-\Db\,\cot^2\beta\right)\,,
\ee
where $\ybp$ are the tree-level Higgs-bottom couplings defined in
eq.~(\ref{couplings}), and, retaining only the $\oas$ contribution
from diagrams with sbottoms and gluinos, the $\tb$-enhanced term $\Db$
reads
\be
\label{deltab}
\Db ~=~ \frac{2\,\as}{3\,\pi}~\frac{\mg\,\mu\,\tb}{\bu-\bd}\,
\left(\frac{\bu}{\bu-\mg^2} \ln \frac{\bu}{\mg^2}
-\frac{\bd}{\bd-\mg^2} \ln \frac{\bd}{\mg^2}\right)\,.
\ee

In the limit $\ma \gg \mz$, where $\cot\alpha \approx -\tb$, the
superparticle contributions encoded in $\Db$ decouple from the
coupling of the lightest scalar, while the couplings of the heaviest
scalar and of the pseudoscalar are both rescaled by a factor
$(1-\Db\,\cot^2\beta)/(1+\Db)$.

In refs.~\cite{effL,GHS} it was shown that, in the calculation of
processes that involve the Higgs-bottom couplings, the $\tb$-enhanced
corrections can be resummed to all orders in the expansion in powers
of $\Db$ by inserting the effective couplings of eq.~(\ref{effcoup})
in the lowest-order amplitude for the considered process. In the case
of gluon fusion, this amounts to using $\ybeffp$ in the bottom
contribution to the one-loop part of the amplitude. However, when this
resummation procedure is combined with the actual calculation of the
superparticle contributions to the one- and two-loop amplitude for
gluon fusion, care must be taken to avoid double counting. To this
effect, we must subtract from the full result for the two-loop
amplitude the contribution obtained by replacing $\ybeffp$ in the
resummed one-loop amplitude with the ${\cal O}(\Db)$ term of the
expansion of $\ybeffp$ in powers of $\Db$. Depending on the choice of
renormalization scheme for the parameters in the sbottom sector,
additional $\tb$-enhanced terms could be induced in the two-loop
amplitude by the counterterm of the Higgs-sbottom coupling that enters
the sbottom contribution to the one-loop amplitude. To avoid the
occurrence of large two-loop corrections, which would put the validity
of the perturbative expansion into question, we employ for the sbottom
sector the OS renormalization scheme described in
ref.~\cite{Degrassi:2010eu}.

An ambiguity in the procedure for the resummation of the $\Db$ terms
concerns the treatment of the Higgs-bottom couplings entering the
two-loop part of the gluon-fusion amplitude. The difference between
the results obtained using either $\ybp$ or $\ybeffp$ in the two-loop
part is formally of higher order, i.e., it amounts to three-loop terms
that are suppressed by a factor $A_b/(\mu\tb)$ with respect to the
dominant three-loop terms of ${\cal O}(\Db^2)$ accounted for by the
resummation.  Nevertheless, in our study we choose to identify the
Higgs-bottom couplings in both the one- and two-loop parts of the
amplitude with $\ybeffp$. We found that this choice allows us to
reproduce -- after an expansion in powers of $\Db$ -- the three-loop
result that can be inferred from ref.~\cite{GHS}, where the
sub-dominant terms proportional to $A_b$ were also resummed in the
effective couplings.

For large values of $\tb$, the factor $\Db$ can even become of order
one, unless the superpotential parameter $\mu$ is suppressed with
respect to the soft SUSY-breaking masses. The effect of the SUSY
correction on the effective Higgs-bottom couplings depends crucially
on the sign of $\Db$. For positive $\Db$ the correction suppresses the
couplings, reducing the overall relevance of the bottom contribution
to gluon fusion. On the other hand, for negative $\Db$ the correction
enhances the couplings, which diverge as $\Db$ approaches $-1$. As a
consequence, when $\Db$ is large and negative the result for the
gluon-fusion cross section is extremely sensitive to the precise value
of $\Db$, and a refined calculation of the latter becomes mandatory to
reduce the uncertainty associated to the bottom contribution.

The first obvious step to improve the calculation of $\Db$ consists in
including other one-loop contributions that are not shown in
eq.~(\ref{deltab}). In particular, the diagrams with stops and
charginos induce a contribution, controlled by the top Yukawa
coupling, that can be comparable in size with the $\oas$ contribution
in eq.~(\ref{deltab}). In our numerical analysis we use by default the
full one-loop result for $\Db$ as computed by \FH, which allows us to
resum in our prediction for the Higgs-production cross section also
the $\tb$-enhanced corrections of electroweak origin.

Another improvement in the calculation would come from the inclusion
of the dominant two-loop contributions to $\Db$, which have been
computed in ref.~\cite{deltab2l} but are not yet implemented in
\FH. Indeed, it was shown in ref.~\cite{deltab2l} that the one-loop
result for $\Db$ is particularly sensitive to changes in the
renormalization scales at which the strong-gauge and top-Yukawa
couplings are expressed, and that the inclusion of the two-loop
contributions stabilizes this scale dependence. In particular, both
the one-loop sbottom-gluino and stop-chargino contributions to $\Db$
vary by roughly $\pm 10\%$ when the renormalization scales are lowered
or raised by a factor of two around their central values, which are
chosen as the average of the masses of the relevant superparticles.
We can therefore estimate the uncertainty of the gluon-fusion cross
section associated to the one-loop computation of $\Db$ by varying by
$\pm 10\%$ the result provided by \FH.

In general, the impact of the uncertainty of $\Db$ on the total
uncertainty of the gluon-fusion cross section depends on the
considered point in the MSSM parameter space. As was the case also for
the scheme and scale dependence of $\ybp$ discussed in the previous
section, the $\Db$ uncertainty can be significant only if the bottom
contribution to the cross section is substantially enhanced with
respect to the SM case.  For illustration, we consider again the
\lss\ with $\ma=130$~GeV and $\tb=40$, where both Higgs scalars have
enhanced couplings to bottom quarks. The superpotential parameter
$\mu$ has positive sign, and the $\Db$ corrections suppress the
effective couplings $\ybeffp$. We find that the cross sections for $h$
and $H$ production in gluon fusion increase by $4\%$ and $7\%$,
respectively, if the value of $\Db$ is reduced by $10\%$, while they
decrease by $4\%$ and $6\%$, respectively, if $\Db$ is increased by
$10\%$. The effect is larger if $\mu$ is taken negative, so that the
$\Db$ corrections enhance the effective couplings. In that case the
dependence on $\Db$ is reversed: if we consider the same point in the
\lss\ but flip the sign of $\mu$, the cross sections for $h$ and $H$
production in gluon fusion decrease by $17\%$ and $16\%$,
respectively, when $|\Db|$ is reduced by $10\%$, while they increase
by $23\%$ and $21\%$, respectively, when $|\Db|$ is increased by
$10\%$.

Finally, we stress that a similar uncertainty affects the cross
section for Higgs production via bottom-quark annihilation, where the
tree-level amplitude is computed in terms of the effective couplings
$\ybeffp$. Also in this case, we can estimate the uncertainty by
varying by $\pm 10\%$ the value of $\Db$ provided by \FH.


\subsection{Uncertainties from the PDFs and $\alpha_s$}
\label{sec:pdfuncertainty}

The prediction for the total cross section at hadron level is affected
by our imperfect knowledge of the proton PDFs.  This uncertainty has
different sources: the PDFs cannot be computed from first principles
but they rather have to be fitted from data, and the experimental
error of the latter affects the outcome of the fit and propagates to
the prediction of any observable.  Also, the choices related to the
fitting methodology and to the mathematical representation of the PDFs
induce an ambiguity in the results, as can be appreciated by comparing
the PDF parameterizations provided by three collaborations that perform
a global fit of low- and high-energy data:
MSTW2008~\cite{Martin:2009iq}, CT10~\cite{Lai:2010vv} and
NNPDF2.3~\cite{Ball:2012cx}. These uncertainties will be discussed in
section~\ref{sec:PDFalphas}, together with the parametric dependence
of the cross section on the value of the strong coupling constant.
Another source of uncertainty is related to the available
perturbative-QCD information on the scattering processes from which
the PDFs are extracted.  Among these perturbative effects, an issue
that is particularly relevant in the case of Higgs production via
bottom-quark annihilation is the consistent inclusion of the
bottom-mass effects in the evolution of the PDFs according to the
DGLAP equations. The transition between four and five active flavors
in the proton occurs at a matching scale that is set equal to the
bottom mass.  The bottom density in the proton depends parametrically
on this matching scale, which in turn affects the predictions for the
cross section.  The phenomenological implications of this issue will
be discussed in detail in section~\ref{sec:pdfbottom}.  A systematic
discussion of further sources of theoretical uncertainty -- such as,
e.g., the dependence of the PDFs on the choice of renormalization and
factorization scale in the matrix elements that are used to perform
the fit -- is not yet available in the literature, and goes beyond the
scope of this paper.

\subsubsection{Combination of PDF and $\alpha_s$ uncertainties}
\label{sec:PDFalphas}
The uncertainty associated to the experimental errors of the data from
which the PDFs are extracted is represented by the PDF collaborations
with the introduction of $\Nrep$ different PDF sets (replicas), all
equivalent from the statistical point of view in the description of
the data.  Any observable has to be computed $\Nrep$ times with the
different sets, and the spread of the results can be interpreted as
the error induced by the PDF due to the data and to the fitting
methodology.  The replicas are determined by the PDF collaborations
following the Hessian (for MSTW2008 and CT10) or the Monte Carlo (for
NNPDF2.3) approaches, and the PDF error has to be computed
accordingly.  In QCD the cross sections are also affected by a
parametric uncertainty associated to the input value of the strong
coupling constant. This dependence is particularly relevant in the
gluon-fusion cross section, which is proportional to $\alpha_s^2$ at
the LO and is subject to very large QCD corrections, of ${\cal
  O}(\alpha_s^3)$, at the NLO.  Each PDF collaboration recommends a
different central value for $\alpha_s(\mz)$, generating a spread of
the central predictions for the Higgs-production cross section.  The
combination of the PDF and $\alpha_s$ uncertainties (henceforth,
\pdfas) and their correlation was first discussed in
ref.~\cite{Demartin:2010er}.  A conservative approach to combine the
different predictions obtained using the MSTW2008, CT10 and NNPDF2.3
PDF sets is known as PDF4LHC recipe, and it amounts to taking the
envelope of the \pdfas\ uncertainty bands of the three collaborations,
where for each group the preferred $\alpha_s(\mz)$ central value is
adopted~\cite{pdf4lhc}.  Following this reference we take
$\Delta\alpha_s=\pm 0.0012$ for the experimental error on the strong
coupling constant.

Due to the very steep behavior of the PDFs for increasing values of
the final-state invariant mass, the gluon-fusion process receives its
dominant contribution from the threshold production region, with a
very important role played by the virtual corrections and by the
universal, factorizable, soft-gluon corrections. Consequently, the
cross section is dominated by the LO-kinematics configurations also at
higher perturbative orders.  At the LO, the gluon-fusion cross section
depends on the rapidity of the Higgs boson only through the PDFs,
therefore the relative size of the \pdfas\ uncertainty does not depend
on the details of the partonic process, but only on the value of the
Higgs-boson mass.  As a consequence, the relative \pdfas\ uncertainty,
for a given value of the Higgs mass, can be read directly from the
tables of the SM predictions reported in the appendix B of the latest
\WG\ report~\cite{Heinemeyer:2013tqa}.  Differences with respect to
the SM predictions may originate from hard, process-dependent
radiative corrections, but their impact on the relative
\pdfas\ uncertainty is at the sub-percent level.

To assess the \pdfas\ uncertainty of the cross section for Higgs
production in bottom-quark annihilation we adopt again the PDF4LHC
recipe.  The bottom density in the proton does not have an intrinsic
component, but it is generated dynamically, via gluon splittings, by
the DGLAP evolution of the PDFs. Therefore, the uncertainties of the
bottom and gluon PDFs are strongly correlated. 

Similarly to the case of gluon fusion, for a given value of the Higgs
mass the relative \pdfas\ uncertainty of the cross section for
bottom-quark annihilation differs very little between the SM and the
MSSM, because the radiative corrections involving SUSY particles
affect the kinematics of the process only at higher
orders.\footnote{In \sushi\ the SUSY corrections to bottom-quark
  annihilation enter only through the effective couplings
  $\,\ybeffp\!$, therefore our estimate of the \pdfas\ uncertainty for
  a given Higgs mass is exactly the same in the SM and in the MSSM.}
We find that the uncertainty has an almost constant behavior when the
mass $m_\phi$ of the produced Higgs boson is lighter than $300$~GeV,
and that it increases for larger mass values: for example, at the
NNLO, the \pdfas\ uncertainty of the cross section for bottom-quark
annihilation amounts to $\pm\, 6/6/8/21\%$ for
$m_\phi=124/300/500/1000$~GeV.


\subsubsection{Bottom-mass dependence of the PDFs}
\label{sec:pdfbottom}

The calculation of hadronic cross sections involves the convolution of
the partonic cross sections with the PDFs, which have an intrinsic
dependence on the bottom mass. For example, the central set of
MSTW2008~\cite{Martin:2009iq}, which we use as default for our
analysis, assumes a pole mass $\mbp=4.75$~GeV. Converted to the
$\msbar$ mass via a three-loop QCD calculation, this corresponds to
$\mb(\mb)=4.00$~GeV, which differs both from the value
recommended by the \WG, $\mb(\mb)=4.16$~GeV~\cite{SMinputs,botmass},
and from the current PDG value,
$\mb(\mb)=4.18$~GeV~\cite{Beringer:1900zz}.

In addition to their dependence through the PDFs, the cross sections
for Higgs production also depend on the bottom mass at the partonic
level, i.e., through the bottom Yukawa coupling, the bottom-quark
propagators and the phase space. In the regions of the MSSM parameter
space where the bottom-quark contributions to Higgs production are
enhanced, it becomes vital to evaluate the partonic cross sections
with the correct input value for the bottom mass, which, as mentioned
above, may not necessarily correspond to the value used in the
PDFs. In this section we will examine the uncertainty that arises when
we choose the bottom mass entering the partonic cross sections
independently from the PDF set.

The MSTW2008 PDFs come in seven sets obtained with $\mbp$ ranging from
$4$ GeV to $5.5$ GeV in steps of $0.25$ GeV. In
ref.~\cite{Martin:2010db} the MSTW collaboration studied the
sensitivity of the PDFs on the value of the bottom mass, showing that
the PDFs for the gluon and for the four lightest quarks are almost
insensitive to $\mbp$, whereas the bottom PDF exhibits quite a strong
dependence. As shown in figure 6 of ref.~\cite{Martin:2010db}, a
variation by $\pm 0.5$~GeV around the central value $\mbp=4.75$~GeV
leads to changes in the bottom PDF that exceed the $90$\%
C.L.~uncertainty, even for the relatively large value of the
factorization scale relevant to Higgs production, $\muF \approx
100$~GeV.

The cross section for Higgs production via gluon fusion is mostly
sensitive to the gluon PDF, and receives only a small contribution,
starting at the NLO, from diagrams with initial-state bottom
quarks. As a result, when we evaluate the gluon-fusion cross section
with the seven PDF sets -- while fixing the bottom mass in the
partonic cross section -- we find that the result changes only at the
per mil level, independently of the phenomenological scenario under
consideration. We conclude that, for this process, the formal
inconsistency of choosing different values for the bottom mass in the
partonic cross section and in the PDFs induces only a negligible
uncertainty.

\begin{figure}[t]
\begin{center}
\includegraphics[width=0.48\textwidth]{%
  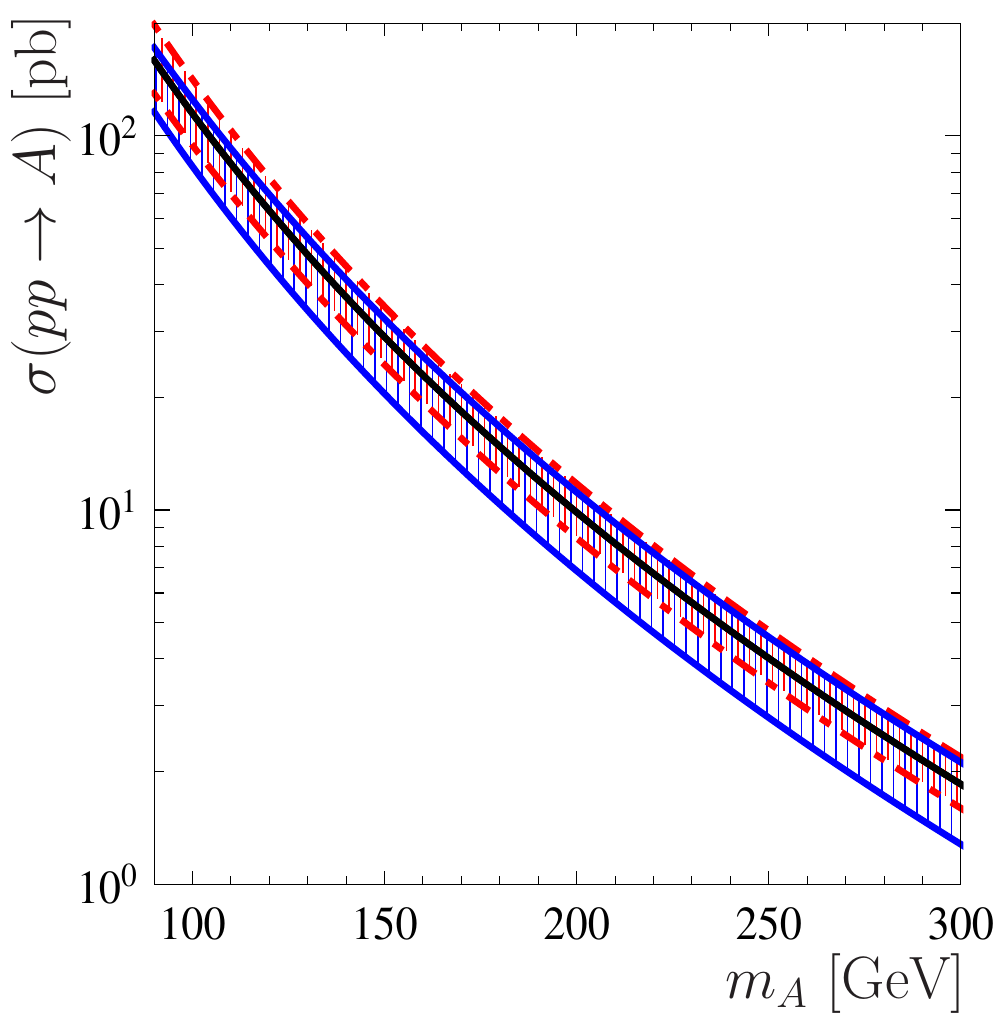}~~~~
\includegraphics[width=0.48\textwidth]{%
  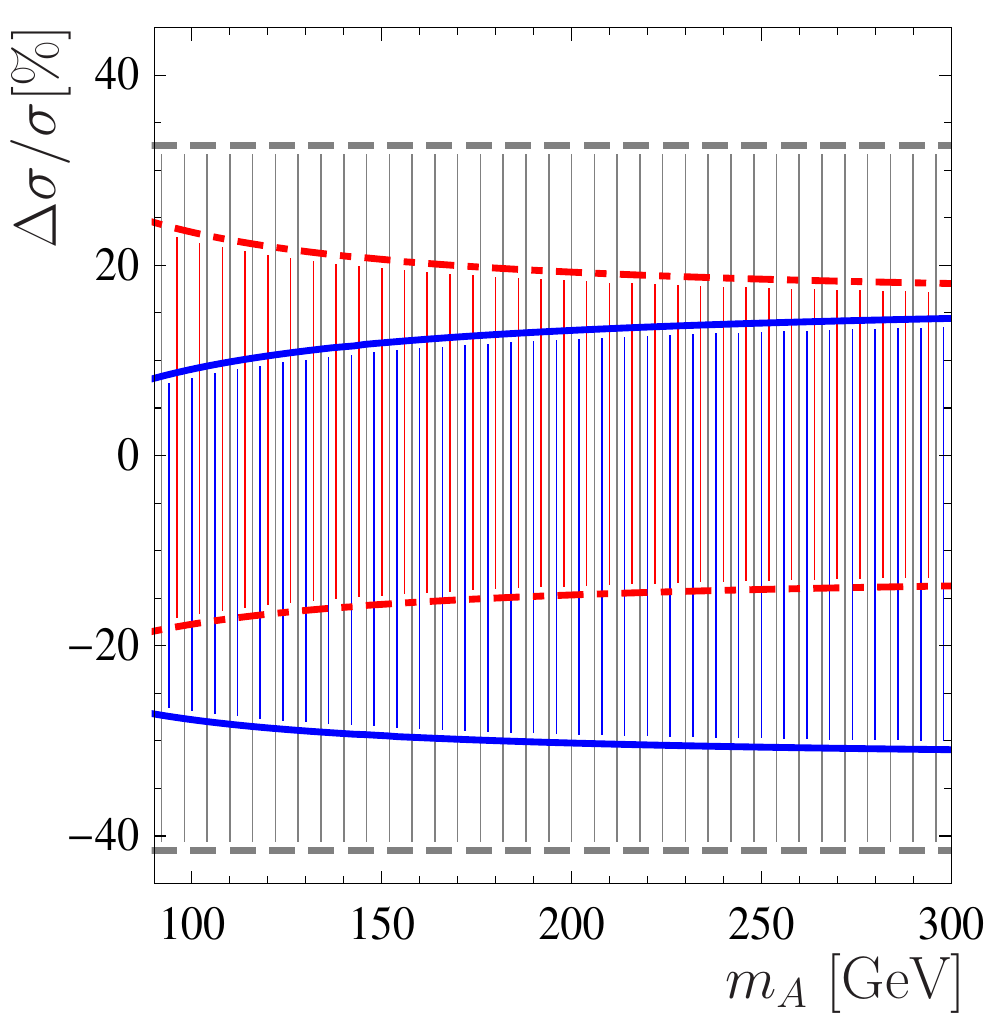}
\end{center}
\vspace*{-3.5mm}
\caption{(Left) Cross section for pseudoscalar Higgs production in
  bottom-quark annihilation as a function of $\ma$ for the \lss\ with
  $\tb=20$. (Right) Relative variation of the cross section for
  different choices of the pole bottom mass used in the PDFs and of the
  running mass used in the partonic cross section. Red: PDF variation,
  $\ybeff^{\smalla}$ fixed; gray: PDF fixed, $\ybeff^{\smalla}$ varies;
  blue: PDF and $\ybeff^{\smalla}$ vary simultaneously.}
\label{fig:bbhmbdep}
\end{figure}

In contrast, the hadronic cross section for Higgs production in
bottom-quark annihilation, when computed in the 5FS, depends directly
on the bottom PDF. As a result, we expect this process to show a
significant dependence on the value of the bottom mass used in the
PDFs, and the issue of consistency with the bottom mass used in the
definition of the bottom Yukawa coupling becomes unavoidable.

In figure~\ref{fig:bbhmbdep} we investigate the bottom-mass dependence
of the hadronic cross section for pseudoscalar production in
bottom-quark annihilation (we find similar behaviors for the
production of the scalars, both light and heavy).  The plot on the
left shows the hadronic cross section as a function of the
pseudoscalar mass $\ma$, in the \lss\ with $\tb=20$. As in section
\ref{sec:xsection}, the renormalization and factorization scales are
set to $\muR = \ma$ and $\muF=\ma/4$. The central (black) solid line
in the left plot is computed with our default settings, namely we use
the PDF set with $\mbp=4.75$~GeV and we relate the Yukawa coupling
$\ybeff^{\smalla}$ to $\mb(\ma)$, which we obtain from the input
$\mb(\mb)=4.16$~GeV via renormalization-group evolution. The plot on
the right of figure~\ref{fig:bbhmbdep} represents the variation of the
cross section relative to this default setting, when we change the
bottom mass in the PDFs and/or in the Yukawa coupling.

In both plots, the red band between dot-dashed lines indicates the
spread in the cross section obtained with the extreme PDF sets --
corresponding to $\mbp=4$ GeV and $\mbp=5.5$ GeV, respectively -- with
$\ybeff^{\smalla}$ fixed to the default value. As expected, the impact
of the bottom mass used in the PDFs is significant: it amounts to
about $(+20/$--$15)\%$ at large $\ma$, with larger values of $\mbp$
corresponding to smaller cross sections. This anti-correlation is a
consequence of the fact that, for larger bottom masses, the reduced
available phase space for the splitting of gluons into bottom pairs
leads to a suppression of the bottom PDF. On the other hand, the
bottom Yukawa coupling is directly correlated with the magnitude of
the cross section. Simultaneously adjusting the bottom mass entering
the bottom Yukawa coupling and the one entering the bottom PDF should
therefore lead to some degree of compensation between these two
effects.

Converting the pole-mass values $\mbp=4$~GeV and $\mbp=5.5$~GeV to the
$\msbar$ scheme at three-loop level, one obtains $\mb(\mb)=3.32$~GeV
and $\mb(\mb)=4.69$~GeV, respectively. Using these numbers to
calculate $\ybeff^{\smalla}$ while fixing the PDF set to the default
(i.e., the set with $\mbp=4.75$\,GeV) results in the gray band between
dashed lines in the right plot of figure~\ref{fig:bbhmbdep}. It turns
out that this band is about twice as large as the red band arising
from PDF variation. However, the gray band is rather asymmetric,
because the pole mass $\mbp=4.75$~GeV for the default PDF set
corresponds at the three-loop level to $\mb(\mb)=4.00$~GeV, which is
significantly smaller than our default input for $\ybeff^{\smalla}$,
i.e.~$\mb(\mb)=4.16$~GeV.  The net effect on the cross section of a
simultaneous variation of the bottom mass in the PDFs and in
$\ybeff^{\smalla}$, shown as a blue band between solid lines in both
the left and the right plots, is thus also asymmetric, and it is of
the order of $(+15/$--$30)\%$ at large $\ma$.

Our procedure to estimate the uncertainty of the cross section for
bottom-quark annihilation arising from the bottom-mass dependence of
the PDFs is similar to the one in ref.~\cite{Baglio:2010ae}. We fix
the bottom Yukawa coupling to the value implied by
$\mb(\mb)=4.16$~GeV, as recommended by the \WG, and we use as
uncertainty the spread in the cross section caused by the variation of
$\mbp$ in the PDFs around the central value of $4.75$~GeV. However,
the full variation of $\pm 0.75$~GeV allowed by the MSTW2008 PDFs,
which would correspond to the red band in figure~\ref{fig:bbhmbdep},
seems overly conservative for our purposes. A variation of $\pm
0.25$~GeV is in fact sufficient to encompass the value
$\mbp=4.92$~GeV, which corresponds at the three-loop level to the
recommended $\msbar$ mass $\mb(\mb)=4.16$~GeV. This variation finally
leads to an estimate of the uncertainty of about $\pm 6\%$.  A similar
estimate is obtained from NNPDF2.1~\cite{Ball:2011mu}, which also
provides PDF sets with different values of $\mbp$.


\subsection{Higher-order \susy\ contributions to gluon fusion}
\label{sec:susyerr}

In this section we discuss two sources of uncertainty affecting the
\susy\ contributions to the cross section for gluon fusion.  The first
is the validity of the expansion in the heavy superparticle masses of
the two-loop \susy\ contributions; the second is the impact of
the three-loop \susy\ contributions that are not included in \sushi.

\subsubsection{Validity of the expansion in the SUSY masses}
\label{sec:msusy}

The results implemented in \sushi\ for the two-loop stop contributions
to lightest-scalar production rely on the VHML, while the results for
the remaining two-loop \susy\ contributions rely on expansions in
inverse powers of the superparticle masses.  The latter include terms
up to ${\cal O}(m_\phi^2/M^2)$, ${\cal O}(\mt^2/M^2)$, ${\cal
  O}(\mb/M)$ and ${\cal O}(\mz^2/M^2)$, where $m_\phi$ denotes a Higgs
mass and $M$ denotes a generic superparticle mass. Therefore, the
validity of the results for the two-loop \susy\ contributions is
limited to the region where the mass of the produced Higgs boson is
smaller than the lowest-lying \susy-particle threshold of the Feynman
diagrams involved. In all of the six benchmark scenarios considered in
our study, the lightest-scalar mass lies comfortably below this
limit. Since we consider $\ma \leq 1$~TeV, the same applies also to
the masses of the heaviest scalar and of the pseudoscalar in the five
scenarios in which the squark masses are themselves of the order of
$1$~TeV. In the \lss, on the other hand, the lowest-lying SUSY
threshold is at $2\,m_{\tilde t_1}\!\approx 650$~GeV, hence our need
to limit our analysis to $\ma \leq 500$~GeV.

To assess the quality of our approximation in the vicinity of the
threshold, we multiply the two-loop stop and sbottom contributions to
the gluon-fusion amplitude by test factors $t_{\tilde q} \equiv {\cal
  A}_{\tilde q_1}^{1\ell}/{\cal A}_{\tilde q_1}^{1\ell,\, {\rm exp}}$,
with $\tilde q = \{\tilde t,\tilde b\}$. Specifically, ${\cal
  A}_{\tilde q_1}^{1\ell}$ is the lightest-squark contribution to the
one-loop part of the scalar-production amplitude including the full
mass dependence, while ${\cal A}_{\tilde q_1}^{1\ell,\, {\rm exp}}$
includes only the leading ${\cal O}(m_{\tilde q_1}^{-2})$ terms in the
expansion in the lightest-squark mass.  Assuming that the expanded
two-loop contributions deviate from the full ones by an amount
comparable to that seen in the one-loop contributions, the variation
in the gluon-fusion cross section resulting from the introduction of
the test factors can be considered as an estimate of the uncertainty
associated to the expansion in the SUSY masses.

\begin{figure}[t]
\begin{center}
\includegraphics[width=0.49\textwidth]{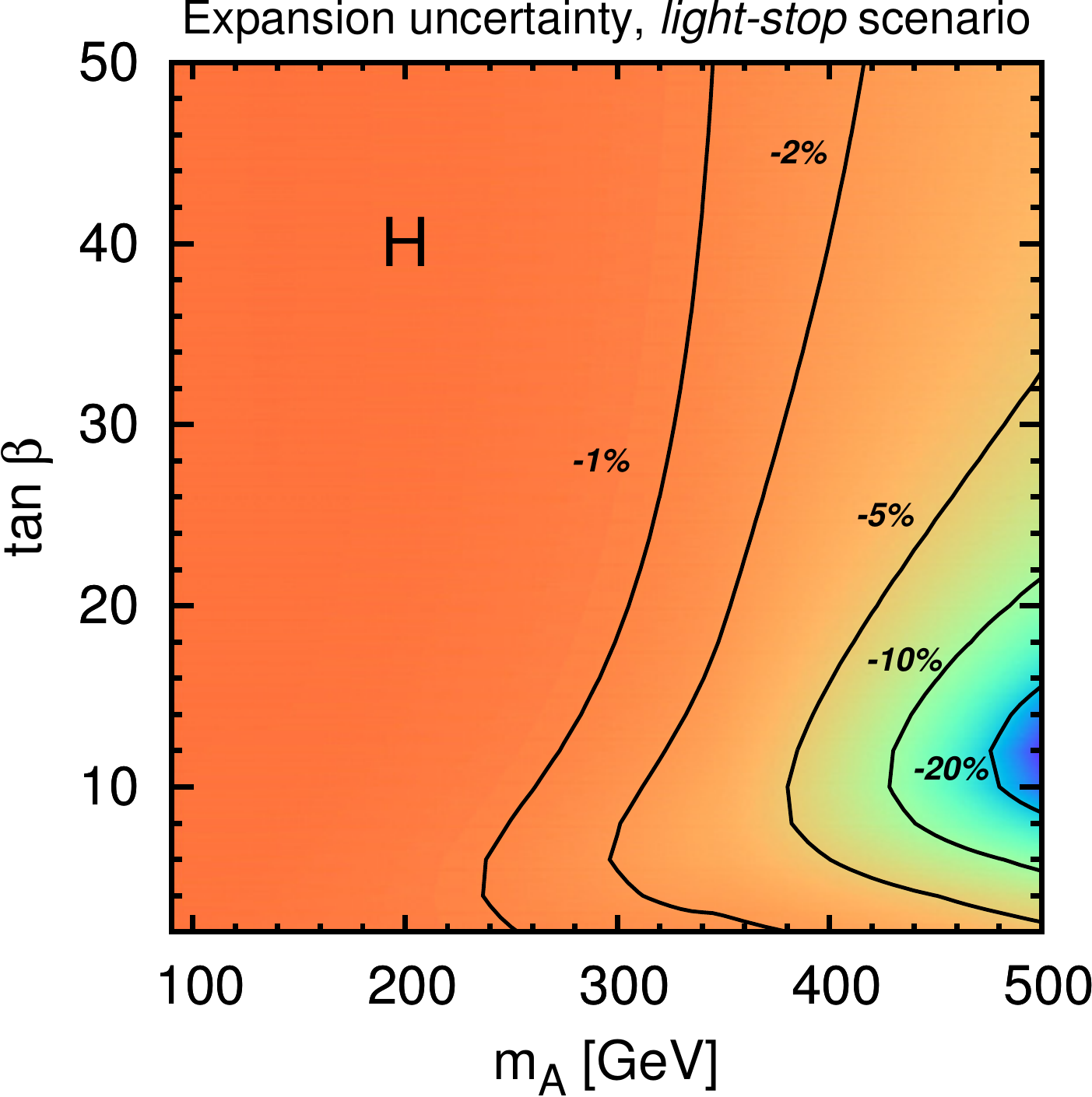}~~~
\includegraphics[width=0.49\textwidth]{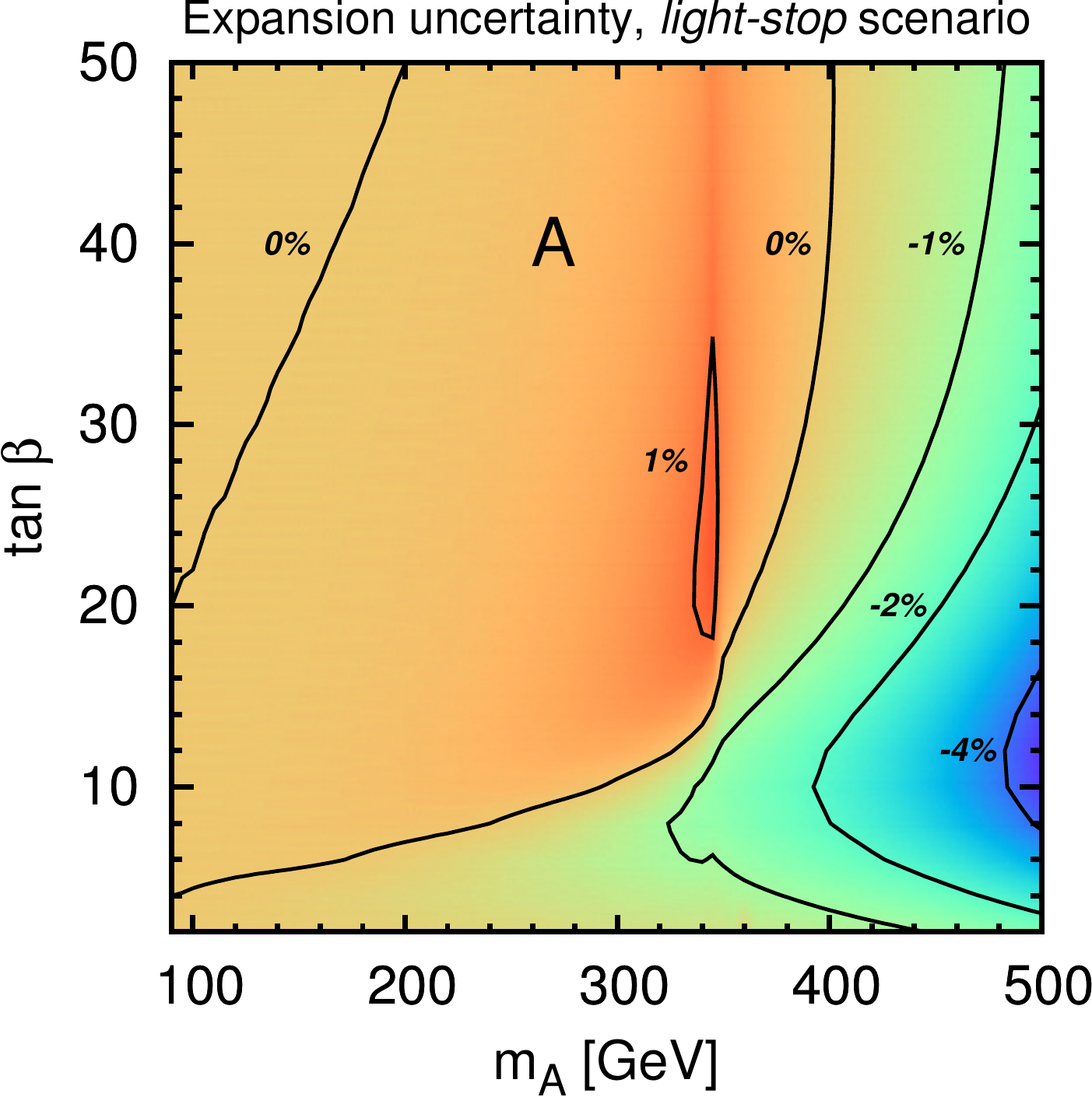}
\end{center}
\vspace*{-0.3cm}
\caption{Variation of the gluon-fusion cross section for the
  production of $H$ (left) and $A$ (right) in the \lss\ when the
  two-loop SUSY contributions are rescaled by $t_{\tilde q} \equiv
  {\cal A}_{\tilde q_1}^{1\ell}/{\cal A}_{\tilde q_1}^{1\ell,\, {\rm
      exp}}$.}
\label{fig:testfactor}
\end{figure}

The contour plots in figure~\ref{fig:testfactor} show the effect of
introducing these test factors on the cross section for the production
of the heaviest scalar (left plot) and of the pseudoscalar (right
plot) in the \lss. In the case of $H$ production, the variation of the
cross section at large $\ma$ amounts to a few percent when $\tb$ is
sufficiently large, but it can exceed $20\%$ when $8 \lsim \tb\lsim
16$. As can be seen in the right plot of
figure~\ref{fig:squark-scalar}, in this region the one-loop quark and
squark contributions to the gluon-fusion amplitude largely cancel each
other, with the result that the total cross section becomes small and
particularly sensitive to variations in the two-loop
contributions. This sensitivity to higher-order effects manifests also
as the large scale uncertainty, up to $50\%$, visible in the right
plot of figure~\ref{fig:ggh-light-heavy-murmuf}. In the case of $A$
production, on the other hand, no such cancellations occur, because
the squarks do not contribute to the one-loop amplitude for gluon
fusion.\footnote{\,For the same reason, we cannot define test factors
  analogous to $t_{\tilde q}$ in terms of the pseudoscalar-production
  amplitude. To estimate the accuracy of the mass expansion for $A$
  production we use the same test factors $t_{\tilde q}$ as for $H$
  production. } The variation of the cross section at large $\ma$ is
therefore limited to a few percent even for moderate $\tb$.

We performed the same analysis on the other five benchmark scenarios,
where the squark masses are of the order of $1$~TeV. As expected, we
found that the effect of rescaling the two-loop SUSY contributions by
test factors $t_{\tilde q}$ is much smaller than in the \lss, and it
is certainly negligible when compared to the scale uncertainty of the
cross section. In particular, in the {\em tau-phobic} scenario --
where the squark contributions to the gluon-fusion amplitude are
enhanced by the large value of the parameter $\mu$ -- the effect on
$H$ production reaches the few-percent level only when $\ma$
approaches $1$~TeV, for moderate $\tb$. In the remaining four
scenarios the effect is even smaller.

\subsubsection{The \susy\ contributions at the NNLO}\label{sec:nnlosusy}

The \qcd\ corrections to the gluon-fusion cross section are large,
typically exceeding $100$\% at the energy of the \lhc. In the \sm, an
excellent approximation to these corrections is obtained in the VHML
(or heavy-top limit)~\cite{H2gQCD2,Anastasiou:2002yz,H2gQCD3}, where a
perturbative $K$-factor is calculated in the effective theory that
results from neglecting the bottom Yukawa coupling and integrating out
the top quark, leaving behind a point-like Higgs-gluon interaction
term ${\cal L}_{ggH} = -(1/4v)\,C(\alpha_s)\,HG_{\mu\nu}G^{\mu\nu}$, with
$v\approx 246$\,GeV. The Wilson coefficient
\begin{equation}
\begin{split}
C(\alpha_s) = C^{(0)} + \api\,C^{(1)}+
  \left(\api\right)^2C^{(2)}
\label{eq:c1}
\end{split}
\end{equation}
accounts for heavy particles that mediate the Higgs-gluon coupling in
the underlying theory. In the SM, this is just the top quark; it is
easy to see, though, that the inclusion of stop squarks (and gluinos)
only affects $C(\alpha_s)$, while the form of ${\cal L}_{ggH}$ remains
unchanged.  A comparison with the full result at the NLO suggests
that, within the SM, the VHML provides a decent approximation of the
NNLO top contributions also for rather large Higgs
masses~\cite{Spira:1997dg,Harlander:2003xy}. Therefore,
\sushi\ includes the NNLO effects in the cross sections for the
production of all three neutral Higgs bosons of the MSSM.

Within the effective theory, the $K$-factor at the \nnlo{} takes the
form
\begin{eqnarray}
\label{eq:sigmannlo}
K &=& 1 + \api\frac{1}{C^{(0)}\,\Sigma^{(0)}}\bigg[
C^{(0)}\,\Sigma^{(1)} + 2\,C^{(1)}\,\Sigma^{(0)}\\\nonumber&+&
\api\,\left( C^{(0)}\,\Sigma^{(2)} +
    2\,C^{(1)}\,\Sigma^{(1)} + (C^{(1)})^2\,\Sigma^{(0)} + 2\,C^{(2)}\,\Sigma^{(0)}
  \right)\bigg]\,,
\end{eqnarray}
where $\Sigma^{(n)}$ is the $n^\text{th}$-order term in the perturbative
expansion of the hadronic cross section based on ${\cal
  L}_{ggH}\big|_{C(\alpha_s)\equiv 1}$.
Note that, in the NNLO part of the $K$-factor in
eq.~(\ref{eq:sigmannlo}), the only genuine three-loop term that
depends on the underlying theory is $C^{(2)}$. This observation was
exploited in ref.~\cite{Harlander:2003kf} to derive an estimate of the
\nnlo\ top/stop contribution to the gluon-fusion cross section in the
\mssm. In particular, it was shown that the final result depends only
very weakly on the numerical value of $C^{(2)}$. Consequently, once
the two-loop stop contributions are included in $C^{(1)}$, the unknown
three-loop stop contributions to $C^{(2)}$ induce an uncertainty in
the cross section much smaller than the residual uncertainty derived
from scale variation. It was suggested to use the top contribution
$C_t^{(2)}$ for the whole $C^{(2)}$, and to estimate the
related uncertainty by varying that coefficient within the interval
$[0,2\, C_t^{(2)}]$.

In ref.~\cite{Harlander:2003kf} the hadronic cross section was
obtained, in analogy to the SM NNLO result, by reweighting its exact
LO expression with the $K$-factor of eq.~(\ref{eq:sigmannlo}):
\begin{equation}
\begin{split}
\label{eq:kfact}
\sigma^{{\rm{\scriptscriptstyle NNLO}}} ~=~ 
K\,|{\cal A}_{t\tilde t}^{1\ell}|^2\,\Sigma_0 \,,
\end{split}
\end{equation}
where ${\cal A}^{1\ell}_{t\tilde t} \equiv {\cal A}_{t}^{1\ell}+{\cal
  A}_{\tilde t}^{1\ell}$ is the one-loop amplitude including both the
top and stop contributions with the exact Higgs-mass dependence (in
particular, ${\cal A}_{t\tilde t}^{1\ell}\to C^{(0)}$ in the VHML,
i.e.~for $m_\phi\!\rightarrow\!0$).
However, as was discussed also in the previous section, there exist
so-called {\em gluophobic} regions of the MSSM parameter space in
which the top and stop contributions to the amplitude can cancel each
other to a large extent. Since the precise values of the MSSM
parameters where this cancellation is maximal differ between the full
calculation and the VHML, the ratio $|{\cal A}_{t\tilde
  t}^{1\ell}|/C^{(0)}$ entering the cross section -- see
eqs.~(\ref{eq:sigmannlo}) and (\ref{eq:kfact}) -- can become
spuriously large when $C^{(0)}\approx0$. In order to evade this
effect, we replace $C^{(0)}$ in eq.~(\ref{eq:c1}) with ${\cal
  A}_{t\tilde t}^{1\ell}\,$.  This leads to the following expression
for the cross section:
\begin{eqnarray}
\sigma^{{\rm{\scriptscriptstyle NNLO}}} &=& |{\cal A}_{t\tilde
  t}^{1\ell}|^2\, \Sigma^{(0)} ~+~ \api\left(
|{\cal A}_{t\tilde  t}^{1\ell}|^2\,\Sigma^{(1)} +
2\,C^{(1)}\,\Sigma^{(0)}\,{\rm Re}\,{\cal A}_{t\tilde t}^{1\ell}
\right) \nonumber\\ &+& \left(\api\right)^2\left[
|{\cal A}_{t\tilde  t}^{1\ell}|^2\,\Sigma^{(2)} +
  2\left(C^{(1)}\,\Sigma^{(1)}+C^{(2)}\,\Sigma^{(0)}\right) {\rm Re
  }\,{\cal A}_{t\tilde t}^{1\ell} +
  (C^{(1)})^2\,\Sigma^{(0)}\right]\,.
\label{eq:newnnlo}
\end{eqnarray}
This formula applies to both MSSM scalars. The effective Lagrangian
for the gluonic interaction of the pseudoscalar involves an additional
operator which contributes at the NNLO~\cite{pseudonnlo}, but it can
be treated in a completely analogous way.

In \sushi, the NNLO top and stop contributions to the gluon-fusion
cross section in the VHML are isolated by subtracting from the
$\sigma^{{\rm{\scriptscriptstyle NNLO}}}$ in eq.~(\ref{eq:newnnlo})
the same quantity truncated at the NLO (and computed with NLO
PDFs). The result is then added to the full NLO cross section, which
accounts also for the bottom and sbottom contributions and for the
known Higgs-mass dependence of the two-loop amplitude. The $6\%$
suppression of the cross section for the production of a SM-like
scalar induced by the NNLO stop contributions in the \lss\ -- see
section \ref{sec:xsec-xsec} -- can be ascribed to the effect of the
term $2\,C^{(1)}\, \Sigma^{(1)}\,{\rm Re}\,{\cal A}_{t\tilde
  t}^{1\ell}$ in the second line of eq.~(\ref{eq:newnnlo}). Indeed,
the large value of the (normalized) NLO term of the cross section in
the effective theory, $\Sigma^{(1)}/\Sigma^{(0)}\approx 26$,
compensates for the suppression by $\alpha_s/\pi$, with the result
that the effect of the two-loop stop contribution to $C^{(1)}$ at the
NNLO is roughly as large as the corresponding effect at the NLO.

To assess the uncertainty arising from the fact that we neglect the
three-loop SUSY contributions to $C^{(2)}$, we make use of a recent
calculation of those contributions in the
VHML~\cite{Pak:2010cu,Pak:2012xr}. The calculation is based on an
expansion of the relevant Feynman diagrams in terms of certain
hierarchies among the different masses, similar to the strategy that
was pursued in ref.~\cite{HS-3l} for the calculation of the 3-loop
corrections to the Higgs mass in the \mssm. The results of
ref.~\cite{Pak:2012xr} are available in the form of a {\tt
  Mathematica} file, which provides the basis for the expansion of
$C^{(2)}$ in various hierarchies of the masses $m_{\stu}$, $m_{\std}$,
$\mgluino$, $\mtop$, and $\msquark$, combined with expansions in
differences of these masses.  Following an algorithm suggested in
ref.~\cite{Pak:2012xr}, these expansions should allow one to derive a
numerical approximation for $C^{(2)}$ in any viable \mssm\ scenario.

Applying this approach to the scenarios defined in
table~\ref{tab:scenarios}, we find that the deviation of the whole
$C^{(2)}$ from the top contribution $C^{(2)}_t$ is rather small, and
the second-order coefficient certainly stays within the range
$[\,0,\,2\,C^{(2)}_t]$.  Varying $C^{(2)}$ within this interval, we
estimate that the effect of the three-loop SUSY contributions to the
gluon-fusion cross section does not exceed $1\%$ in all of the
scenarios considered in this paper. It is therefore a viable strategy
to follow ref.~\cite{Harlander:2003kf} and set $C^{(2)} = C^{(2)}_t$,
attributing an uncertainty of $\pm 1\%$ to the final result for the
cross section.


\vfill
\newpage

\section{Conclusions}
\label{sec:conclusions}

A precise prediction of the cross sections for Higgs-boson production,
as well as a detailed understanding of the associated uncertainties,
are of vital importance to interpret the recent discovery of a Higgs
boson in the context of the MSSM. In this paper we used the public
code \sushi~\cite{Harlander:2012pb}, which computes the cross sections
for gluon fusion and bottom-quark annihilation, to study the
production of scalar and pseudoscalar Higgs bosons in a set of MSSM
scenarios compatible with the LHC results. We showed how the cross
sections can substantially differ from the SM prediction, and how
their qualitative behavior over the MSSM parameter space depends
mainly on the relative importance of the contributions involving top
and bottom quarks. We also emphasized that, in a scenario with
relatively light squarks which is not yet constrained by the LHC, the
contributions to the gluon-fusion process that involve superparticles
can significantly suppress the cross section for scalar production.

Next, we studied the different sources of uncertainty that affect our
predictions for the Higgs-production cross sections. Some of these
uncertainties, namely the ones associated to the choice of
renormalization and factorization scales, to the PDF parameterization
and to the input value for the strong coupling constant, are relevant
also for the production of the SM Higgs, although their size may
differ in the case of the production of non-standard Higgs bosons. In
contrast, the uncertainties associated to the definition of the bottom
mass and Yukawa coupling are practically negligible in the SM -- where
the bottom-quark contributions amount only to a few percent of the
total cross section -- but they can become dominant in regions of the
MSSM parameter space where the couplings of the Higgs bosons to bottom
quarks are enhanced. In the particular case of heavy-scalar and
pseudoscalar production at large $\tb$, we found that legitimate
variations in the renormalization scheme and scale of the bottom
Yukawa coupling can suppress the gluon-fusion cross section by more
than $60\%$, due to the presence of large QCD corrections enhanced by
logarithms of the ratio $m_\phi^2/\mb^2\,$. Luckily, in this case the
total cross section is dominated by the contribution of bottom-quark
annihilation, which is subject to a considerably smaller scale
uncertainty. Finally, we studied the uncertainties associated to our
implementation of the SUSY contributions to gluon fusion at the NLO
and, partially, at the NNLO. With the exception of a gluophobic region
in the \lss, these uncertainties are generally small, reflecting the
sub-dominant nature of the SUSY contributions themselves for values of
the squark masses compatible with the LHC bounds.

Future improvements in the accuracy of the predictions for the
Higgs-production cross sections in the MSSM could come from different
directions. First of all, any progress in the SM calculation will
eventually trickle down to the MSSM calculation, at least where the
production of a SM-like scalar is concerned. In addition, a
resummation of the QCD corrections enhanced by $\ln(m_\phi^2/\mb^2)$,
analogous to the one performed in ref.~\cite{resum} for the Higgs
decay to photons, will be necessary to reduce the large uncertainty in
the production of non-standard Higgs bosons via gluon fusion
(incidentally, such calculation would benefit all models with enhanced
Higgs couplings to bottom quarks, whether they are supersymmetric or
not). Implementing the existing results for the two-loop contributions
to $\Db$~\cite{deltab2l}, in both the Higgs mass and cross-section
calculations, will also reduce the uncertainty in scenarios where the
bottom contributions are relevant. Finally, it could be worthwhile to
improve the calculation of the gluon-fusion cross section by taking
into account the full Higgs-mass dependence of the two-loop
squark-gluon\,\footnote{As well as the two-loop quark-squark-gluino
  contributions~\cite{babis2,spiraDb}, when they become available.}
contributions~\cite{babis1,ABDV,MS} -- to cover scenarios in which the
non-standard Higgs bosons are heavier than the third-generation
squarks -- and by including the genuine three-loop
effects~\cite{Pak:2010cu,Pak:2012xr}.

For the time being, however, we believe that \sushi\ provides the most
sophisticated calculation of the Higgs-production cross sections in the
MSSM available to the physics community. Differently from the case of
the SM, where all the relevant inputs are now known and a definite
prediction for the total cross section can be made solely as a
function of the collision energy, in the MSSM the predictions for the
cross sections -- and the relevance of the different sources of
uncertainty -- depend crucially on a number of yet-undetermined
parameters. In the appendix we collect predictions for the
Higgs-production cross sections via gluon fusion and bottom-quark
annihilation, and their respective uncertainties, in a few
representative points of the six benchmark scenarios described in
section \ref{sec:benchmarks}. However, the tables in the appendix
should be regarded as having an illustrative purpose only: we
encourage the readers to take both \sushi\ and our recipes for the
uncertainties directly in their own hands, and use them to analyze
their favorite corners of the MSSM parameter space. Our results should
prove useful for ruling out scenarios that are incompatible with the
current experimental bounds. We also hope that, when the time comes,
they will help interpret within the MSSM the discovery of new
particles at the LHC.


\section*{Acknowledgments}

This work was initiated in the context of the activities of the \WG,
and some of our preliminary findings were presented in section 14 of
ref.~\cite{Heinemeyer:2013tqa}.
We thank the authors of ref.~\cite{Carena:2013qia} and the SUSY
convenors of the ATLAS and CMS collaborations for helpful
communication about the experimental constraints on the MSSM benchmark
scenarios, and the authors of ref.~\cite{Pak:2012xr} for helpful
communication about their results. We also thank M.~Spira for useful
discussions about the uncertainties of the cross-section calculations.

This work was supported in part by the Research Executive Agency (REA)
of the European Commission under the Grant Agreements
PITN-GA-2010-264564 (LHCPhenoNet), PITN-GA-2012-315877 (MCnet) and
PITN-GA-2012-316704 (HiggsTools), and by the European Research Council
(ERC) under the Advanced Grant ERC-2012-ADG\_20120216-321133
(Higgs@LHC).
The work of E.~B.~and P.~S.~at LPTHE is supported in part by French
state funds managed by the ANR (ANR-11-IDEX-0004-02) in the context of
the ILP LABEX (ANR-10-LABX-63).
S.~L.~acknowledges support by the DFG (SFB~676 ``Particles, Strings
and the Early Universe'').
S.~L., H.~M.~and R.~H.~were also supported by the DFG grant HA
2990/5-1.
A.~V.~was supported in part by an Italian PRIN2009 grant and a
European Investment Bank EIBURS grant.


\vfill
\newpage

\appendix
\section*{Appendix: Cross sections and uncertainties}

In this appendix we include eighteen tables, listing the cross
sections and uncertainties for the production at the LHC of the three
neutral Higgs bosons in the six MSSM scenarios defined in
table~\ref{tab:scenarios}. We use version {\tt 1.3.0} of \sushi, and
provide separate results for gluon fusion and bottom-quark
annihilation. Input files for the six scenarios can be found on the
code's website~\cite{Harlander:2012pb}. We set $\sqrt{s}=8$~TeV, $\mt
= 173.2$~GeV and $\mb(\mb) = 4.16$~GeV, and we choose thirty
combinations of the parameters $\ma$ and $\tb$ for each scenario.  The
predictions for the scalar masses are obtained with version {\tt
  2.10.0} of {\tt FeynHiggs}. The uncertainties provided in the tables
are computed as follows:

\begin{itemize}

\item
The renormalization- and factorization-scale uncertainties are
summarized in the quantities $\Delta_\mu^\pm$, defined as in
section~\ref{sec:uncert-gluon-fusion}, eq.~(\ref{eq:deltamu}).  For
gluon fusion we consider the seven combinations obtained from
$\muR=\{m_\phi/4,\,m_\phi/2,\,m_\phi\}$ and
$\muF=\{m_\phi/4,\,m_\phi/2,\,m_\phi\}$, where we discard the two
pairs with the largest variation of the ratio $\muR/\muF$ with respect
to the central choice.  For bottom-quark annihilation we proceed
accordingly, using $\muR=\{m_\phi/2,m_\phi,2\,m_\phi\}$ and
$\muF=\{m_\phi/8,m_\phi/4,m_\phi/2\}$.

\item
The uncertainty $\delta Y_b$ of the gluon-fusion process, related to
the definition of the bottom Yukawa coupling and discussed in
section~\ref{sec:mub}, is computed as the relative difference between
the cross section calculated with $\ybp \propto \mb(m_\phi/2)$ and the
cross section calculated with $\ybp \propto \mbp$. In the case of
bottom-quark annihilation, the scale dependence of $\ybp$ is included
in the computation of $\Delta_\mu^\pm$.

\item
The uncertainty $\delta \Delta_b$, stemming from the resummation of
$\tb$-enhanced corrections to $\ybp$ and described in
section~\ref{sec:deltab}, is computed by adding an uncertainty of $\pm
10\%$ to the value of $\Delta_b$ obtained from {\tt FeynHiggs}.

\end{itemize}

The PDF uncertainties are not included in the tables, but they were
extensively discussed in section~\ref{sec:pdfuncertainty}.  In
section~\ref{sec:PDFalphas} we pointed out that the relative size of
the \pdfas\ uncertainty depends mainly on the value of the Higgs mass,
thus it can be taken over directly from the existing estimates for the
production of the SM Higgs.  Apart from the \pdfas\ uncertainty, in
the case of bottom-quark annihilation an additional uncertainty of
$\pm 6\%$ has to be added due to the dependence of the bottom-quark
PDF on the pole bottom mass (see section~\ref{sec:pdfbottom}). The
uncertainties associated to our incomplete knowledge of the SUSY
contributions to the gluon-fusion cross section can become sizeable
only in the \lss, especially in the case of $H$ production at large
$\ma$ and moderate $\tb$. We do not include them in the tables,
pointing the reader to the discussion in section~\ref{sec:susyerr}.

We show the results for the $\mh^{\rm max}$ scenario in
tables~\ref{tab:LH_mhmax}$-$\ref{tab:A_mhmax}.  Rather similar results
for the $\mh^{\rm mod+}$ and $\mh^{\rm mod-}$ scenarios are given in
tables~\ref{tab:LH_mhmodp}$-$\ref{tab:A_mhmodp} and
\ref{tab:LH_mhmodm}$-$\ref{tab:A_mhmodm}, respectively.  The {\em
  light-stop} and {\em light-stau} scenarios are presented in
tables~\ref{tab:LH_lightstop}$-$\ref{tab:A_lightstop} and
\ref{tab:LH_lightstau}$-$\ref{tab:A_lightstau}, respectively.
Finally, cross sections and uncertainties for the {\em tau-phobic}
scenario are shown in
tables~\ref{tab:LH_tauphobic}$-$\ref{tab:A_tauphobic}, which are
limited to $\tb\leq 40$ due to a drop in the lightest-scalar mass for
larger $\tb$. 

\newcommand{\firstlinelight}{
$\ma$\,\footnotesize{[GeV]} & $\tan\beta$ & $\mh$\,\footnotesize{[GeV]} &
$\sigma_{ggh}$\,\footnotesize{[pb]} & $\Delta_\mu^\pm$\,\footnotesize{[\%]} & $\delta Y_b$\,\footnotesize{[\%]} & $\delta \Delta_b$\,\footnotesize{[\%]} &
$\sigma_{bbh}$\,\footnotesize{[pb]} & $\Delta_\mu^\pm$\,\footnotesize{[\%]} & $\delta \Delta_b$\,\footnotesize{[\%]}\\
}

\newcommand{\firstlineheavy}{
$\ma$\,\footnotesize{[GeV]} & $\tan\beta$ & $\mH$\,\footnotesize{[GeV]} &
$\sigma_{gg\smallH}$\,\footnotesize{[pb]} & $\Delta_\mu^\pm$\,\footnotesize{[\%]} & $\delta Y_b$\,\footnotesize{[\%]} & $\delta \Delta_b$\,\footnotesize{[\%]} &
$\sigma_{bb\smallH}$\,\footnotesize{[pb]} & $\Delta_\mu^\pm$\,\footnotesize{[\%]} & $\delta \Delta_b$\,\footnotesize{[\%]}\\
}

\newcommand{\firstlinepseudo}{
$\ma$\,\footnotesize{[GeV]} & $\tan\beta$ &
$\sigma_{gg\smalla}$\,\footnotesize{[pb]} & $\Delta_\mu^\pm$\,\footnotesize{[\%]} & $\delta Y_b$\,\footnotesize{[\%]} & $\delta \Delta_b$\,\footnotesize{[\%]} &
$\sigma_{bb\smalla}$\,\footnotesize{[pb]} & $\Delta_\mu^\pm$\,\footnotesize{[\%]} & $\delta \Delta_b$\,\footnotesize{[\%]}\\
}

\newcommand{\uncer}[2]{\scriptsize{$\begin{array}{r}#2\\[-3pt]#1\end{array}$}}
\newcommand{\shexp}{{\raisebox{1pt}{$\scriptstyle{\times 10^{\text{-}3}}$}}}
\newcommand{\shexpt}{{\raisebox{1pt}{$\scriptstyle{\times 10^{\text{-}6}}$}}}

\newcommand{\zeroset}{\text{---}}
\newcommand{\zeroup}{\text{---}}
\newcommand{\zerodown}{\text{---}}

\newcommand{\appendixtablecap}[2]{Cross sections and uncertainties 
for #1 production in the #2 scenario for
$\sqrt{s}=8$~TeV. Uncertainties below $0.1\%$ are not listed
($\zeroset$).  For the PDF uncertainties see text.}


\begin{table}[htp]
\begin{center}
\begin{tabular}{|c|c|c||c|r|r|r||c|r|r|}
\hline \firstlinelight \hline\hline
$100$ & $5$ & $92.2$ & $15.08$ & \uncer{-14.5}{+17.3} & \footnotesize{$-41.6$} & \uncer{-0.3}{+0.3} & $12.47$ & \uncer{-17.6}{+9.3} & \uncer{-0.3}{+0.3}\\[-1.5pt]\hdashline[1.5pt/1.5pt]
$100$ & $10$ & $97.6$ & $42.08$ & \uncer{-14.2}{+16.9} & \footnotesize{$-53.5$} & \uncer{-0.7}{+0.7} & $42.73$ & \uncer{-16.3}{+8.8} & \uncer{-0.7}{+0.7}\\[-1.5pt]\hdashline[1.5pt/1.5pt]
$100$ & $15$ & $98.9$ & $90.51$ & \uncer{-13.9}{+16.4} & \footnotesize{$-53.8$} & \uncer{-1.1}{+1.1} & $91.10$ & \uncer{-16.0}{+8.7} & \uncer{-1.0}{+1.1}\\[-1.5pt]\hdashline[1.5pt/1.5pt]
$100$ & $20$ & $99.3$ & $156.0$ & \uncer{-13.8}{+16.1} & \footnotesize{$-53.5$} & \uncer{-1.4}{+1.5} & $155.5$ & \uncer{-15.8}{+8.6} & \uncer{-1.4}{+1.4}\\[-1.5pt]\hdashline[1.5pt/1.5pt]
$100$ & $30$ & $99.7$ & $329.5$ & \uncer{-13.7}{+15.9} & \footnotesize{$-53.2$} & \uncer{-2.0}{+2.1} & $325.5$ & \uncer{-15.8}{+8.6} & \uncer{-2.0}{+2.0}\\[-1.5pt]\hdashline[1.5pt/1.5pt]
$100$ & $50$ & $99.9$ & $806.0$ & \uncer{-13.6}{+15.7} & \footnotesize{$-53.0$} & \uncer{-3.2}{+3.3} & $792.8$ & \uncer{-15.7}{+8.6} & \uncer{-3.1}{+3.2}\\[-1.5pt]\hdashline[1.5pt/1.5pt]
$120$ & $5$ & $105.6$ & $10.69$ & \uncer{-12.6}{+13.9} & \footnotesize{$-16.3$} & \uncer{-0.1}{+0.1} & $6.418$ & \uncer{-14.5}{+8.1} & \uncer{-0.3}{+0.3}\\[-1.5pt]\hdashline[1.5pt/1.5pt]
$120$ & $10$ & $114.2$ & $16.02$ & \uncer{-13.5}{+15.8} & \footnotesize{$-43.5$} & \uncer{-0.6}{+0.6} & $20.57$ & \uncer{-13.0}{+7.5} & \uncer{-0.7}{+0.7}\\
\hline
\end{tabular}
\end{center}
\caption{\appendixtablecap{lightest-scalar}{$\mh^{\rm max}$}}
\label{tab:LH_mhmax}
\end{table}
\begin{table}[htp]
\begin{center}
\begin{tabular}{|c|c|c||c|r|r|r||c|r|r|}
\hline \firstlineheavy \hline\hline
$100$ & $5$ & $136.8$ & $16.65$ & \uncer{-9.3}{+7.8} & \footnotesize{$-3.9$} & \uncer{\zerodown}{\zeroup} & $0.530$ & \uncer{-10.1}{+6.3} & \uncer{-0.6}{+0.6}\\[-1.5pt]\hdashline[1.5pt/1.5pt]
$100$ & $10$ & $131.9$ & $19.86$ & \uncer{-9.5}{+8.0} & \footnotesize{$-5.0$} & \uncer{\zerodown}{\zeroup} & $0.941$ & \uncer{-10.6}{+6.5} & \uncer{-1.0}{+1.0}\\[-1.5pt]\hdashline[1.5pt/1.5pt]
$100$ & $15$ & $130.8$ & $20.92$ & \uncer{-9.5}{+8.1} & \footnotesize{$-5.3$} & \uncer{-0.1}{+0.1} & $1.077$ & \uncer{-10.7}{+6.5} & \uncer{-1.5}{+1.5}\\[-1.5pt]\hdashline[1.5pt/1.5pt]
$100$ & $20$ & $130.4$ & $21.33$ & \uncer{-9.5}{+8.1} & \footnotesize{$-5.4$} & \uncer{-0.2}{+0.2} & $1.124$ & \uncer{-10.8}{+6.6} & \uncer{-1.9}{+2.0}\\[-1.5pt]\hdashline[1.5pt/1.5pt]
$100$ & $30$ & $130.2$ & $21.60$ & \uncer{-9.5}{+8.1} & \footnotesize{$-5.4$} & \uncer{-0.3}{+0.3} & $1.143$ & \uncer{-10.8}{+6.6} & \uncer{-2.8}{+2.9}\\[-1.5pt]\hdashline[1.5pt/1.5pt]
$100$ & $50$ & $130.3$ & $21.62$ & \uncer{-9.5}{+8.1} & \footnotesize{$-5.4$} & \uncer{-0.4}{+0.4} & $1.134$ & \uncer{-10.8}{+6.6} & \uncer{-4.3}{+4.6}\\[-1.5pt]\hdashline[1.5pt/1.5pt]
$200$ & $5$ & $205.9$ & $1.521$ & \uncer{-8.8}{+7.1} & \footnotesize{$-17.5$} & \uncer{-0.1}{+0.1} & $0.743$ & \uncer{-5.6}{+4.4} & \uncer{-0.4}{+0.4}\\[-1.5pt]\hdashline[1.5pt/1.5pt]
$200$ & $10$ & $201.7$ & $1.838$ & \uncer{-10.2}{+9.7} & \footnotesize{$-41.9$} & \uncer{-0.5}{+0.5} & $3.337$ & \uncer{-5.8}{+4.4} & \uncer{-0.7}{+0.7}\\
\hline
\end{tabular}
\end{center}
\caption{\appendixtablecap{heaviest-scalar}{$\mh^{\rm max}$}}
\label{tab:HH_mhmax}
\end{table}
\begin{table}[htp]
\begin{center}
\begin{tabular}{|c|c||c|r|r|r||c|r|r|}
\hline \firstlinepseudo \hline\hline
$100$ & $5$ & $10.55$ & \uncer{-13.5}{+15.6} & \footnotesize{$-47.9$} & \uncer{-0.4}{+0.4} & $10.74$ & \uncer{-15.7}{+8.6} & \uncer{-0.4}{+0.4}\\[-1.5pt]\hdashline[1.5pt/1.5pt]
$100$ & $10$ & $43.02$ & \uncer{-13.4}{+15.4} & \footnotesize{$-53.9$} & \uncer{-0.8}{+0.8} & $41.46$ & \uncer{-15.7}{+8.6} & \uncer{-0.7}{+0.7}\\[-1.5pt]\hdashline[1.5pt/1.5pt]
$100$ & $15$ & $97.28$ & \uncer{-13.3}{+15.2} & \footnotesize{$-53.3$} & \uncer{-1.1}{+1.1} & $90.03$ & \uncer{-15.7}{+8.6} & \uncer{-1.1}{+1.1}\\[-1.5pt]\hdashline[1.5pt/1.5pt]
$100$ & $20$ & $169.6$ & \uncer{-13.2}{+15.0} & \footnotesize{$-53.0$} & \uncer{-1.4}{+1.5} & $154.5$ & \uncer{-15.7}{+8.6} & \uncer{-1.4}{+1.4}\\[-1.5pt]\hdashline[1.5pt/1.5pt]
$100$ & $30$ & $360.3$ & \uncer{-13.2}{+14.9} & \footnotesize{$-52.7$} & \uncer{-2.0}{+2.1} & $324.7$ & \uncer{-15.7}{+8.6} & \uncer{-2.0}{+2.1}\\[-1.5pt]\hdashline[1.5pt/1.5pt]
$100$ & $50$ & $881.7$ & \uncer{-13.1}{+14.8} & \footnotesize{$-52.6$} & \uncer{-3.2}{+3.3} & $791.8$ & \uncer{-15.7}{+8.6} & \uncer{-3.1}{+3.2}\\[-1.5pt]\hdashline[1.5pt/1.5pt]
$200$ & $5$ & $0.566$ & \uncer{-9.9}{+9.2} & \footnotesize{$-4.2$} & \uncer{\zerodown}{\zeroup} & $0.922$ & \uncer{-5.9}{+4.5} & \uncer{-0.4}{+0.4}\\[-1.5pt]\hdashline[1.5pt/1.5pt]
$200$ & $10$ & $0.951$ & \uncer{-11.7}{+12.3} & \footnotesize{$-57.7$} & \uncer{-0.8}{+0.8} & $3.560$ & \uncer{-5.9}{+4.5} & \uncer{-0.7}{+0.7}\\
\hline
\end{tabular}
\end{center}
\caption{\appendixtablecap{pseudoscalar}{$\mh^{\rm max}$}}
\label{tab:A_mhmax}
\end{table}

\begin{table}[htp]
\begin{center}
\begin{tabular}{|c|c|c||c|r|r|r||c|r|r|}
\hline \firstlinelight \hline\hline
$100$ & $5$ & $91.0$ & $16.46$ & \uncer{-14.4}{+17.2} & \footnotesize{$-38.3$} & \uncer{-0.2}{+0.2} & $12.60$ & \uncer{-18.0}{+9.5} & \uncer{-0.3}{+0.3}\\[-1.5pt]\hdashline[1.5pt/1.5pt]
$100$ & $10$ & $97.0$ & $42.48$ & \uncer{-14.3}{+17.1} & \footnotesize{$-52.8$} & \uncer{-0.6}{+0.6} & $42.91$ & \uncer{-16.4}{+8.8} & \uncer{-0.6}{+0.6}\\[-1.5pt]\hdashline[1.5pt/1.5pt]
$100$ & $15$ & $98.5$ & $91.10$ & \uncer{-14.0}{+16.5} & \footnotesize{$-53.6$} & \uncer{-0.9}{+0.9} & $91.98$ & \uncer{-16.0}{+8.7} & \uncer{-0.9}{+0.9}\\[-1.5pt]\hdashline[1.5pt/1.5pt]
$100$ & $20$ & $99.1$ & $157.8$ & \uncer{-13.8}{+16.2} & \footnotesize{$-53.4$} & \uncer{-1.2}{+1.3} & $157.9$ & \uncer{-15.9}{+8.6} & \uncer{-1.2}{+1.2}\\[-1.5pt]\hdashline[1.5pt/1.5pt]
$100$ & $30$ & $99.6$ & $337.2$ & \uncer{-13.7}{+16.0} & \footnotesize{$-53.1$} & \uncer{-1.8}{+1.8} & $333.8$ & \uncer{-15.8}{+8.6} & \uncer{-1.7}{+1.8}\\[-1.5pt]\hdashline[1.5pt/1.5pt]
$100$ & $50$ & $99.8$ & $841.1$ & \uncer{-13.6}{+15.8} & \footnotesize{$-52.8$} & \uncer{-2.8}{+2.9} & $827.2$ & \uncer{-15.7}{+8.6} & \uncer{-2.7}{+2.8}\\[-1.5pt]\hdashline[1.5pt/1.5pt]
$120$ & $5$ & $103.5$ & $12.56$ & \uncer{-12.5}{+13.8} & \footnotesize{$-13.4$} & \uncer{\zerodown}{\zeroup} & $6.377$ & \uncer{-14.9}{+8.2} & \uncer{-0.3}{+0.3}\\[-1.5pt]\hdashline[1.5pt/1.5pt]
$120$ & $10$ & $112.7$ & $16.82$ & \uncer{-13.4}{+15.5} & \footnotesize{$-37.5$} & \uncer{-0.4}{+0.4} & $19.42$ & \uncer{-13.2}{+7.6} & \uncer{-0.6}{+0.6}\\
\hline
\end{tabular}
\end{center}
\caption{\appendixtablecap{lightest-scalar}{$\mh^{\rm mod+}$}}
\label{tab:LH_mhmodp}
\end{table}
\begin{table}[htp]
\begin{center}
\begin{tabular}{|c|c|c||c|r|r|r||c|r|r|}
\hline \firstlineheavy \hline\hline
$100$ & $5$ & $134.7$ & $17.43$ & \uncer{-9.4}{+8.0} & \footnotesize{$-4.5$} & \uncer{\zerodown}{\zeroup} & $0.693$ & \uncer{-10.3}{+6.4} & \uncer{-0.5}{+0.5}\\[-1.5pt]\hdashline[1.5pt/1.5pt]
$100$ & $10$ & $129.3$ & $21.67$ & \uncer{-9.6}{+8.3} & \footnotesize{$-6.3$} & \uncer{\zerodown}{\zeroup} & $1.428$ & \uncer{-10.9}{+6.6} & \uncer{-0.8}{+0.9}\\[-1.5pt]\hdashline[1.5pt/1.5pt]
$100$ & $15$ & $128.0$ & $23.41$ & \uncer{-9.7}{+8.3} & \footnotesize{$-7.0$} & \uncer{-0.2}{+0.2} & $1.806$ & \uncer{-11.1}{+6.7} & \uncer{-1.2}{+1.2}\\[-1.5pt]\hdashline[1.5pt/1.5pt]
$100$ & $20$ & $127.5$ & $24.27$ & \uncer{-9.7}{+8.4} & \footnotesize{$-7.4$} & \uncer{-0.2}{+0.2} & $2.039$ & \uncer{-11.1}{+6.7} & \uncer{-1.6}{+1.6}\\[-1.5pt]\hdashline[1.5pt/1.5pt]
$100$ & $30$ & $127.2$ & $25.16$ & \uncer{-9.7}{+8.4} & \footnotesize{$-8.0$} & \uncer{-0.3}{+0.3} & $2.361$ & \uncer{-11.2}{+6.7} & \uncer{-2.2}{+2.3}\\[-1.5pt]\hdashline[1.5pt/1.5pt]
$100$ & $50$ & $127.3$ & $26.08$ & \uncer{-9.7}{+8.5} & \footnotesize{$-9.0$} & \uncer{-0.6}{+0.6} & $2.889$ & \uncer{-11.1}{+6.7} & \uncer{-3.4}{+3.6}\\[-1.5pt]\hdashline[1.5pt/1.5pt]
$200$ & $5$ & $206.0$ & $1.526$ & \uncer{-8.9}{+7.2} & \footnotesize{$-17.4$} & \uncer{-0.1}{+0.1} & $0.750$ & \uncer{-5.6}{+4.4} & \uncer{-0.3}{+0.3}\\[-1.5pt]\hdashline[1.5pt/1.5pt]
$200$ & $10$ & $201.8$ & $1.860$ & \uncer{-10.3}{+9.7} & \footnotesize{$-41.6$} & \uncer{-0.5}{+0.5} & $3.369$ & \uncer{-5.8}{+4.4} & \uncer{-0.6}{+0.6}\\
\hline
\end{tabular}
\end{center}
\caption{\appendixtablecap{heaviest-scalar}{$\mh^{\rm mod+}$}}
\label{tab:HH_mhmodp}
\end{table}
\begin{table}[htp]
\begin{center}
\begin{tabular}{|c|c||c|r|r|r||c|r|r|}
\hline \firstlinepseudo \hline\hline
$100$ & $5$ & $10.61$ & \uncer{-13.5}{+15.6} & \footnotesize{$-47.8$} & \uncer{-0.3}{+0.3} & $10.80$ & \uncer{-15.7}{+8.6} & \uncer{-0.3}{+0.3}\\[-1.5pt]\hdashline[1.5pt/1.5pt]
$100$ & $10$ & $43.51$ & \uncer{-13.4}{+15.4} & \footnotesize{$-53.7$} & \uncer{-0.7}{+0.7} & $41.90$ & \uncer{-15.7}{+8.6} & \uncer{-0.6}{+0.6}\\[-1.5pt]\hdashline[1.5pt/1.5pt]
$100$ & $15$ & $98.90$ & \uncer{-13.3}{+15.2} & \footnotesize{$-53.1$} & \uncer{-1.0}{+1.0} & $91.44$ & \uncer{-15.7}{+8.6} & \uncer{-0.9}{+0.9}\\[-1.5pt]\hdashline[1.5pt/1.5pt]
$100$ & $20$ & $173.3$ & \uncer{-13.3}{+15.1} & \footnotesize{$-52.7$} & \uncer{-1.2}{+1.3} & $157.7$ & \uncer{-15.7}{+8.6} & \uncer{-1.2}{+1.2}\\[-1.5pt]\hdashline[1.5pt/1.5pt]
$100$ & $30$ & $371.6$ & \uncer{-13.2}{+15.0} & \footnotesize{$-52.5$} & \uncer{-1.8}{+1.8} & $334.4$ & \uncer{-15.7}{+8.6} & \uncer{-1.7}{+1.8}\\[-1.5pt]\hdashline[1.5pt/1.5pt]
$100$ & $50$ & $925.5$ & \uncer{-13.2}{+14.9} & \footnotesize{$-52.4$} & \uncer{-2.8}{+2.9} & $829.0$ & \uncer{-15.7}{+8.6} & \uncer{-2.7}{+2.8}\\[-1.5pt]\hdashline[1.5pt/1.5pt]
$200$ & $5$ & $0.567$ & \uncer{-9.9}{+9.2} & \footnotesize{$-4.3$} & \uncer{\zerodown}{\zeroup} & $0.927$ & \uncer{-5.9}{+4.5} & \uncer{-0.3}{+0.3}\\[-1.5pt]\hdashline[1.5pt/1.5pt]
$200$ & $10$ & $0.962$ & \uncer{-11.7}{+12.4} & \footnotesize{$-57.5$} & \uncer{-0.7}{+0.7} & $3.598$ & \uncer{-5.9}{+4.5} & \uncer{-0.6}{+0.6}\\
\hline
\end{tabular}
\end{center}
\caption{\appendixtablecap{pseudoscalar}{$\mh^{\rm mod+}$}}
\label{tab:A_mhmodp}
\end{table}

\begin{table}[htp]
\begin{center}
\begin{tabular}{|c|c|c||c|r|r|r||c|r|r|}
\hline \firstlinelight \hline\hline
$100$ & $5$ & $91.5$ & $16.24$ & \uncer{-14.5}{+17.3} & \footnotesize{$-38.9$} & \uncer{\zerodown}{\zeroup} & $12.92$ & \uncer{-17.8}{+9.4} & \uncer{\zerodown}{\zeroup}\\[-1.5pt]\hdashline[1.5pt/1.5pt]
$100$ & $10$ & $97.3$ & $45.45$ & \uncer{-14.2}{+17.0} & \footnotesize{$-51.6$} & \uncer{\zerodown}{\zeroup} & $45.95$ & \uncer{-16.3}{+8.8} & \uncer{\zerodown}{\zeroup}\\[-1.5pt]\hdashline[1.5pt/1.5pt]
$100$ & $15$ & $98.7$ & $101.7$ & \uncer{-13.9}{+16.4} & \footnotesize{$-52.0$} & \uncer{\zerodown}{\zeroup} & $102.1$ & \uncer{-16.0}{+8.7} & \uncer{\zerodown}{\zeroup}\\[-1.5pt]\hdashline[1.5pt/1.5pt]
$100$ & $20$ & $99.3$ & $182.6$ & \uncer{-13.8}{+16.2} & \footnotesize{$-51.7$} & \uncer{-0.1}{+0.1} & $181.3$ & \uncer{-15.9}{+8.6} & \uncer{-0.1}{+0.1}\\[-1.5pt]\hdashline[1.5pt/1.5pt]
$100$ & $30$ & $99.7$ & $416.8$ & \uncer{-13.7}{+16.0} & \footnotesize{$-51.4$} & \uncer{-0.2}{+0.2} & $409.6$ & \uncer{-15.8}{+8.6} & \uncer{-0.2}{+0.2}\\[-1.5pt]\hdashline[1.5pt/1.5pt]
$100$ & $50$ & $99.9$ & $1182.$ & \uncer{-13.6}{+15.9} & \footnotesize{$-51.1$} & \uncer{-0.3}{+0.3} & $1153.$ & \uncer{-15.7}{+8.6} & \uncer{-0.3}{+0.3}\\[-1.5pt]\hdashline[1.5pt/1.5pt]
$120$ & $5$ & $104.1$ & $12.11$ & \uncer{-12.6}{+13.9} & \footnotesize{$-14.3$} & \uncer{\zerodown}{\zeroup} & $6.521$ & \uncer{-14.8}{+8.2} & \uncer{\zerodown}{\zeroup}\\[-1.5pt]\hdashline[1.5pt/1.5pt]
$120$ & $10$ & $113.1$ & $17.34$ & \uncer{-13.5}{+15.7} & \footnotesize{$-38.9$} & \uncer{\zerodown}{\zeroup} & $20.88$ & \uncer{-13.2}{+7.5} & \uncer{\zerodown}{\zeroup}\\
\hline
\end{tabular}
\end{center}
\caption{\appendixtablecap{lightest-scalar}{$\mh^{\rm mod-}$}}
\label{tab:LH_mhmodm}
\end{table}
\begin{table}[htp]
\begin{center}
\begin{tabular}{|c|c|c||c|r|r|r||c|r|r|}
\hline \firstlineheavy \hline\hline
$100$ & $5$ & $134.3$ & $17.53$ & \uncer{-9.5}{+8.1} & \footnotesize{$-4.4$} & \uncer{\zerodown}{\zeroup} & $0.704$ & \uncer{-10.3}{+6.4} & \uncer{\zerodown}{\zeroup}\\[-1.5pt]\hdashline[1.5pt/1.5pt]
$100$ & $10$ & $129.1$ & $21.64$ & \uncer{-9.7}{+8.4} & \footnotesize{$-6.1$} & \uncer{\zerodown}{\zeroup} & $1.410$ & \uncer{-10.9}{+6.6} & \uncer{\zerodown}{\zeroup}\\[-1.5pt]\hdashline[1.5pt/1.5pt]
$100$ & $15$ & $127.8$ & $23.23$ & \uncer{-9.7}{+8.4} & \footnotesize{$-6.8$} & \uncer{\zerodown}{\zeroup} & $1.747$ & \uncer{-11.1}{+6.7} & \uncer{-0.1}{+0.1}\\[-1.5pt]\hdashline[1.5pt/1.5pt]
$100$ & $20$ & $127.4$ & $23.97$ & \uncer{-9.8}{+8.5} & \footnotesize{$-7.3$} & \uncer{\zerodown}{\zeroup} & $1.941$ & \uncer{-11.1}{+6.7} & \uncer{-0.2}{+0.2}\\[-1.5pt]\hdashline[1.5pt/1.5pt]
$100$ & $30$ & $127.2$ & $24.67$ & \uncer{-9.8}{+8.5} & \footnotesize{$-7.9$} & \uncer{\zerodown}{\zeroup} & $2.187$ & \uncer{-11.2}{+6.7} & \uncer{-0.3}{+0.3}\\[-1.5pt]\hdashline[1.5pt/1.5pt]
$100$ & $50$ & $127.5$ & $25.20$ & \uncer{-9.8}{+8.5} & \footnotesize{$-8.8$} & \uncer{\zerodown}{\zeroup} & $2.512$ & \uncer{-11.1}{+6.7} & \uncer{-0.4}{+0.4}\\[-1.5pt]\hdashline[1.5pt/1.5pt]
$200$ & $5$ & $205.8$ & $1.541$ & \uncer{-9.0}{+7.3} & \footnotesize{$-17.2$} & \uncer{\zerodown}{\zeroup} & $0.784$ & \uncer{-5.6}{+4.4} & \uncer{\zerodown}{\zeroup}\\[-1.5pt]\hdashline[1.5pt/1.5pt]
$200$ & $10$ & $201.7$ & $1.963$ & \uncer{-10.3}{+9.8} & \footnotesize{$-41.0$} & \uncer{\zerodown}{\zeroup} & $3.631$ & \uncer{-5.8}{+4.4} & \uncer{\zerodown}{\zeroup}\\
\hline
\end{tabular}
\end{center}
\caption{\appendixtablecap{heaviest-scalar}{$\mh^{\rm mod-}$}}
\label{tab:HH_mhmodm}
\end{table}
\begin{table}[htp]
\begin{center}
\begin{tabular}{|c|c||c|r|r|r||c|r|r|}
\hline \firstlinepseudo \hline\hline
$100$ & $5$ & $11.00$ & \uncer{-13.6}{+15.6} & \footnotesize{$-47.0$} & \uncer{\zerodown}{\zeroup} & $11.22$ & \uncer{-15.7}{+8.6} & \uncer{\zerodown}{\zeroup}\\[-1.5pt]\hdashline[1.5pt/1.5pt]
$100$ & $10$ & $46.87$ & \uncer{-13.4}{+15.4} & \footnotesize{$-52.2$} & \uncer{\zerodown}{\zeroup} & $44.98$ & \uncer{-15.7}{+8.6} & \uncer{\zerodown}{\zeroup}\\[-1.5pt]\hdashline[1.5pt/1.5pt]
$100$ & $15$ & $110.1$ & \uncer{-13.3}{+15.2} & \footnotesize{$-51.5$} & \uncer{\zerodown}{\zeroup} & $101.5$ & \uncer{-15.7}{+8.6} & \uncer{\zerodown}{\zeroup}\\[-1.5pt]\hdashline[1.5pt/1.5pt]
$100$ & $20$ & $199.6$ & \uncer{-13.3}{+15.1} & \footnotesize{$-51.2$} & \uncer{-0.1}{+0.1} & $181.0$ & \uncer{-15.7}{+8.6} & \uncer{-0.1}{+0.1}\\[-1.5pt]\hdashline[1.5pt/1.5pt]
$100$ & $30$ & $457.1$ & \uncer{-13.2}{+15.0} & \footnotesize{$-50.9$} & \uncer{-0.2}{+0.2} & $409.9$ & \uncer{-15.7}{+8.6} & \uncer{-0.2}{+0.2}\\[-1.5pt]\hdashline[1.5pt/1.5pt]
$100$ & $50$ & $1294.$ & \uncer{-13.2}{+15.0} & \footnotesize{$-50.7$} & \uncer{-0.3}{+0.3} & $1153.$ & \uncer{-15.7}{+8.6} & \uncer{-0.3}{+0.3}\\[-1.5pt]\hdashline[1.5pt/1.5pt]
$200$ & $5$ & $0.570$ & \uncer{-9.9}{+9.3} & \footnotesize{$-5.4$} & \uncer{\zerodown}{\zeroup} & $0.963$ & \uncer{-5.9}{+4.5} & \uncer{\zerodown}{\zeroup}\\[-1.5pt]\hdashline[1.5pt/1.5pt]
$200$ & $10$ & $1.039$ & \uncer{-11.7}{+12.4} & \footnotesize{$-56.6$} & \uncer{\zerodown}{\zeroup} & $3.863$ & \uncer{-5.9}{+4.5} & \uncer{\zerodown}{\zeroup}\\
\hline
\end{tabular}
\end{center}
\caption{\appendixtablecap{pseudoscalar}{$\mh^{\rm mod-}$}}
\label{tab:A_mhmodm}
\end{table}

\begin{table}[htp]
\begin{center}
\begin{tabular}{|c|c|c||c|r|r|r||c|r|r|}
\hline \firstlinelight \hline\hline
$100$ & $5$ & $88.1$ & $18.27$ & \uncer{-14.6}{+17.4} & \footnotesize{$-37.3$} & \uncer{-0.8}{+0.8} & $12.88$ & \uncer{-18.8}{+9.8} & \uncer{-1.1}{+1.1}\\[-1.5pt]\hdashline[1.5pt/1.5pt]
$100$ & $10$ & $94.3$ & $41.55$ & \uncer{-14.4}{+17.3} & \footnotesize{$-52.2$} & \uncer{-2.1}{+2.2} & $39.99$ & \uncer{-17.1}{+9.1} & \uncer{-2.1}{+2.2}\\[-1.5pt]\hdashline[1.5pt/1.5pt]
$100$ & $15$ & $95.8$ & $81.16$ & \uncer{-14.1}{+16.7} & \footnotesize{$-53.6$} & \uncer{-3.1}{+3.3} & $79.26$ & \uncer{-16.7}{+9.0} & \uncer{-3.0}{+3.1}\\[-1.5pt]\hdashline[1.5pt/1.5pt]
$100$ & $20$ & $96.4$ & $130.2$ & \uncer{-13.9}{+16.3} & \footnotesize{$-53.7$} & \uncer{-4.0}{+4.2} & $126.6$ & \uncer{-16.6}{+8.9} & \uncer{-3.8}{+4.0}\\[-1.5pt]\hdashline[1.5pt/1.5pt]
$100$ & $30$ & $96.8$ & $243.2$ & \uncer{-13.7}{+15.9} & \footnotesize{$-53.5$} & \uncer{-5.4}{+5.8} & $235.0$ & \uncer{-16.4}{+8.9} & \uncer{-5.2}{+5.6}\\[-1.5pt]\hdashline[1.5pt/1.5pt]
$100$ & $50$ & $97.0$ & $484.8$ & \uncer{-13.5}{+15.5} & \footnotesize{$-53.4$} & \uncer{-7.5}{+8.4} & $468.1$ & \uncer{-16.4}{+8.8} & \uncer{-7.2}{+8.1}\\[-1.5pt]\hdashline[1.5pt/1.5pt]
$120$ & $5$ & $101.2$ & $12.97$ & \uncer{-12.4}{+13.5} & \footnotesize{$-13.9$} & \uncer{-0.3}{+0.3} & $6.357$ & \uncer{-15.4}{+8.4} & \uncer{-1.0}{+1.0}\\[-1.5pt]\hdashline[1.5pt/1.5pt]
$120$ & $10$ & $110.5$ & $16.39$ & \uncer{-13.3}{+15.3} & \footnotesize{$-35.5$} & \uncer{-1.4}{+1.5} & $17.64$ & \uncer{-13.6}{+7.7} & \uncer{-2.0}{+2.1}\\
\hline
\end{tabular}
\end{center}
\caption{\appendixtablecap{lightest-scalar}{{\em light-stop}}}
\label{tab:LH_lightstop}
\end{table}
\begin{table}[htp]
\begin{center}
\begin{tabular}{|c|c|c||c|r|r|r||c|r|r|}
\hline \firstlineheavy \hline\hline
$100$ & $5$ & $130.9$ & $13.85$ & \uncer{-8.4}{+5.9} & \footnotesize{$-5.0$} & \uncer{-0.2}{+0.2} & $0.650$ & \uncer{-10.7}{+6.5} & \uncer{-1.7}{+1.8}\\[-1.5pt]\hdashline[1.5pt/1.5pt]
$100$ & $10$ & $125.9$ & $17.64$ & \uncer{-8.7}{+6.4} & \footnotesize{$-6.5$} & \uncer{-0.4}{+0.4} & $1.163$ & \uncer{-11.3}{+6.8} & \uncer{-3.0}{+3.2}\\[-1.5pt]\hdashline[1.5pt/1.5pt]
$100$ & $15$ & $124.7$ & $19.01$ & \uncer{-8.8}{+6.6} & \footnotesize{$-6.7$} & \uncer{-0.5}{+0.5} & $1.287$ & \uncer{-11.5}{+6.9} & \uncer{-4.3}{+4.5}\\[-1.5pt]\hdashline[1.5pt/1.5pt]
$100$ & $20$ & $124.3$ & $19.53$ & \uncer{-8.8}{+6.6} & \footnotesize{$-6.7$} & \uncer{-0.6}{+0.7} & $1.283$ & \uncer{-11.5}{+6.9} & \uncer{-5.4}{+5.8}\\[-1.5pt]\hdashline[1.5pt/1.5pt]
$100$ & $30$ & $124.0$ & $19.79$ & \uncer{-8.8}{+6.6} & \footnotesize{$-6.7$} & \uncer{-0.8}{+0.9} & $1.184$ & \uncer{-11.6}{+6.9} & \uncer{-7.4}{+8.1}\\[-1.5pt]\hdashline[1.5pt/1.5pt]
$100$ & $50$ & $123.9$ & $19.64$ & \uncer{-8.8}{+6.6} & \footnotesize{$-6.8$} & \uncer{-1.1}{+1.2} & $0.982$ & \uncer{-11.6}{+6.9} & \uncer{-10.5}{+12.0}\\[-1.5pt]\hdashline[1.5pt/1.5pt]
$200$ & $5$ & $203.7$ & $0.919$ & \uncer{-7.3}{+4.6} & \footnotesize{$-23.8$} & \uncer{-0.5}{+0.5} & $0.715$ & \uncer{-5.7}{+4.4} & \uncer{-1.3}{+1.3}\\[-1.5pt]\hdashline[1.5pt/1.5pt]
$200$ & $10$ & $199.9$ & $1.266$ & \uncer{-9.7}{+8.9} & \footnotesize{$-48.9$} & \uncer{-1.9}{+2.0} & $2.943$ & \uncer{-5.9}{+4.5} & \uncer{-2.2}{+2.3}\\
\hline
\end{tabular}
\end{center}
\caption{\appendixtablecap{heaviest-scalar}{{\em light-stop}}}
\label{tab:HH_lightstop}
\end{table}
\begin{table}[htp]
\begin{center}
\begin{tabular}{|c|c||c|r|r|r||c|r|r|}
\hline \firstlinepseudo \hline\hline
$100$ & $5$ & $9.685$ & \uncer{-13.5}{+15.5} & \footnotesize{$-47.8$} & \uncer{-1.2}{+1.2} & $9.841$ & \uncer{-15.7}{+8.6} & \uncer{-1.2}{+1.2}\\[-1.5pt]\hdashline[1.5pt/1.5pt]
$100$ & $10$ & $36.48$ & \uncer{-13.5}{+15.5} & \footnotesize{$-54.1$} & \uncer{-2.3}{+2.4} & $35.29$ & \uncer{-15.7}{+8.6} & \uncer{-2.2}{+2.3}\\[-1.5pt]\hdashline[1.5pt/1.5pt]
$100$ & $15$ & $76.79$ & \uncer{-13.3}{+15.2} & \footnotesize{$-53.4$} & \uncer{-3.2}{+3.4} & $71.44$ & \uncer{-15.7}{+8.6} & \uncer{-3.1}{+3.2}\\[-1.5pt]\hdashline[1.5pt/1.5pt]
$100$ & $20$ & $125.2$ & \uncer{-13.2}{+15.0} & \footnotesize{$-53.1$} & \uncer{-4.0}{+4.2} & $114.8$ & \uncer{-15.7}{+8.6} & \uncer{-3.9}{+4.1}\\[-1.5pt]\hdashline[1.5pt/1.5pt]
$100$ & $30$ & $235.2$ & \uncer{-13.1}{+14.7} & \footnotesize{$-52.9$} & \uncer{-5.4}{+5.8} & $214.0$ & \uncer{-15.7}{+8.6} & \uncer{-5.2}{+5.6}\\[-1.5pt]\hdashline[1.5pt/1.5pt]
$100$ & $50$ & $467.6$ & \uncer{-12.9}{+14.3} & \footnotesize{$-52.9$} & \uncer{-7.5}{+8.4} & $426.8$ & \uncer{-15.7}{+8.6} & \uncer{-7.2}{+8.1}\\[-1.5pt]\hdashline[1.5pt/1.5pt]
$200$ & $5$ & $0.534$ & \uncer{-9.1}{+7.8} & \footnotesize{$-4.0$} & \uncer{-0.2}{+0.2} & $0.845$ & \uncer{-5.9}{+4.5} & \uncer{-1.2}{+1.2}\\[-1.5pt]\hdashline[1.5pt/1.5pt]
$200$ & $10$ & $0.808$ & \uncer{-11.9}{+12.7} & \footnotesize{$-57.8$} & \uncer{-2.4}{+2.4} & $3.030$ & \uncer{-5.9}{+4.5} & \uncer{-2.2}{+2.3}\\
\hline
\end{tabular}
\end{center}
\caption{\appendixtablecap{pseudoscalar}{{\em light-stop}}}
\label{tab:A_lightstop}
\end{table}

\begin{table}[htp]
\begin{center}
\begin{tabular}{|c|c|c||c|r|r|r||c|r|r|}
\hline \firstlinelight \hline\hline
$100$ & $5$ & $90.2$ & $16.60$ & \uncer{-14.5}{+17.4} & \footnotesize{$-38.3$} & \uncer{-0.5}{+0.5} & $12.49$ & \uncer{-18.2}{+9.5} & \uncer{-0.7}{+0.7}\\[-1.5pt]\hdashline[1.5pt/1.5pt]
$100$ & $10$ & $96.3$ & $40.47$ & \uncer{-14.3}{+17.1} & \footnotesize{$-52.7$} & \uncer{-1.4}{+1.5} & $40.40$ & \uncer{-16.6}{+8.9} & \uncer{-1.4}{+1.4}\\[-1.5pt]\hdashline[1.5pt/1.5pt]
$100$ & $15$ & $97.8$ & $82.65$ & \uncer{-14.0}{+16.5} & \footnotesize{$-53.6$} & \uncer{-2.1}{+2.2} & $82.90$ & \uncer{-16.2}{+8.8} & \uncer{-2.0}{+2.1}\\[-1.5pt]\hdashline[1.5pt/1.5pt]
$100$ & $20$ & $98.4$ & $137.3$ & \uncer{-13.8}{+16.2} & \footnotesize{$-53.5$} & \uncer{-2.7}{+2.8} & $136.7$ & \uncer{-16.1}{+8.7} & \uncer{-2.6}{+2.7}\\[-1.5pt]\hdashline[1.5pt/1.5pt]
$100$ & $30$ & $98.9$ & $272.2$ & \uncer{-13.7}{+15.9} & \footnotesize{$-53.2$} & \uncer{-3.8}{+4.0} & $268.7$ & \uncer{-15.9}{+8.7} & \uncer{-3.7}{+3.9}\\[-1.5pt]\hdashline[1.5pt/1.5pt]
$100$ & $50$ & $99.2$ & $596.9$ & \uncer{-13.5}{+15.6} & \footnotesize{$-53.0$} & \uncer{-5.5}{+6.0} & $587.1$ & \uncer{-15.9}{+8.6} & \uncer{-5.4}{+5.9}\\[-1.5pt]\hdashline[1.5pt/1.5pt]
$120$ & $5$ & $103.2$ & $12.37$ & \uncer{-12.6}{+14.0} & \footnotesize{$-13.9$} & \uncer{-0.2}{+0.2} & $6.361$ & \uncer{-15.0}{+8.3} & \uncer{-0.6}{+0.7}\\[-1.5pt]\hdashline[1.5pt/1.5pt]
$120$ & $10$ & $112.3$ & $16.25$ & \uncer{-13.5}{+15.6} & \footnotesize{$-37.6$} & \uncer{-1.0}{+1.0} & $18.67$ & \uncer{-13.3}{+7.6} & \uncer{-1.3}{+1.4}\\
\hline
\end{tabular}
\end{center}
\caption{\appendixtablecap{lightest-scalar}{{\em light-stau}}}
\label{tab:LH_lightstau}
\end{table}
\begin{table}[htp]
\begin{center}
\begin{tabular}{|c|c|c||c|r|r|r||c|r|r|}
\hline \firstlineheavy \hline\hline
$100$ & $5$ & $135.4$ & $16.97$ & \uncer{-9.3}{+7.9} & \footnotesize{$-4.2$} & \uncer{\zerodown}{\zeroup} & $0.614$ & \uncer{-10.2}{+6.3} & \uncer{-1.1}{+1.1}\\[-1.5pt]\hdashline[1.5pt/1.5pt]
$100$ & $10$ & $130.0$ & $20.91$ & \uncer{-9.5}{+8.1} & \footnotesize{$-5.8$} & \uncer{-0.2}{+0.2} & $1.228$ & \uncer{-10.8}{+6.6} & \uncer{-2.0}{+2.0}\\[-1.5pt]\hdashline[1.5pt/1.5pt]
$100$ & $15$ & $128.6$ & $22.49$ & \uncer{-9.6}{+8.2} & \footnotesize{$-6.5$} & \uncer{-0.3}{+0.3} & $1.523$ & \uncer{-11.0}{+6.6} & \uncer{-2.8}{+2.9}\\[-1.5pt]\hdashline[1.5pt/1.5pt]
$100$ & $20$ & $128.1$ & $23.25$ & \uncer{-9.6}{+8.2} & \footnotesize{$-6.8$} & \uncer{-0.4}{+0.5} & $1.689$ & \uncer{-11.0}{+6.7} & \uncer{-3.5}{+3.7}\\[-1.5pt]\hdashline[1.5pt/1.5pt]
$100$ & $30$ & $127.7$ & $23.98$ & \uncer{-9.6}{+8.3} & \footnotesize{$-7.3$} & \uncer{-0.6}{+0.7} & $1.890$ & \uncer{-11.1}{+6.7} & \uncer{-4.9}{+5.2}\\[-1.5pt]\hdashline[1.5pt/1.5pt]
$100$ & $50$ & $127.4$ & $24.77$ & \uncer{-9.7}{+8.3} & \footnotesize{$-8.2$} & \uncer{-1.0}{+1.1} & $2.199$ & \uncer{-11.1}{+6.7} & \uncer{-7.0}{+7.7}\\[-1.5pt]\hdashline[1.5pt/1.5pt]
$200$ & $5$ & $205.7$ & $1.494$ & \uncer{-8.6}{+6.7} & \footnotesize{$-17.1$} & \uncer{-0.2}{+0.2} & $0.713$ & \uncer{-5.6}{+4.4} & \uncer{-0.8}{+0.8}\\[-1.5pt]\hdashline[1.5pt/1.5pt]
$200$ & $10$ & $201.5$ & $1.739$ & \uncer{-10.0}{+9.3} & \footnotesize{$-41.2$} & \uncer{-1.1}{+1.1} & $3.090$ & \uncer{-5.8}{+4.5} & \uncer{-1.5}{+1.5}\\
\hline
\end{tabular}
\end{center}
\caption{\appendixtablecap{heaviest-scalar}{{\em light-stau}}}
\label{tab:HH_lightstau}
\end{table}
\begin{table}[htp]
\begin{center}
\begin{tabular}{|c|c||c|r|r|r||c|r|r|}
\hline \firstlinepseudo \hline\hline
$100$ & $5$ & $10.13$ & \uncer{-13.5}{+15.6} & \footnotesize{$-47.7$} & \uncer{-0.8}{+0.8} & $10.30$ & \uncer{-15.7}{+8.6} & \uncer{-0.8}{+0.8}\\[-1.5pt]\hdashline[1.5pt/1.5pt]
$100$ & $10$ & $39.73$ & \uncer{-13.5}{+15.5} & \footnotesize{$-53.8$} & \uncer{-1.5}{+1.6} & $38.34$ & \uncer{-15.7}{+8.6} & \uncer{-1.5}{+1.5}\\[-1.5pt]\hdashline[1.5pt/1.5pt]
$100$ & $15$ & $86.71$ & \uncer{-13.3}{+15.2} & \footnotesize{$-53.2$} & \uncer{-2.2}{+2.2} & $80.38$ & \uncer{-15.7}{+8.6} & \uncer{-2.1}{+2.2}\\[-1.5pt]\hdashline[1.5pt/1.5pt]
$100$ & $20$ & $146.1$ & \uncer{-13.2}{+15.0} & \footnotesize{$-52.9$} & \uncer{-2.7}{+2.9} & $133.4$ & \uncer{-15.7}{+8.6} & \uncer{-2.7}{+2.8}\\[-1.5pt]\hdashline[1.5pt/1.5pt]
$100$ & $30$ & $291.4$ & \uncer{-13.2}{+14.9} & \footnotesize{$-52.6$} & \uncer{-3.8}{+4.0} & $263.4$ & \uncer{-15.7}{+8.6} & \uncer{-3.7}{+3.9}\\[-1.5pt]\hdashline[1.5pt/1.5pt]
$100$ & $50$ & $638.7$ & \uncer{-13.1}{+14.7} & \footnotesize{$-52.5$} & \uncer{-5.5}{+6.0} & $576.3$ & \uncer{-15.7}{+8.6} & \uncer{-5.4}{+5.9}\\[-1.5pt]\hdashline[1.5pt/1.5pt]
$200$ & $5$ & $0.550$ & \uncer{-9.5}{+8.5} & \footnotesize{$-4.2$} & \uncer{-0.1}{+0.1} & $0.884$ & \uncer{-5.9}{+4.5} & \uncer{-0.8}{+0.8}\\[-1.5pt]\hdashline[1.5pt/1.5pt]
$200$ & $10$ & $0.879$ & \uncer{-11.8}{+12.5} & \footnotesize{$-57.6$} & \uncer{-1.6}{+1.6} & $3.292$ & \uncer{-5.9}{+4.5} & \uncer{-1.5}{+1.5}\\
\hline
\end{tabular}
\end{center}
\caption{\appendixtablecap{pseudoscalar}{{\em light-stau}}}
\label{tab:A_lightstau}
\end{table}

\begin{table}[htp]
\begin{center}
\begin{tabular}{|c|c|c||c|r|r|r||c|r|r|}
\hline \firstlinelight \hline\hline
$100$ & $5$ & $85.5$ & $19.70$ & \uncer{-14.5}{+17.4} & \footnotesize{$-52.8$} & \uncer{-1.8}{+1.9} & $15.57$ & \uncer{-19.7}{+10.1} & \uncer{-1.7}{+1.8}\\[-1.5pt]\hdashline[1.5pt/1.5pt]
$100$ & $10$ & $86.5$ & $62.66$ & \uncer{-14.3}{+16.9} & \footnotesize{$-52.9$} & \uncer{-3.3}{+3.4} & $49.53$ & \uncer{-19.3}{+10.0} & \uncer{-3.1}{+3.3}\\[-1.5pt]\hdashline[1.5pt/1.5pt]
$100$ & $15$ & $86.0$ & $121.0$ & \uncer{-14.3}{+16.9} & \footnotesize{$-52.9$} & \uncer{-4.5}{+4.8} & $95.38$ & \uncer{-19.5}{+10.0} & \uncer{-4.3}{+4.6}\\[-1.5pt]\hdashline[1.5pt/1.5pt]
$100$ & $20$ & $85.7$ & $187.9$ & \uncer{-14.2}{+16.8} & \footnotesize{$-52.9$} & \uncer{-5.5}{+6.0} & $147.3$ & \uncer{-19.6}{+10.1} & \uncer{-5.3}{+5.7}\\[-1.5pt]\hdashline[1.5pt/1.5pt]
$100$ & $30$ & $85.2$ & $329.9$ & \uncer{-14.1}{+16.6} & \footnotesize{$-52.8$} & \uncer{-7.2}{+8.1} & $257.0$ & \uncer{-19.8}{+10.2} & \uncer{-6.9}{+7.7}\\[-1.5pt]\hdashline[1.5pt/1.5pt]
$100$ & $40$ & $84.9$ & $468.4$ & \uncer{-14.0}{+16.3} & \footnotesize{$-52.8$} & \uncer{-8.5}{+9.7} & $364.3$ & \uncer{-19.9}{+10.2} & \uncer{-8.1}{+9.2}\\[-1.5pt]\hdashline[1.5pt/1.5pt]
$120$ & $5$ & $105.6$ & $7.186$ & \uncer{-13.9}{+16.4} & \footnotesize{$-37.7$} & \uncer{-1.2}{+1.3} & $7.355$ & \uncer{-14.5}{+8.1} & \uncer{-1.6}{+1.7}\\[-1.5pt]\hdashline[1.5pt/1.5pt]
$120$ & $10$ & $109.0$ & $20.00$ & \uncer{-13.2}{+15.1} & \footnotesize{$-54.5$} & \uncer{-3.3}{+3.5} & $23.99$ & \uncer{-13.9}{+7.8} & \uncer{-3.1}{+3.3}\\
\hline
\end{tabular}
\end{center}
\caption{\appendixtablecap{lightest-scalar}{{\em tau-phobic}}}
\label{tab:LH_tauphobic}
\end{table}
\begin{table}[htp]
\begin{center}
\begin{tabular}{|c|c|c||c|r|r|r||c|r|r|}
\hline \firstlineheavy \hline\hline
$100$ & $5$ & $124.1$ & $21.05$ & \uncer{-9.6}{+8.1} & \footnotesize{$-0.6$} & \uncer{\zerodown}{\zeroup} & $20.5\shexp$ & \uncer{-11.6}{+6.9} & \uncer{-7.5}{+7.9}\\[-1.5pt]\hdashline[1.5pt/1.5pt]
$100$ & $10$ & $124.9$ & $20.60$ & \uncer{-9.6}{+8.1} & \footnotesize{$-1.1$} & \uncer{-0.2}{+0.2} & $55.1\shexp$ & \uncer{-11.4}{+6.8} & \uncer{-9.3}{+10.1}\\[-1.5pt]\hdashline[1.5pt/1.5pt]
$100$ & $15$ & $125.8$ & $21.16$ & \uncer{-9.5}{+8.0} & \footnotesize{$-3.2$} & \uncer{-0.4}{+0.4} & $0.394$ & \uncer{-11.3}{+6.8} & \uncer{-7.4}{+8.1}\\[-1.5pt]\hdashline[1.5pt/1.5pt]
$100$ & $20$ & $126.3$ & $21.94$ & \uncer{-9.5}{+8.0} & \footnotesize{$-5.3$} & \uncer{-0.7}{+0.7} & $0.913$ & \uncer{-11.3}{+6.8} & \uncer{-7.8}{+8.6}\\[-1.5pt]\hdashline[1.5pt/1.5pt]
$100$ & $30$ & $126.6$ & $23.85$ & \uncer{-9.5}{+8.0} & \footnotesize{$-9.1$} & \uncer{-1.3}{+1.5} & $2.251$ & \uncer{-11.2}{+6.8} & \uncer{-9.0}{+10.1}\\[-1.5pt]\hdashline[1.5pt/1.5pt]
$100$ & $40$ & $125.5$ & $26.57$ & \uncer{-9.6}{+8.1} & \footnotesize{$-13.0$} & \uncer{-2.1}{+2.4} & $4.121$ & \uncer{-11.4}{+6.8} & \uncer{-9.9}{+11.4}\\[-1.5pt]\hdashline[1.5pt/1.5pt]
$200$ & $5$ & $194.9$ & $1.152$ & \uncer{-8.0}{+6.0} & \footnotesize{$-23.5$} & \uncer{-0.8}{+0.8} & $0.831$ & \uncer{-6.1}{+4.6} & \uncer{-1.9}{+1.9}\\[-1.5pt]\hdashline[1.5pt/1.5pt]
$200$ & $10$ & $193.5$ & $1.303$ & \uncer{-10.1}{+9.6} & \footnotesize{$-50.4$} & \uncer{-2.9}{+3.1} & $3.044$ & \uncer{-6.2}{+4.6} & \uncer{-3.2}{+3.4}\\
\hline
\end{tabular}
\end{center}
\caption{\appendixtablecap{heaviest-scalar}{{\em tau-phobic}}}
\label{tab:HH_tauphobic}
\end{table}
\begin{table}[htp]
\begin{center}
\begin{tabular}{|c|c||c|r|r|r||c|r|r|}
\hline \firstlinepseudo \hline\hline
$100$ & $5$ & $9.055$ & \uncer{-13.4}{+15.4} & \footnotesize{$-47.8$} & \uncer{-1.7}{+1.8} & $9.194$ & \uncer{-15.7}{+8.6} & \uncer{-1.8}{+1.9}\\[-1.5pt]\hdashline[1.5pt/1.5pt]
$100$ & $10$ & $32.17$ & \uncer{-13.5}{+15.5} & \footnotesize{$-54.3$} & \uncer{-3.4}{+3.6} & $31.25$ & \uncer{-15.7}{+8.6} & \uncer{-3.2}{+3.4}\\[-1.5pt]\hdashline[1.5pt/1.5pt]
$100$ & $15$ & $64.46$ & \uncer{-13.3}{+15.1} & \footnotesize{$-53.7$} & \uncer{-4.6}{+4.9} & $60.29$ & \uncer{-15.7}{+8.6} & \uncer{-4.4}{+4.7}\\[-1.5pt]\hdashline[1.5pt/1.5pt]
$100$ & $20$ & $100.6$ & \uncer{-13.2}{+14.9} & \footnotesize{$-53.4$} & \uncer{-5.6}{+6.1} & $92.87$ & \uncer{-15.7}{+8.6} & \uncer{-5.4}{+5.8}\\[-1.5pt]\hdashline[1.5pt/1.5pt]
$100$ & $30$ & $175.3$ & \uncer{-13.0}{+14.5} & \footnotesize{$-53.2$} & \uncer{-7.2}{+8.1} & $161.2$ & \uncer{-15.7}{+8.6} & \uncer{-7.0}{+7.8}\\[-1.5pt]\hdashline[1.5pt/1.5pt]
$100$ & $40$ & $246.7$ & \uncer{-12.8}{+14.1} & \footnotesize{$-53.2$} & \uncer{-8.5}{+9.7} & $227.6$ & \uncer{-15.7}{+8.6} & \uncer{-8.2}{+9.3}\\[-1.5pt]\hdashline[1.5pt/1.5pt]
$200$ & $5$ & $0.510$ & \uncer{-8.5}{+6.7} & \footnotesize{$-3.7$} & \uncer{-0.2}{+0.2} & $0.790$ & \uncer{-5.9}{+4.5} & \uncer{-1.8}{+1.9}\\[-1.5pt]\hdashline[1.5pt/1.5pt]
$200$ & $10$ & $0.712$ & \uncer{-12.0}{+12.9} & \footnotesize{$-58.0$} & \uncer{-3.4}{+3.6} & $2.683$ & \uncer{-5.9}{+4.5} & \uncer{-3.2}{+3.4}\\
\hline
\end{tabular}
\end{center}
\caption{\appendixtablecap{pseudoscalar}{{\em tau-phobic}}}
\label{tab:A_tauphobic}
\end{table}



\end{document}